\documentclass{aa}  
\usepackage{graphicx}
\usepackage{grffile}

\usepackage{txfonts}
\usepackage{hyperref}
\usepackage[section]{placeins}
\usepackage{float}
\begin{document} 

   \title{Radio source-component association for the LOFAR Two-metre Sky Survey with region-based convolutional neural networks}
\authorrunning{Mostert et al.}
\titlerunning{Automated radio source-component association}
   
\author{Rafa\"el I.J. Mostert\thanks{E-mail: mostert@strw.leidenuniv.nl}\inst{1,2,3} \and
Kenneth J. Duncan\inst{1,4}\and
Lara Alegre\inst{4}\and
Huub J.A. R\"{o}ttgering\inst{1}\and
Wendy L. Williams\inst{1}\and
Philip N. Best\inst{4}\and
Martin J. Hardcastle\inst{5}\and
Raffaella Morganti\inst{2,6}}

\institute{Leiden Observatory, Leiden University, PO Box 9513, NL-2300 RA Leiden, The Netherlands \and
ASTRON, the Netherlands Institute for Radio Astronomy, Oude Hoogeveensedijk 4, 7991 PD Dwingeloo, The Netherlands \and
Leiden Institute of Advanced Computer Science, Niels Bohrweg 1, Leiden, The Netherlands \and
SUPA, Institute for Astronomy, Royal Observatory, Blackford Hill, Edinburgh, EH9 3HJ, UK \and
Centre for Astrophysics Research, Department of Physics, Astronomy and Mathematics, University of Hertfordshire, College Lane, Hatfield AL10 9AB, UK  \and
Kapteyn Astronomical Institute, University of Groningen, P.O. Box 800, 9700 AV Groningen, The Netherlands }

   \date{Received 04/03/2022; accepted 12/09/2022}

  \abstract
   {Radio loud active galactic nuclei (RLAGNs) are often morphologically complex objects that can consist of multiple, spatially separated, components.
   Only when the spatially separated radio components are correctly grouped together can we start to look for the corresponding optical host galaxy and infer physical parameters such as the size and luminosity of the radio object.
Existing radio detection software to group these spatially separated components together is either experimental or based on assumptions that do not hold for current generation surveys, such that, in practise, astronomers often rely on visual inspection to resolve radio component association.
      However, applying visual inspection to all the hundreds of thousands of well-resolved RLAGNs that appear in the images from the Low Frequency Array (LOFAR) Two-metre Sky Survey (LoTSS) at $144$ MHz, is a daunting, time-consuming process, even with extensive manpower.}
   {Using a machine learning approach, we aim to automate the radio component association of large ($> 15$ arcsec)  radio components.}
   {We turned the association problem into a classification problem and trained an adapted Fast region-based convolutional neural network to mimic the expert annotations from the first LoTSS data release. We implemented a rotation data augmentation to reduce overfitting and simplify the component association by removing unresolved radio sources that are likely unrelated to the large and bright radio components that we consider using predictions from an existing gradient boosting classifier.}
   {For  large ($> 15$ arcsec) and bright ($> 10$ mJy) radio components in the LoTSS first data release, our model provides the same associations  for $85.3\%\pm0.6$ of the cases as those derived when astronomers perform the association manually.
   When the association is done through public crowd-sourced efforts, a result similar to that of our model is attained.}
   {Our method is able to efficiently carry out manual radio-component association for huge radio surveys and can serve as a basis for either automated radio morphology classification or automated optical host identification. 
   This opens up an avenue to study the completeness and reliability of samples of radio sources with extended, complex morphologies.}
   

   \maketitle
%

\section{Introduction}
\label{sec:intro} 
In the low-frequency radio regime, most objects we observe are either radio-loud active galactic nuclei (RLAGNs) or star forming galaxies \citep[e.g.][]{wilman2008semi}.
The RLAGNs are often morphologically complex objects that can consist of multiple components, such as a core, jets, hotspots, and lobes \citep[e.g. ][]{Miley1980,hardcastle2020radio}.
Due to the spectral properties of the components, we observe large variations in frequency in the relative apparent brightness of the components \citep[e.g.][]{alexander1987ageing, Harwood2013}. 
As a result, we do not always observe all components of a RLAGN, and different components of the same RLAGN can appear spatially separated on the sky.
The steep-spectrum lobes of an edge-brightened or Fanaroff-Riley type II radio source \citep[FRII;][]{Fanaroff1974} might, for example, be observable, while the flat-spectrum emission connecting the two lobes, from the jets and the core falls below the noise. These separate radio RLAGN components need to be grouped together before we can start to look for the corresponding host galaxy in the optical or infrared and infer the radio object's physical parameters including the proper size and luminosity \citep{Williams2019}. 

Commonly used (radio) source detection software is not designed to group spatially separated components together.
Existing source finders -- such as the Python Blob Detection and Source Finder\footnote{\url{https://github.com/lofar-astron/PyBDSF}} \citep[\textit{PyBDSF};][]{Mohan2015}, \textit{AEGEAN} \citep[][]{hancock2012compact,2018PASA...35...11H}, and \textit{ProFound} \citep[][]{robotham2018profound} -- are designed as robust rule-based algorithms to detect patches of contiguous (radio) emission that surpass the local noise by a certain threshold and deblend them if necessary. 
If two related patches of radio emission (for example, two lobes originating from a single RLAGN) are spatially separated and the connecting radio emission (from the jets and or lobes) falls below a user-defined signal-to-noise threshold, they will be treated as two different radio sources. 
Even when the emission connecting different components of a resolved radio object does not fall below the noise level, components are sometimes erroneously deblended into multiple sources. 
This is a conscious trade-off used in existing source detection software to prevent the association of spurious unrelated radio emission, and it can be partly overcome through subsequent manual visual inspection.
However, applying this visual inspection to hundreds of thousands of well-resolved RLAGNs is a daunting and time-consuming process, even when delegated to multiple astronomers.
For the Low Frequency Array \citep[LOFAR;][]{vanHaarlem2013} Two-metre Sky Survey first data release \citep[LoTSS-DR1;][]{Shimwell2017}, $\sim15,000$ extended sources ($4.9\%$ of the total number of sources in LoTSS-DR1) were manually associated through visual inspection over the course of eight months by 66 astronomers who were part of the LOFAR collaboration \citep{Williams2019}.
Over the course of the inspection of thousands of objects, mistakes are easily made and certain radio emission can be faint or complex to untangle.
Therefore, the LOFAR consortium required each source to be annotated by five different astronomers, thereby increasing the required time investment of all involved.

Even a public crowd-sourced annotation process with thousands of volunteers to associate radio components has its limitations.
The sky area and the number of sources requiring visual inspection increased tenfold for the second LoTSS data release \citep[LoTSS-DR2;][]{Shimwell2022} and internal manual classification was deemed infeasible because of this.
A public version of the crowd-sourced manual annotation platform used for LoTSS-DR1 was therefore created (Hardcastle et al. in prep.). 
In six months, $\sim80,000$ sources will have been annotated by five different people.
Fully annotating LoTSS-DR2 will likely take more than a year.
Unsurprisingly, the time it takes for the public to annotate sources (and for the astronomers to guide the project) is hard to predict. 
Furthermore, the annotation quality is hard to monitor directly.
It can be improved indirectly, by enhancing the tutorial and introduction on the platform or by requiring more views per radio component, but the latter comes at the cost of a decreased rate of completed annotations \citep[e.g.][]{spacewarp,Williams2019}. 
For further LOFAR data releases and future large-scale sky surveys from the Square Kilometre Array \citep[SKA;][]{2015aska.confE.174B} and its pathfinders, relying solely on crowd-sourced annotation is clearly unsustainable and undesirable. 
An automated approach is needed.
In the future, automated radio source component association will be an essential step in the study of the completeness and reliability of extended objects such as FRI and FRII in existing and upcoming large-scale sky surveys.

In this work, we aim to create an automated pipeline that works well on most large ($> 15$ arcsec) radio components at MHz frequencies. 
Of all large radio components, $28\%$ are part of a multi-component source.
Large and bright ($>10$ mJy) radio components are more often part of a multi-component source than large and faint radio components ($41\%$ versus $21\%,$ respectively).
Of all sources that \textit{PyBDSF} finds in LoTSS-DR1, $94\%$ are smaller than $15$ arcsec.
Of these small sources, $95.8\%$ are unresolved, and $3.7\%$ are slightly resolved but correctly associated by \textit{PyBDSF}. 
Only $0.5\%$ of the sources smaller than $15$ arcsec are part of a multi-component radio source, and in many such cases ($48.7\%$) one of the other components of that source is larger than $15$ arcsec, and so the source can be associated with our pipeline.

The specific $15$ arcsec cut coincides with the cut above which \citet{Williams2019} required most components to be manually associated by professional astronomers and therefore provides us with a large training set.  
We procedurally combine the possibility to learn these manual component associations using a convolutional neural network as demonstrated by Chen \citet[][see Section \ref{sec:existing}]{Wu2019}, with the better completeness of a rule-based emission detection algorithm and various augmentations that make the network more suited to associating radio components. 
It is unrealistic to expect any algorithm to perform perfectly on all extended radio emission, as the observed radio emission in LoTSS can be faint and too complex even for a radio astronomer to associate and our method does not make use of potential optical host information available to researchers during visual inspection. 
We expect our method to work less well for sources with low signal-to-noise ratios and sources located in cluster environments, as radio-lobes in those environments can be greatly distorted and superposed onto a cluster halo or relic emission. 

In the next section, we highlight past automated approaches to the radio component association problem. In Section \ref{sec:data}, we lay out the radio survey and the manual annotation process on which our automated method explained in Section \ref{sec:methods} is based.  
We present our results in Section \ref{sec:results} and we elaborate on the scope and limitations of our method in Section \ref{sec:discussion}. 
Finally, in the same section, we discuss how our pipeline can aid the development of automated morphological classification pipelines and cross-identification pipelines.

\section{Existing automated radio component association approaches}
\label{sec:existing}
In the past, there have been multiple rule-based attempts to perform the task of associating radio components.
\citet{van2015nature} assumed that FRII radio sources at $z \approx 1$ and $1.4$ GH, appear as two unresolved point sources in images of the Faint Images of the Radio Sky at Twenty-cm \citep[FIRST;][]{White1997} survey. 
They proceeded to match all radio blobs in the FIRST catalogue with a minimum separation of $18$ arcsec and a maximum separation of $1$ arcmin. The authors noted that this method only works for FRII with large ($>100$ kpc) radio lobes. The association of unrelated, chance-aligned radio sources using this method is deemed unavoidable. 
\citet{fan2015matching} combined source association with cross-identification using Bayesian hypothesis testing. 
They searched for radio components in the Australia Telescope Large Area Survey \citep[ATLAS;][]{norris2006deep} that lie within $2$ arcmin of a source in the Spitzer Wide-Area Infrared Extragalactic Survey \citep[SWIRE;][]{lonsdale2003swire} and tested the likelihood of the association of the radio components and cross-identification with the infrared source. 
They assumed a radio source to consist either of a core with a pair of lobes, just a core, or just a pair of lobes, and adopted a Rayleigh distribution with a mean of $9$ arcsec as the prior probability distribution function of possible core-lobe distances. As expected, the likelihood method works best for cross-identifying infrared sources with single-core objects  (finding $536$ out of $558$ such sources in common with a manual cross-identification), reasonably well for triplets ($9$ out of $10$), and less well for doublet radio sources ($19$ out of $27$). 
Unfortunately, the assumption of unresolved point-like source-components required by these simple parametric models does not hold for extended sources in LoTSS. 
The higher resolution of LoTSS at $\sim100$ MHz frequencies \citep{Shimwell2017} and the dense core of the LOFAR antennas allow for better surface brightness sensitivity, revealing complex morphologies for extended sources.

More recent attempts to perform the association task involve unsupervised and supervised machine learning.
In the domain of unsupervised machine learning, the introduction of a rotation and flipping invariant self-organised maps (SOM) code by \citet[PINK;][]{Polsterer2016} spurred work on morphological clustering of extragalactic radio sources \citep[e.g. ][]{galvin2019radio, ralph2019radio, Mostert2020}. 
Both \citet{galvin2019radio} and \citet{Mostert2020} speculated on the ability to associate the radio-emission from the simplest resolved radio objects on the sky: non-bent, double-lobed RLAGNs. 
\citet{galvin2020cataloging} trained an SOM with $40\times40$ neurons on 
images from FIRST and accompanying infrared images from the Widefield Infrared Survey Explorer \citep[WISE;][]{2010AJ....140.1868W} and demonstrated the ability to combine component-association and infrared cross-identification.
\citet{galvin2020cataloging} turned the common radio-emission morphologies, modelled in the neurons of an SOM, into segmented images. Each segmented image is manually annotated: the authors judge which segments in the image are likely to belong to the central radio component and which are not. 
In the inference phase, an image (from outside the training dataset) centred on a particular radio component is matched to the neuron that is morphologically most similar. 
The radio components of the image that fall within the neuron segments that were judged to belong together will be associated with each other. 
Although this approach is promising, \citet{galvin2020cataloging} did not quantify the performance of the radio component association. 
Changes to the SOM parameters or applying this technique to a different set of surveys, for example on LoTSS and the Panoramic Survey Telescope and Rapid Response System 1 $3\pi$ sterradian survey \citep[Pan-STARRS1;][]{panstarrs}, requires the retraining of the SOM. 
Moreover, on every such occasion one is required to manually re-annotate each of the neurons ($1600$ in the case of a $40\times40$ SOM) for the approach to work, making this method less appealing to us. 

Chen \citet{Wu2019} arguably produced the most promising supervised deep learning approach towards the association task, as their method is not based on a template and in theory allows for the association of a wide variety of differently shaped radio sources.
They show the possibility to detect radio emission by predicting rectangular boxes around radio emission contours based on a combined Stokes-I radio image from FIRST with an optical image from WISE  using a Faster region-based convolutional neural network \citep[Faster R-CNN;][]{Ren2015}. 
This is a supervised training process as the neural network improves and validates its performance based on a given `ground truth' region, which is drawn around the radio components that volunteers of the citizen science project Radio Galaxy Zoo \citep{Banfield2015} considered to belong together.
The approach was intended to replace rule-based detection software such as \textit{PyBDSF}, \textit{AEGEAN,} or \textit{ProFound} and detect both unresolved and resolved radio emission for the SKA data challenge.
Apart from radio emission detection and association, their network also predicts the number of components and the number of brightness peaks for each detected source.
However, the association of multi-component radio objects specifically still poses a challenge, according to the authors.
The part of their test set that contains more than one source per input achieves a mean average precision\footnote{Mean average precision is a single performance metric that combines precision and recall.} of $0.74$ out of $1$ for the $487$ single-component sources, $0.28$ out of $1$ for the $13$ dual-component sources, and $0.89$ out of $1$ for the five triple-component sources in the test set.
Dual-component sources with more than two brightness peaks, triple-component sources with more than three brightness peaks, and all sources consisting of more than three components were excluded from their training and test sets.
 This approach is fine for the FIRST survey, as different sources are often separated and decomposed in only two or three components.
For LoTSS, the source density is higher than that in FIRST, leading to more closely neighbouring unrelated emission, and the surface brightness sensitivity of LoTSS is also higher than that of FIRST, resulting in the detection of more components and more brightness peaks per source.

More work focussing specifically on the large and extended objects is thus needed.
Indeed, in a recent review of the techniques used in the first SKA data challenge \citep{bonaldi2018square}, \citet{Bonaldi2021} concluded that the ability to deal with highly resolved radio sources in surveys as effectively as the unresolved source population is an outstanding challenge.

\section{Data}
\label{sec:data}

The survey we use in this work, LoTSS, is being carried out using the high-band (120-168 MHz) antennas of LOFAR and will eventually cover the entire northern sky.
The first data release covers $424$ square degrees (right ascension 10h45m00s to 15h30m00s and declination $45^{\circ}$ to $57^{\circ}$),
with a resolution of $6\arcsec$ and a median sensitivity of $S_{144MHz} = 71 \mu$Jy beam$^{-1}$.
It is accompanied by multiple radio catalogues. 
\citet{Williams2019} describe how the \textit{PyBDSF} source detection software was applied to the radio intensity images in LoTSS-DR1 to find $325,694$ radio components (the majority of which are unresolved). 
The subset of components deemed to require manual association were manually associated by a group of 66 radio astronomers (see Section \ref{sec:manual_process}). 
After association, the final source catalogue contained $318,520$ radio sources.

The second data release \citep[LoTSS-DR2;][]{Shimwell2022}\footnote{LoTSS-DR2 release page: \url{https://lofar-surveys.org/dr2_release.html}.} includes the full LoTSS-DR1 region and covers an area of $5,720$ square degrees. 
It consists of two discrete fields that avoid both the Milky Way and low declinations, denoting the 0h and 13h fields with a resolution of $6\arcsec$ and a median sensitivity of $83 \mu$Jy beam$^{-1}$.
The dynamic range of images in LoTSS-DR2 is approximately two times better than those in LoTSS-DR1 \citep{Shimwell2022}. 
As with LoTSS-DR1, the second data release is accompanied by a radio component catalogue. 
A subset of the $4,395,448$ detected radio components (outside of the LoTSS-DR1 region) will be associated with unique radio sources by a crowd of lay volunteers, this manual association process is ongoing and will be described in a future publication.
Therefore, we made use of the LoTSS-DR1 catalogue and the LoTSS-DR2 images, for training, testing and validation. The LoTSS-DR1 observed region is divided over 58 observed pointings, from which we randomly picked 38 to be used for training our network; we used ten different pointings as `validations' to assess and choose different design implementations and settings, and we used ten more pointings for `testing' to assess the final performance of our trained network. 
We also randomly selected ten LoTSS-DR2 pointings (outside of the LoTSS-DR1 area) to assess the performance of our trained network when compared to the publicly crowd-sourced DR2 catalogue for these pointings (see Section \ref{sec:public_zoo}). 

\subsection{Selection of source components}
\label{sec:manual_scope}
As most radio sources observed in large-scale sky surveys are unresolved and isolated, it follows that most radio components do not require (manual) association with other radio components.
For filtering, which for the $325,694$ detected radio components in LoTSS-DR1 would require manual association, \citet{Williams2019} used a hand-crafted decision tree.
Essential criteria used inside this tree are total flux density, apparent angular size, distance to the nearest neighbouring radio component, and the number of Gaussians fitted to each radio component by \textit{PyBDSF}.
Following this decision tree, $15,806$ of the radio components in LoTSS-DR1 ($4.9\%$ of $325,694$) required further manual inspection through Zooniverse, an online platform that enables and simplifies crowd-sourced annotation processes.\footnote{\url{https://zooniverse.org}}

In this paragraph, we summarise the relevant steps taken in their decision tree.
First, they reduced the number of imaging artefacts that mostly occur around bright compact sources.
They did so by considering all components brighter than $5$ mJy and smaller than $15$ arcsec and selecting the neighbours within $10$ arcsec of these components, which are $1.5$ times larger.
They visually confirmed $733$ of these $884$ candidates to be artefacts and removed them from the catalogue.
Next, they filtered out $223$ radio components that correspond to apparently large star forming galaxies by associating the radio components that lie within the ellipse of a $\geq60$ arcsec source from the Two Micron All Sky Survey \citep[2MASS;][]{2MASS} extended source catalogue \citep[2MASSX;][]{2MASSX}.
Then, the decision tree splits based on the size of the radio components; components are considered `large' when their major axis exceeds $15$ arcsec.
In this work, we did not consider small components, but using our method small components may still be associated with a neighbouring large radio component.
For the large components, the \citet{Williams2019} decision tree splits again based on the brightness of the components; components are considered `bright' when their total flux density is $>10$ mJy.
The $6,981$ large and bright radio components ($44.2\%$ of $15,806$) went to the Zooniverse platform.
It is these high signal-to-noise ratio components (of which we can be relatively certain of the association) that we want to use to train our neural network.
For the $13,321$ large and faint radio components, \citet{Williams2019} used pre-filtering (visual inspection by a single expert) to decide whether these radio components required manual association using the Zooniverse platform. 
We did not use the large and faint radio components for training, but we did estimate the accuracy of our automated component association on the large and faint components described in Section \ref{sec:discussion_faint}.

For our training set, we roughly emulated the tree in \citet{Williams2019} up to the large and bright components\footnote{For the corresponding script, see \url{https://github.com/RafaelMostert/lofar_frcnn_prepro/blob/main/imaging_scripts/multi_field_decision_tree.py}.}. 
We simply discarded all $876$ components that met the artefact candidate criteria and discarded all $458$ radio components belonging to nearby star forming galaxies (a higher number than that reported by \citet{Williams2019} as we use a circle instead of an ellipse in the cross-match process). 
This left us with $6,930$ large and bright radio components.
\citet{Williams2019} published a component catalogue that links the names of \textit{PyBDSF}-detected components to the value-added source catalogue that includes manual source component associations.
Using this component catalogue, we were able to link $6,573$ of the $6,930$ components to their final source in the the value-added catalogue from \citet{Williams2019}, which we needed in order to create training labels. 
Subsequently, we discarded $260$ LoTSS-DR1 components that contained no five-sigma emission in the LoTSS-DR2 images (due to the improved calibration).
We also discarded the components for which we were not able to extract large enough cutouts and those that contained NaNs, which left us with $6,158$ radio components.
A random split of the dataset based on $38$ pointings for training, ten for validation, and ten for testing leads to $3,983$ components for training, $1,054$ for validation, and $1,121$ for testing.

The observed pointings do partly overlap, but each unique component will only appear once in the LoTSS-DR1 catalogue that we use.
If a unique component is observed in multiple pointings, it is listed as appearing in the pointing for which it is closest to the pointing centre.
The validation and testing dataset still partly overlap with the training dataset if some components from multi-component sources in the training dataset are closer to a pointing center of a validation or test pointing.
This is the case for three components in the validation set and six components in the test set, leading us to overestimate the accuracy on these sets with at most $0.28\%$ and $0.54\%,$ respectively.

\subsection{Manual association process}
\label{sec:manual_process}
The manual association using the Zooniverse platform for DR1 is described in detail by \citet{Williams2019}, but we briefly recapitulate the process below.
This manual process was completed by 66 astronomers from the LOFAR collaboration and was not available to the public.
For each radio component, each platform user was informed about the component and its surroundings through three figures showing radio contours on a background of Pan-STARRS1 and WISE images.
The users then identified other components that they associate with this specific component as part of the same physical source.  
After each component was viewed by five different users, the resulting judgements on which components belonged to the same physical sources were centrally aggregated to form a consensus-based radio source catalogue. 
Each radio component associated with another radio component by at least one user was grouped into a `set'. All sets for which more than two-thirds of the users agreed were inserted into the catalogue as candidate sources. For sets that were subsets of larger sets, the largest set with a two-third consensus was chosen. In the end, radio components that belonged to multiple conflicting sets were manually resolved through visual inspection by a single expert using the LoTSS-DR1 images, the DR1 component-locations, and the corresponding WISE and Pan-STARRS1 images.

The LoTSS-DR1 manual associations were performed using LoTSS-DR1 images, while in this paper we use LoTSS-DR2 images with improved calibration, which improves dynamic range and reduces the number of artefacts \citep{Shimwell2022}.\footnote{The estimation and correction of the effect that instrumental errors have on the observed visibilities (known as calibration in radio synthesis imaging), can locally amplify or reduce the noise or signal to the point of generating spurious source components \citep[e.g.][]{Vidal2008,Grobler2014}.}
We used the LoTSS-DR2 images for training and inference because we aim to use our association for the future LoTSS data-releases, which will all use this same improved calibration.
As a result of the improved calibration, more of the connecting structure of extended sources is visible (such as emission bridging two lobes or tails extending farther out), making accurate component association easier.
The new images spurred us to manually improve the associations for the large and bright radio components that we used for training, testing, and validation.
For this manuscript, the authors manually sorted all images into the categories `association seems correct', `association is hard to judge', and `association seems incorrect'. 
We did so through visual inspection by a single expert using the LoTSS-DR2 images, the DR1 component-locations and associations, and the corresponding WISE and Pan-STARRS1 images.
For the `association is hard to judge' category, we deem it hard or impossible to infer a correct association beyond reasonable doubt given the LoTSS-DR2 image and overlaid WISE and Pan-STARRS1 source locations. 
We judged $88.02\%$ and $3.3\%$ of the radio components to be in the `association seems correct' and `association is hard to judge' category, respectively. These associations will not be altered. 
We did manually correct the associations for the $8.68\%$ of the components that we judged to be in the `association seems incorrect' category. 
Appendix \ref{app:correct} shows examples of manually corrected associations, of which we distinguished three sub-categories: the initial association seems incorrect in light of the improved calibration ($58\%$ of $8.68\%$), the initial association left an artefact unassociated and unflagged ($26\%$ of $8.68\%$), and the association seems incorrect due to human error ($16\%$ of $8.68\%$).
The manually corrected catalogue, and all other data products used for training, can be accessed online.\footnote{\url{lofar-surveys.org/radio_association.html}}

Image artefacts (due to imperfect calibration of the observations) that enter the final catalogue as individual sources have an impact on subsequent statistical analysis. Source density counts will artificially be increased and nearest-neighbour distances will artificially decrease. In the Zooniverse project, image artefacts were supposed to be flagged as artefacts. However, if more than a third of the volunteers did not flag the artefact, the dataset was entered as an individual radio source. In our manual correction of the Zooniverse associations, we opted to associate image artefacts with the bright sources from which they originate. Ideally, image artefacts would be automatically detected as a separate object class and entirely removed from the final catalogue, but that is beyond the scope of this paper. 

For the association of radio-components in LoTSS-DR2 and the identification of corresponding host galaxies, a public LoTSS radio galaxy Zooniverse project was set up; this ongoing project will be the subject of a separate publication.\footnote{\url{https://lofargalaxyzoo.nl}}
We invite the reader to consult Appendix \ref{app:lgz2} for an example of its interface.
The required number of views per component and the aggregation of these clicks is exactly the same for the internal LoTSS-DR1 Zooniverse project as for the public LoTSS-DR2 Zooniverse project.
The public LoTSS-DR2 Zooniverse project is different in the following ways: the optical background images shown are from the more sensitive Legacy Surveys \citep{legacy2019} instead of Pan-STARRS1; FIRST contours are not shown overlapping the LoTSS contours; markers for WISE and PAN-STARRS1 objects are not shown; the LoTSS data uses improved calibration \citep{Shimwell2022}; and users are shown a LoTSS intensity image to help them interpret the LoTSS contour lines in the other panels.
As of August 2 2022, $10,854$ volunteers are part of the public LoTSS Zooniverse project.
As we suspect that astronomers do a better job at associating radio components than the public volunteers, we did not use public volunteer labels to train our network.
In Section \ref{sec:public_zoo}, we discuss the quality difference between the associations in the internal and the public LoTSS Zooniverse project.

\section{Methods}
\label{sec:methods}
We propose using a neural network to replicate the manual radio source-component associations performed by a group of radio astronomers. 
For this purpose, motivated by the work of Chen \citet{Wu2019}, we adapted a type of neural network known as a region-based detector \citep{liu2020deep} or region-based convolutional neural network (R-CNN). 
This type of network is designed to detect instances of objects of a certain class within an image. Their output is a set of regions within an image, and for each region the predicted probability (or class score) for its most likely object class (example outputs can be found further below in Fig. \ref{fig:good} and Fig. \ref{fig:bad} of the results section).  

We looked for a single class of object instances: radio sources. 
For a given radio intensity image that is centred on a single radio component, we used an R-CNN to predict which rectangular region (or `bounding box') exclusively encompasses the radio source components that belong to the centred radio component. 
In practise, the R-CNN evaluates multiple regions and predicts a likelihood (or predicted class score) for each of these.
We associated the radio source components for which the central coordinates fall within the predicted region that includes the centred radio source component and has the highest predicted class score. 
Radio components appearing in multiple associated groups were only assigned to the group inside the largest region.

Our neural network must be trained to give a high probability (predicted class score) to regions of a radio image that contain likely radio component combinations and low probability to regions of a radio image that contain an unlikely combination of radio components. The neural network architecture that we used is described in Section \ref{sec:architectures} and our training and inference processes are given in Section \ref{sec:train}.
We describe how our input images were created in Section \ref{sec:prepro}, how we created pre-computed regions in Section \ref{sec:pre_computed_roi}, and we explain how we removed unresolved or barely resolved sources that are likely unrelated to our large and bright radio components in Section \ref{sec:remove}. 
Finally, we show how we implemented rotation data augmentation to prevent the network from overfitting in Section \ref{sec:augmentation_method}.

\subsection{The R-CNN architectures}
\label{sec:architectures}
\begin{figure*}
\centering
\includegraphics[width=0.8\textwidth]{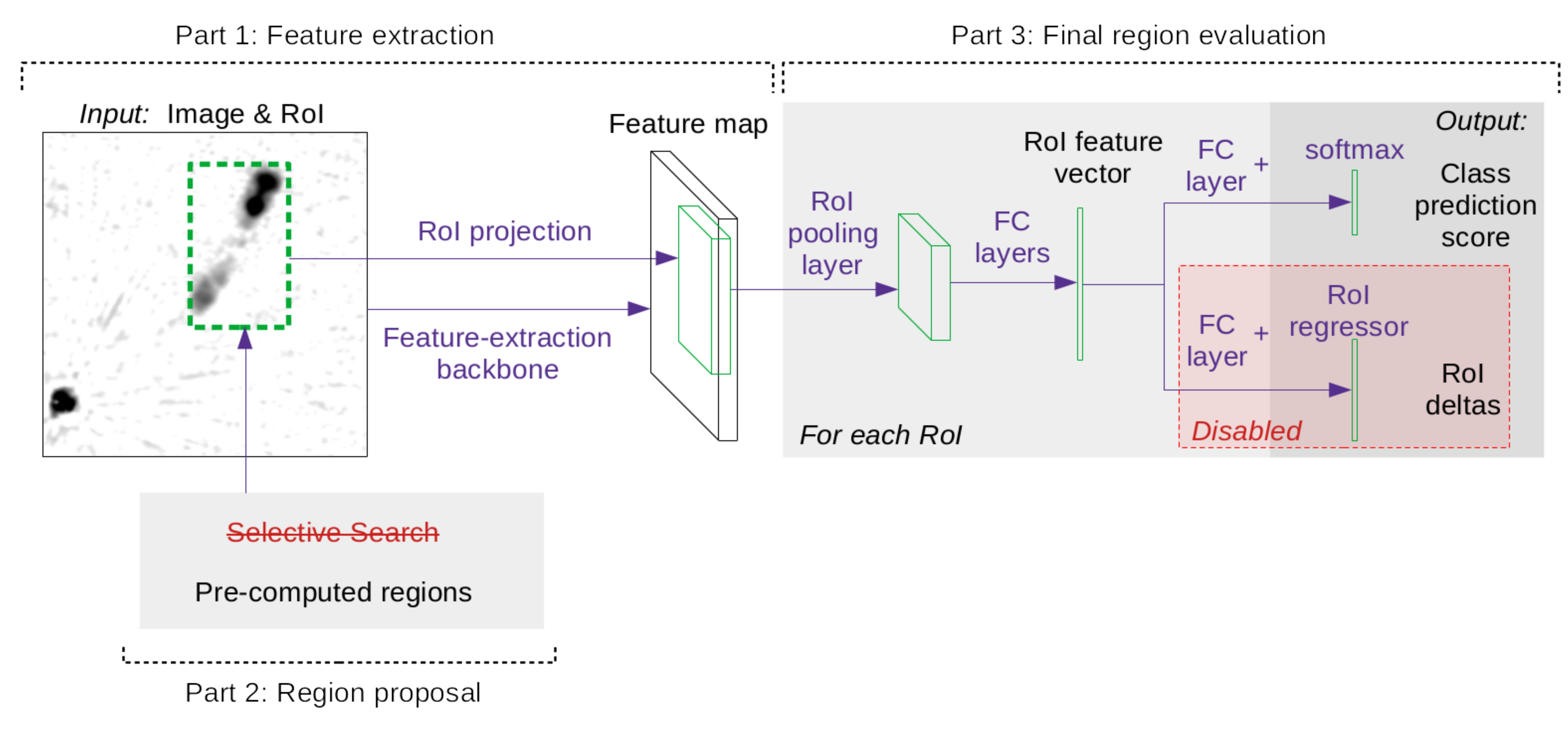}
      \caption{Adapted Fast R-CNN diagram, where convolutional layer is abbreviated as `Conv layer', fully connected layer is abbreviated
as `FC layer', and region of interest is abbreviated
as `RoI'.
      In the original Fast R-CNN, \textit{Selective Search} \citep{uijlings2013selective} is a general way to generate region proposals by exhaustively sampling regions in any image based on hierarchical image segmentation. Instead, we pre-computed our own region proposals as source detection software provides us with the exact locations of significant blobs of radio emission (see Section \ref{sec:architectures}).
      This means we can also disable the part of the Fast R-CNN that is designed to refine the location and dimension of proposed regions.}
      \label{fig:fast-rcnn}
\end{figure*}
The region based convolutional neural network that we use, an adapted Fast R-CNN \citep[][see Fig. \ref{fig:fast-rcnn}]{girshick2015fast},  consists of three consecutive parts:
a first part that extracts image features, a second part that generates region proposals, and a third part that classifies the proposed regions and suggests improvements to the location and shape of the proposed region.
We cover the workings of all three parts below.

The part of the architecture that provides the feature image is a neural network that is generally used for image classification (we refer to it as the `feature-extraction backbone' hereafter).
Image classification is a task that is performed based on discernible features within an image. 
For automated image classification, features may be hand-crafted based on a heuristic function or template. 
For example, to detect an FRII, we may code up a template that looks for two edge-darkened, aligned, elongated emission patches (radio lobes) with a small round emission patch in the middle of the two (the radio core). 
However, it is hard to create templates that generalise well. 
Once the data deviates from our pre-conceived template, we might not extract our desired features.
Convolutional neural networks (CNNs) are a template-free method to extract features from images through subsequent convolutional and pooling layers.\footnote{See \citet{Goodfellow2016}, \citet{murphy2012machine}, or \citet{dumoulin2016guide} for an introduction to convolutional neural networks.}
Subsequent convolutional layers create features with progressively higher abstractions from the original image. 
In between convolutional layers, the feature maps are commonly downsized to reduce the number of trainable parameters in the model, a process known as `pooling'. 
The subsequent convolutional and pooling layers reduce our Stokes I radio image to a multi-dimensional array known as a `feature image'.
Which features are extracted by the convolution layers depends on the parameters of the convolutional layers. 
During training (Section \ref{sec:train}), these parameters are optimised to extract the features that are crucial to detect the specific objects for the task at hand (radio sources in our case).

We used the Detectron2 \citep[Yuxin][]{wu2019detectron2} framework, which implements the Fast R-CNN in PyTorch and enables us to swap different feature-extraction backbones for our R-CNN.
We tested two industry standard feature-extraction backbones of the feature pyramid network type \citep[FPN;][]{lin2017feature}, specifically FPN-ResNet  \citep[ResNet;][]{resnet} and FPN-ResNeXt \citep[ResNeXt;][]{resnext}.
Feature pyramid networks improve convolutional networks for object detection by outputting feature maps at different resolutions from different stages of the ResNet, thereby improving object detection for objects at multiple size scales \citep{dollar2014fast,lin2017feature}.
The backbones can have an arbitrary `size', by which one indicates the number of convolutional and pooling layers, and we tested two common sizes for each backbone.

After the feature extraction, an R-CNN requires initial guesses of plausible regions of the image that enclose the objects of interest.
In the architectures used in this work, these regions are always rectangular regions -- and referred to as regions of interest (RoI). 
This means it is not always possible to include only a single object of interest and avoid interlopers. Section \ref{sec:remove} partly addresses this issue.
For each region, an `objectness' score is produced, which predicts the likelihood of the region mostly overlapping with an object or with the background. 
Overlap is always measured using the intersection over union (IoU), which is the overlap between the predicted region and the ground truth region divided by their area of union.
Additionally, for each object class (we chose a single `radio object' class), changes to the location, width, and height of the region are predicted that improve the region and its overlap with the encompassed object.

For training, the number of proposed initial regions is filtered down to a smaller, balanced set of regions with both a high objectness score (regions likely to contain radio components) and a low objectness score (regions likely to contain mostly background noise).
Finally, only a few regions that are most likely to tightly encompass the right radio sources are returned.
Technically, this means that for regions with a high IoU with respect to each other, the regions with lower objectness scores are discarded; this process is referred to as non-maximum suppression (NMS).

The Fast R-CNN relies on external algorithms for region proposal generation. 
The one used in the original Fast R-CNN by \citet{girshick2015fast} is \textit{Selective Search} \citep{uijlings2013selective}, which attempts to provide (a few thousand) initial regions for different kinds of everyday objects in common contexts.
However, regions can in principle be tailor-made depending on the field of application. 
In radio astronomy, the detection of emission is a well-studied problem for which a number of robust algorithms, such as \textit{PyBDSF}, \textit{ProFound,} and \textit{AEGEAN}, have already been developed. 
We can thus use a Fast R-CNN in combination with pre-computed regions that tightly enclose all combinations of radio components detected by either one of these emission-detection algorithms. 
We describe the creation of our pre-computed regions in Section \ref{sec:pre_computed_roi}.

For the final stage of an R-CNN, the neural network branches into two parts. 
Both parts take the same input: the region of interest (or RoI) pooled part of the feature image.
One branch predicts the object class probability of the considered RoI.
The other branch suggests transformations of the shapes of the foreground regions such that they better enclose the objects located in the foreground region.
This latter branch can be omitted for our adapted Fast R-CNN, as we provide regions that tightly enclose the radio emission exceeding five sigma.
For reproducibility, the adapted Fast R-CNN used in this paper, and the image extraction, image pre-processing, and pre-computed RoI code are available online.\footnote{\url{lofar-surveys.org/radio_association.html}}

\begin{figure}
\centering
\fbox{\includegraphics[width=0.99\columnwidth]{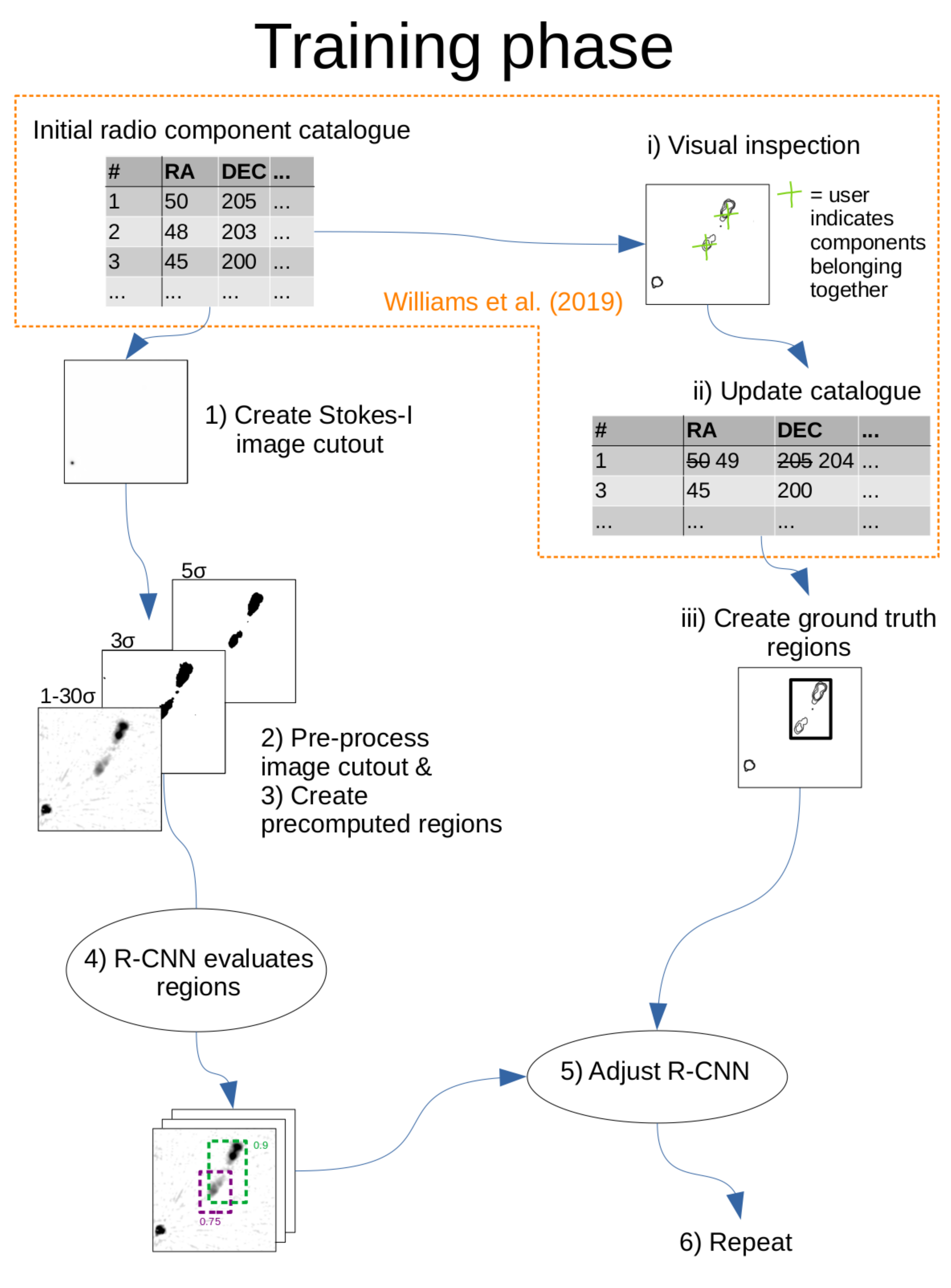}}
      \caption{Diagram of training phase. We start from a radio component catalogue created by \textit{PyBDSF}. i) Users indicate which radio components belong together via crowd-sourced visual inspection. ii) This information is used to create an improved source catalogue (see Section \ref{sec:data}). iii) This improved source catalogue and the component catalogue are used to draw ground truth regions. 1) We create an image cutout centred on a radio component from the initial catalogue (if it is included in our training set). 2) We pre-process the image for the R-CNN. 3) We pre-compute regions (see Section \ref{sec:pre_computed_roi}). 4) The R-CNN  evaluates the regions and predicts corresponding class scores based on the image (known as a ‘forward pass’). 5) We update the network parameters using stochastic gradient descent such that subsequent predicted regions have greater overlap with the ground truth region (known as ‘backpropagation’). 6) Steps 1-4 are repeated for all radio components in our training dataset.}
      \label{fig:train}
\end{figure}
\begin{figure}
\centering
\fbox{\includegraphics[width=0.99\columnwidth]{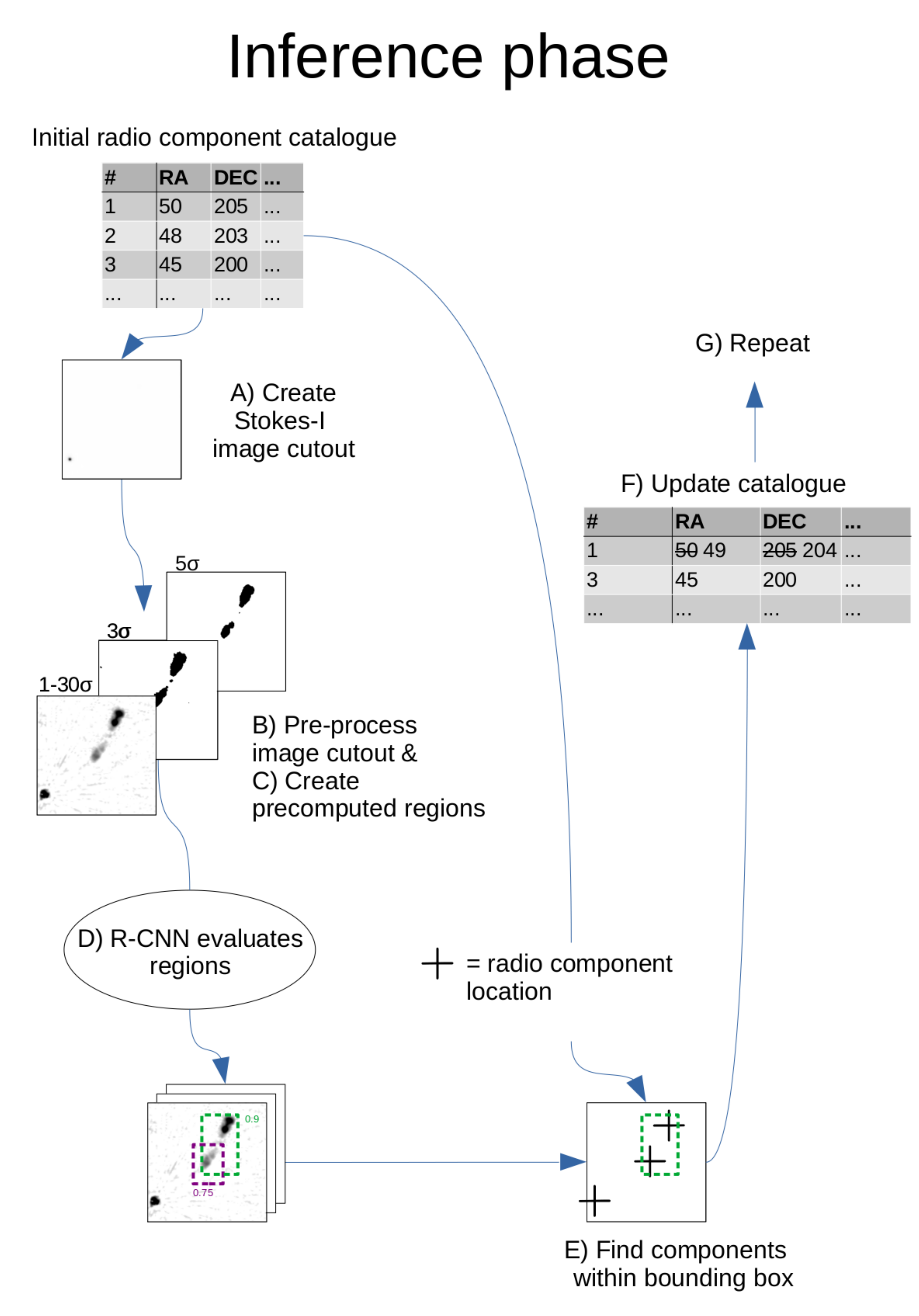}}
      \caption{Diagram of inference phase. We start from a radio component catalogue created by \textit{PyBDSF}.  A) We create an image cutout centred on a radio component from the initial catalogue. B) We pre-process the image for the R-CNN. C) We pre-compute regions (see Section \ref{sec:pre_computed_roi}). D) The R-CNN  predicts several regions and corresponding prediction scores based on the image (known as a ‘forward pass’). E) We select the region that covers the central radio component and has the highest prediction score. We then look for the radio component coordinates that lie within this region. F) These radio components will enter the updated radio source catalogue combined into a single entry. G) Steps A-E are repeated for all radio components in our inference dataset.}
      \label{fig:inference}
\end{figure}

\subsection{Training and inference phase}
\label{sec:train}
Figure \ref{fig:train} and Fig. \ref{fig:inference} show a schematic view of our pipeline in the training phase and in the prediction phase (also known as inference phase), respectively. 
Given an image and the pre-computed regions for the Fast R-CNN (step 3 in Fig. \ref{fig:train}), the network will classify all RoIs with a score between 0 and 1, where a higher score means the prediction is more likely to have a high overlap with the ground truth region. 

Subsequently (step 4 in Fig. \ref{fig:train}), the network parameters are updated using stochastic gradient descent\footnote{See \citet{Goodfellow2016} or \citet{murphy2012machine} for an introduction to stochastic gradient descent.} such that predicted regions that barely overlap the ground truth region from the manual association have a higher chance of receiving the `background' class label, while predicted regions that largely overlap have a higher chance of receiving the `radio source' class label. 
Specifically, we use stochastic gradient descent with a learning rate of $0.0003$, a momentum \citep{sutskever2013importance} of $0.9,$ and a weight decay \citep{hanson1988comparing} of $0.0001$.\footnote{The training configuration files are available in our git repository: \url{https://github.com/RafaelMostert/detectron2/tree/master/configs/lofar_detection}}
We classify regions into 'radio source' or 'background' using a softmax function and quantify the training error using a cross-entropy loss function.

At the inference phase (step C in Fig. \ref{fig:inference}), the neural network will again predict the likelihood that the suggested regions of the feature-image contain a radio source.
This time, the network will not be updated using the ground truth region as we are looking at new data for which there is no manual association, or because we are interested in measuring the performance of the network on images not used during training (processes known as `validation' and `testing').
Due to our pre-computed regions, both training and inference is relatively fast. 
On the NVIDIA Tesla P100-SXM2 graphics card that we use, training takes one hour and 37 minutes for 50k iterations, and inference takes $0.043$ seconds per radio component.

We remind the reader that for training, we will use the LoTSS-DR2 images within the LoTSS-DR1 area of the sky, as expert associations are available for this part of the sky.
We did manually correct these expert annotations as they were initially done using the LoTSS-DR1 images, which had a worse dynamic range and more image artefacts (see Section \ref{sec:manual_process}). 

\subsection{Pre-processing the images and labels}
\label{sec:prepro}
We create image cutouts centred on the sources in the LoTSS-DR1 \textit{PyBDSF}-created radio source catalogue provided by \citet{Williams2019} as described in Section \ref{sec:data}. 
Per image, the neural network is also given the coordinates of the `ground truth': a rectangular region that tightly fits around the five-sigma radio emission of all radio components that, according to earlier performed manual association (see Section \ref{sec:manual_process}), belong together.

We take the size of the cutouts to be $300 \times 300$ arcsec\textsuperscript{2}, resulting in $200 \times 200$ pixel images for the $1.5$ arcsec pixel angular resolution of LoTSS.
This size ensures that $99.30\%$ ($93.36\%$) of the large and bright sources fully fit inside the cutout if the focussed component is located in the centre (on the edge) of the associated source, based on the angular sizes of the sources in the manually associated LoTSS-DR1 catalogue.

We want to trigger the same prediction for faint as for bright sources with similar morphology.
However, the predictions of a neural network are dependent on the magnitude of the input parameters \citep{cs231n}. The Detectron2 framework was originally built for regular three-channel images (red, green, and blue).
As our radio images span a broad contrast range,
we tested a number of different ways to encode the radio intensity image into a three-channel image.
For the first channel, we encode the radio emission square-root stretched between 1 and 30 sigma, the second channel sets all radio emission above three sigma to one, and all radio emission below that value to zero. The third channel sets all radio emission above five sigma to one and all radio emission below that value to zero. The choice for the third channel is meant to guide the network to regions enclosing the five sigma emission (see step 1 of Fig. \ref{fig:train} for an example Stokes-I radio image and step 2 of the same figure for the three corresponding channels).

In total, the pre-processing for training, including the creation of rotation augmentations and pre-computed regions, takes $1.9$ seconds per radio component on a single cpu.
Pre-processing for inference takes $1.0$ seconds per radio component on a single cpu.

\subsection{Pre-computed regions}
\label{sec:pre_computed_roi}
We created our pre-computed regions by drawing regions around each combination of radio-components in the image that includes the centred radio component. These regions will be tightly drawn around the five-sigma-level contours of these radio components.
The resulting number of pre-computed regions is then equal to $2^{n-1}$ where $n$ is the number of radio-components in the image. 
\cite{girshick2015fast} showed that a Fast R-CNN performs best at recognising everyday objects when the number of pre-computed regions during training is of the order of a few thousand. 
This means that $n$ can easily be as large as 12 ($=2,048$ proposals), while values above 14 ($=8,192$ proposals) will start to slow down the network. 
In practice, not every combination of radio components produces a unique region. We discard all duplicate regions, reducing the number of proposals to evaluate.  
For simplicity, we discard all sources with more than $12$ \textit{PyBDSF}-detected components within the cutout from our datasets. Using our cutout size of 300 arcsec, this results in the removal of only 17 out of 6192 cutouts ($0.27\%$) from our dataset. 
These excluded cutouts tend to be focussed on radio components in complex, clustered environments worthy of manual visual inspection and association.
R-CNNs work with a lower and an upper IoU threshold for classifying proposals as good (containing an object) or bad (containing mostly background) during training. 
We set the lower IoU threshold to $0.5$, this means that proposals that have an IoU with the ground truth lower than $0.5$ will be considered as background (or negative examples) during training.
We set the upper IoU threshold to $0.8$, meaning that proposals with a higher IoU with the ground truth than $0.8$ will be considered as 'foreground' (or positive examples) during training.
As we want to evaluate all our pre-computed regions, we disabled NMS.

\subsection{Removing unresolved sources}
\label{sec:remove}
\begin{figure}\begin{center}
\includegraphics[width=0.95\columnwidth]{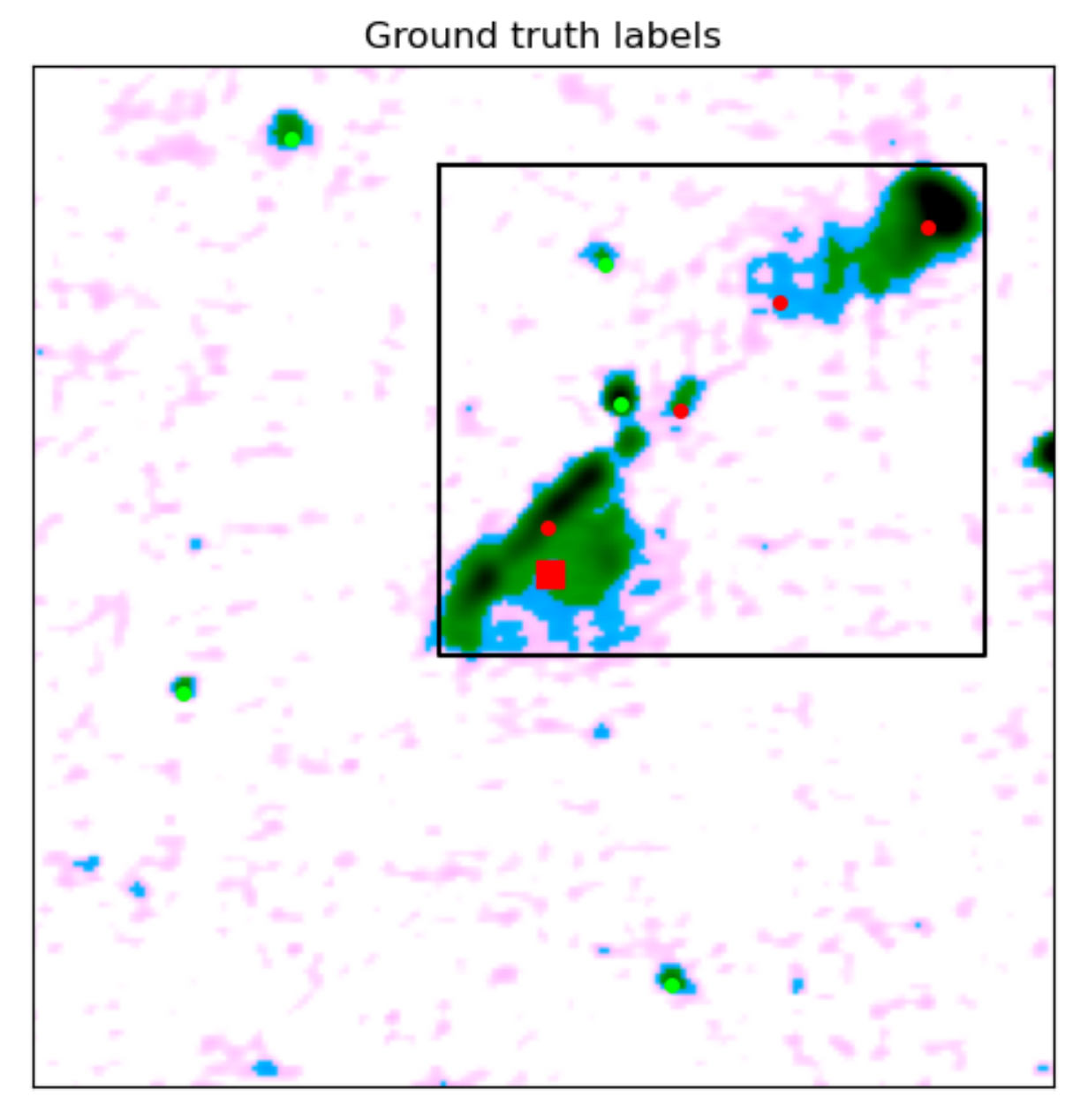}
\includegraphics[width=0.95\columnwidth]{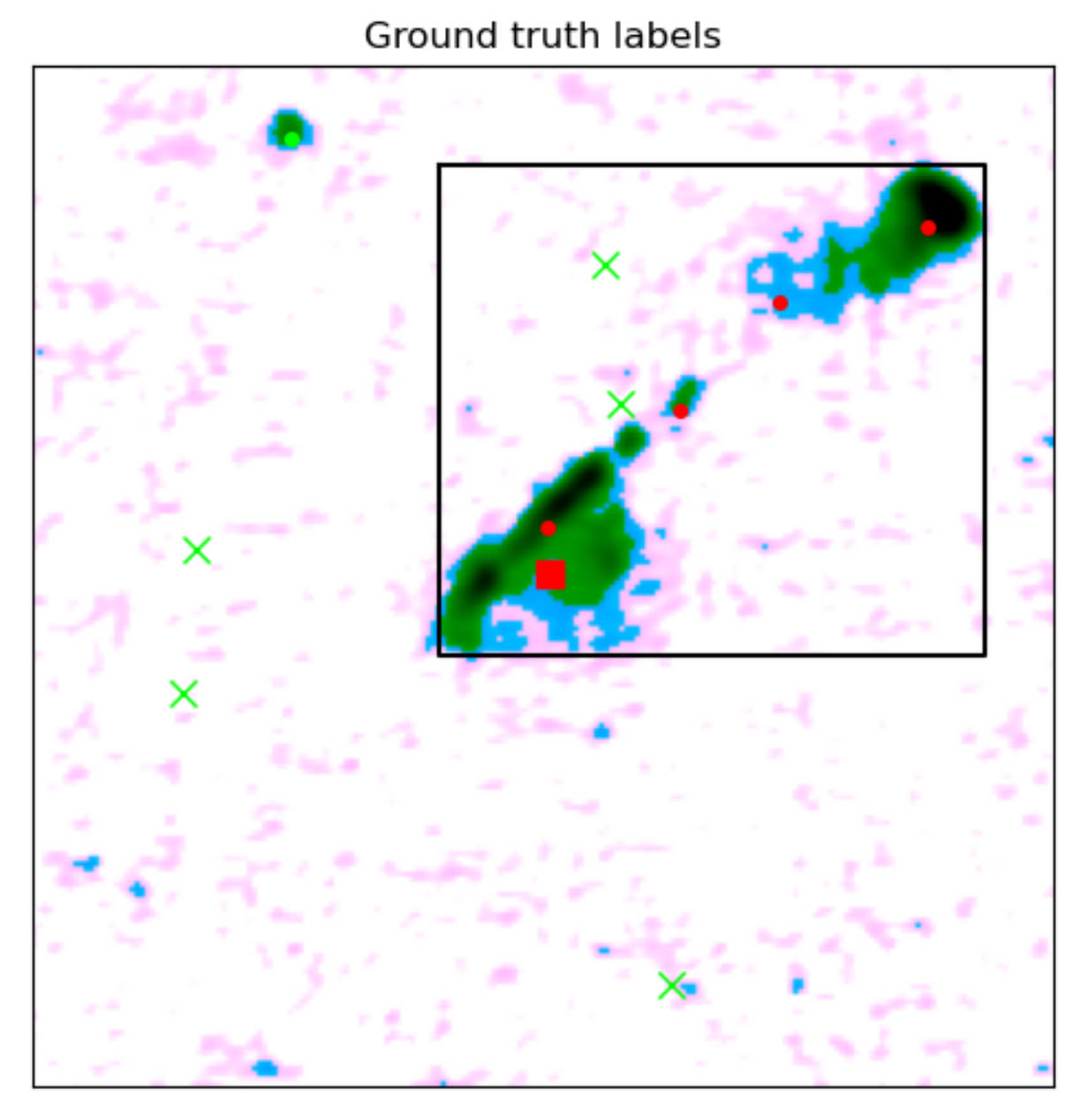}
\caption{Two figures demonstrating the simplification achieved by removing sources that are likely to not require manual association. The background shows LoTSS-DR2 intensity images, the black rectangle indicates our ground truth region encompassing the focussed radio component (red square) and its related components (red dots). The green dots indicate the locations of unrelated radio components. In the second figure, components removed by the GBC are shown as `x's.}
\label{fig:removed}
\end{center}\end{figure}

Independent of the association technique, the component association of large and bright RLAGNs is also complicated by the chance alignment of unrelated radio sources. 
We simplify the association task by removing unresolved and barely resolved sources that are likely unrelated, before feeding the images and labels to our neural network. 
We do so by using the predictions of a gradient boosting classifier (GBC) trained by \citet{Alegre2022} to detect whether a source can be directly associated with an underlying optical or infrared host galaxy using a likelihood ratio or if manual association and cross-identification is required. 
If the GBC decides that a source can be matched to an underlying source using the likelihood ratio (at value $<0.20$), the source major axis is smaller than $9$ arcsec ($1.5$ times the synthesised beam), and the ratio of the major axis over the minor axis is smaller than $1.5$, then we remove the source. 
The focussed radio component in a cutout will never be removed.
These chosen values are tradeoffs between removing as many background sources as possible while keeping the removal of foreground radio components low. 
Specifically, at a likelihood ratio below $0.20$, \citet{Alegre2022} estimate that $0.8\%$ of the radio components that should have been combined into another radio component are removed.
Removing radio components that are likely unrelated to the focussed radio component from our images (both during training and at inference), greatly reduces the number of pre-computed regions that require evaluation and simplifies the association task (see Fig. \ref{fig:removed}). 
All details and considerations for the training of this GBC can be found in \citet{Alegre2022}, the catalogue with GBC predictions that we use can also be found online.\footnote{See \url{lofar-surveys.org/radio_association.html}, specifically, we use the \texttt{0.20} column.}

\begin{figure}\begin{center}
\includegraphics[width=0.95\columnwidth]{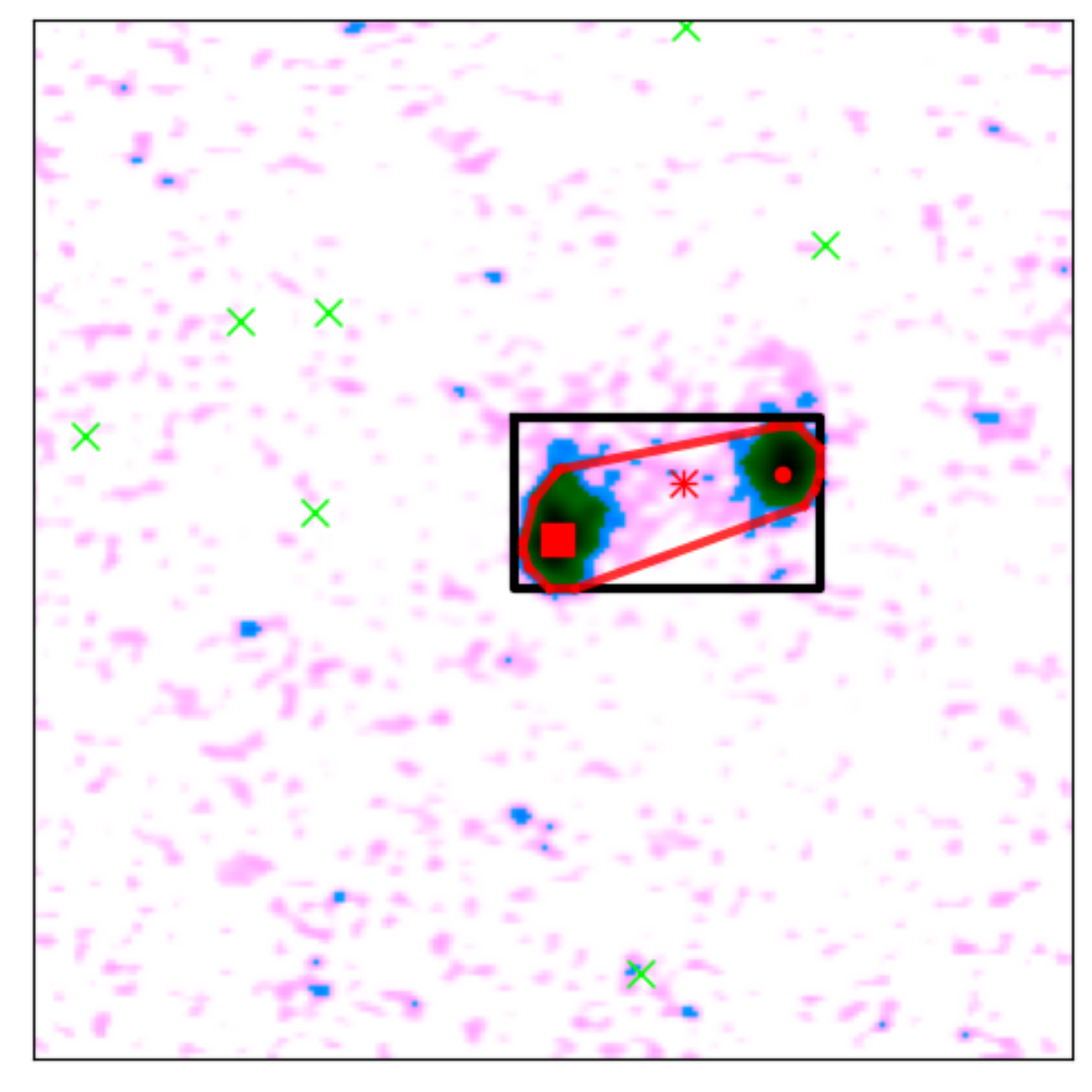}
\caption{Demonstrating reinsertion of a removed unresolved source. The background shows a LoTSS-DR2 intensity image, the black rectangle indicates our ground truth region encompassing the focussed radio component (red square) and its related components (red). The green markers indicate the locations of unrelated radio components. Components removed by the GBC are shown as `x's. Sources that fall within the convex hull (solid red line) around the five-sigma emission of the radio components within the predicted region will be reinserted (a `+' on top of an `x').}
\label{fig:reinserted}
\end{center}\end{figure}

For certain large and bright RLAGNs that show two spatially separated lobes and a compact radio core, the GBC can erroneously identify this core as a source that does not require manual association and manual optical identification, prompting us to remove this component before inference. 
To accommodate for this behaviour, we first remove the components at pre-processing, then secondly run the R-CNN to identify radio components that need to be associated -- in this case, the two radio lobes will be associated. 
Thirdly, we draw a convex hull around the five-sigma contours of the to-be-associated components and reinsert all radio components that lie within this convex hull and were removed earlier (see Fig. \ref{fig:reinserted} for an example).
In the cases where a compact core is removed, the R-CNN cannot use this core to help it infer which components should be associated. However, as the LoTSS data contains many double-lobed RLAGN without a detected compact core, this should not be problematic.

\subsection{Data augmentation through rotation}
\label{sec:augmentation_method}

The radio component association has to perform well, irrespective of the orientation of the radio sources on the sky plane presented to our automated pipeline. 
The features extracted by our neural network are not inherently invariant to rotation or flipping. 
The neural network framework by Yuxin \citet{wu2019detectron2} that we adapt has built in on-the-fly flipping augmentation; every training image and the ground-truth regions are randomly presented to the network in its original orientation and flipped. 
To prevent the network from over-fitting on the specific orientations of the images it is shown during training, we implemented rotation data augmentation. 
Data augmentation is a common practise in deep learning to increase the size of the training dataset with slightly modified copies of the initial training dataset. We insert copies rotated at angles of 25, 50, and 100 degree.

We implemented rotation augmentation in our pre-processing stage instead of `on-the-fly' at the training time.  This allows us to recalculate the tightest rectangular region enclosing the relevant five-sigma emission for every new orientation instead of simply drawing a larger rectangle around the rotated original region (note that our network requires rectangular regions with edges parallel to the edges of the input image).
This form of data augmentation comes at the cost of longer training times (as all rotated versions of the training images need to be evaluated), which scales linearly with the chosen number of additional rotation angles.

\section{Results}
\label{sec:results}
Before examining our results, we define suitable quantitative performance metrics.
If our predicted region uniquely encompasses the central coordinates of the (non-removed or reinserted) radio-components in accordance with the manual association, we have a true positive (TP).\footnote{Only the central coordinate of a component needs to fall within the predicted region to be counted as TP or FP. The final radio catalogue that we create with our pipeline includes all \textit{PyBDSF}-detected radio emission, as we only use our predictions to combine certain entries in the existing radio component catalogue.}
If the region does not encompass all of the radio components that belong together, we have a false positive (FP). 
If the region encompasses all the radio components that belong together, but also encompasses additional unrelated radio components, that also counts as a FP.
If there is no region covering the central coordinate of the focussed radio component with a score surpassing the user-set threshold we have a false negative (FN). A true negative (TN) is the absence of a region where this is indeed warranted.
True negatives should not appear in our data, as we only consider radio images centred on radio components with a signal-to-noise ratio surpassing five.
So, for example, a single component source that has a bounding box that only encloses this one component is a TP. 
If the bounding box does not enclose this component, it will be a FN. 
If the bounding box includes more than this one component it will be a FP.
The metrics only consider if the bounding box does or does not enclose the central coordinates of the components that compose a source. 
This is all we need as \textit{PyBDSF} carries the rest of the morphological and flux  information of these components.

We will use catalogue accuracy, defined as $accuracy = \frac{TP + TN}{TP + TN + FP + FN}$ , as our main metric, although in our case this reduces to $accuracy = \frac{TP}{TP + FP + FN}$ as our $TN$ is always zero.
Statements in the results section about the quality of the predictions  (such as the predicted region encompasses too few or too many radio components) are always with respect to the manual associations as described in Section \ref{sec:manual_process}.
We aim to maximise the single metric of catalogue accuracy in our experiments. 

We explored a number of design implementations of our pipeline, as discussed in the methods section. 
The reported results from Section \ref{sec:backbone_experiments} onwards are the mean and standard deviations of the catalogue accuracy on our validation dataset, which were obtained by training our network for three independent runs with three different random initialisation seeds.

\subsection{Baseline and upper-boundary performance}
\label{sec:baseline}
We begin with a catalogue without any radio-component association: a catalogue for which we assume that each \textit{PyBDSF}-detected radio component is a single (unique) radio source. Comparing this catalogue to our corrected LoTSS-DR1 catalogue (described in Section \ref{sec:manual_process}) we obtain a baseline catalogue accuracy of $62.1\%$ for the large and bright sources.
This baseline tells us that almost two-thirds of the radio components in the large and bright source cut are stand-alone unique radio sources and do not require association with other radio components. The accuracy of any component association technique would have to surpass this baseline to be useful.

We also set a more advanced baseline by training a random forest to determine for each component $A$ whether it belongs to another component $B$. 
To keep a class imbalance in check, we only considered components $A$ that are within a $100$ arcsec radius of component $B$ (the larger the radius, the higher the fraction of components $A$ that do not belong to component $B$).
The random forest has access to seven features: component $A$'s major axis length in arcsec (feature: \texttt{Maj}), the ratio between the major axis length of $A$ to that of $B$  (feature: \texttt{Maj\_ratio}), the total flux density of component $A$ in mJy (feature: \texttt{Total\_flux}), the ratio between the total flux density of component $A$ to that of $B$ (feature: \texttt{Total\_flux\_ratio}), the peak flux of component $A$ in mJy (feature: \texttt{Peak\_flux}), the ratio between the peak flux of component $A$ to that of $B$ (feature: \texttt{Peak\_flux\_ratio}), and finally, the angular on-sky separation between component $A$ and $B$ in degrees (feature: \texttt{Separation}).
The random forest was trained using the same components in the training pointings that our Fast R-CNN uses.
We set our random forest, taken from the scikit-learn Python package, to use an ensemble of 1000 trees. As a result of a grid search whereby we evaluated the performance on the validation dataset, we take the maximum number of features to be $0.4$ and the maximum depth of the trees to be $10$. We adopt the default settings for all further hyper-parameters.\footnote{See \url{https://scikit-learn.org/0.24/modules/generated/sklearn.ensemble.RandomForestClassifier.html} for these default hyper-parameters.}
The trained random forest is then evaluated using the same components in the test pointings that our Fast R-CNN uses.

The relative predictive power of each feature in this random forest turns out to be $0.25$ for \texttt{Maj}, $0.23$ for \texttt{Separation}, $0.18$ for \texttt{Maj\_ratio}, $0.13$ for \texttt{Total\_flux\_ratio}, $0.11$ for \texttt{Total\_flux}, $0.05$ for \texttt{Peak\_flux\_ratio}, and $0.03$ for \texttt{Peak\_flux}.
The resulting catalogue accuracy on the test dataset for the large and bright radio components is $69.0\%$.
We repeated this process for the large ($>15$ arcsec) and faint ($<10$ mJy) radio components, which resulted in a catalogue accuracy of $80.1\%$. 
However, we note that this last number might be off by a few percentage points for two reasons.
First, it is generally harder to associate sources with a low signal-to-noise and this is not reflected in the (crowd-sourced) labels. 
Second, we did not manually check and correct the associations for the large and faint components as we did for the large and bright components (see section \ref{sec:discussion_faint}).

Moving on to our Fast R-CNN method, there is also an upper boundary to our association performance, since it is based on combining components within a rectangular region. 
The tightest rectangle around a set of related radio components will sometimes encompass unrelated radio sources, leading to an upper-limit in catalogue accuracy of $98.5\%$ for the large and bright sources. 
This upper limit is lowered to $96.6\%$ when source-removal at pre-processing is applied, partly because the gradient boosting tree erroneously removes related components, and partly due to the unwarranted reinsertion of a component as a result of our convex-hull method. We observe that in practise, the source removal has a positive effect on the attained accuracy (Section \ref{sec:ablation}).

\subsection{Classification backbone and learning rate experiments}
\label{sec:backbone_experiments}

\begin{table}
\caption{Maximum catalogue accuracy attained by different CNN backbones on the large and bright source components. The reported numbers in this and further tables are the mean and standard deviations of three training runs with different random seeds and otherwise equal set-ups.}             
\label{table:backbones}      
\centering                          
\begin{tabular}{c c c}        
\hline\hline                 
CNN backbone & Train accuracy & Val. accuracy \\    
\hline                        
ResNet 50 layers & $89.0\%\pm0.3$ & $84.3\%\pm0.4$ \\
ResNeXt 50 layers & $87.7\%\pm0.3$ & $82.8\%\pm0.2$ \\
ResNet 101 layers  & $88.5\%\pm0.2$ & $83.1\%\pm0.9$ \\
ResNeXt 101 layers & $87.7\%\pm0.5$ & $82.8\%\pm0.5$ \\
\hline                                   
\end{tabular}
\end{table}

We trained our adapted Fast R-CNN with a constant learning rate with two different, state of the art, residual convolutional neural network backbones and for each we tested two commonly used model sizes.\footnote{For reproducibility, the full configuration files for all runs are available in our git repository: \url{https://github.com/RafaelMostert/detectron2/tree/master/configs/lofar_detection}.}
We tested the FPN-ResNet CNN \citep{resnet,lin2017feature} and the FPN-ResNeXt CNN \citep{resnext,lin2017feature}, each at a depth of 50 and 101 layers. 
The experiments we performed include source removal (Section \ref{sec:remove}) and rotation augmentation (Section  \ref{sec:augmentation_method}).

The catalogue accuracy on the validation set peaked at around 20k iterations for the FPN-ResNet-50, but to allow the networks with more parameters to fully train, we trained these networks with up to 50k iterations. 
We evaluated the accuracy after every 10k iterations, and for each
interval we reported the mean and standard deviation of three runs with different random seeds and all other hyper-parameters kept fixed. 
Table \ref{table:backbones} shows the maximum attained mean accuracy values for each backbone.
Given this result of four models with insignificant differences in performance, we proceeded in our experiments with the least complex model, which is the FPN-ResNet-50 model. 

\begin{table}
\caption{Maximum catalogue accuracy attained on the large and bright source components by using different learning rate (decay) schemes.}             
\label{table:learning_rate}      
\centering                          
\begin{tabular}{c c c}        
\hline\hline                 
Learning rate & Train accuracy & Val. accuracy \\    
\hline                        
Constant & $89.0\%\pm0.3$ & $84.3\%\pm0.4$ \\
Step-wise & $88.7\%\pm0.2$ & $83.7\%\pm0.4$ \\
Cosine & $88.7\%\pm0.2$ & $83.4\%\pm0.2$ \\
\hline                                   
\end{tabular}
\end{table}

Next, we explore the effect of the learning-rate hyper-parameter on the accuracy we attain. 
The learning-rate hyper-parameter sets the magnitude of the effect of each update of the model parameters during training. 
Using the adapted Fast R-CNN with the FPN-ResNet-50 backbone, we tested different learning-rate decay schemes with the idea that a high learning rate might overshoot our global (or even local) minimum.
Apart from a constant learning rate, we tested a step-wise learning rate that stays constant but drops to a tenth of its former value after 30k and 40k iterations, respectively. We also tested a cosine learning rate, which starts at the same value as the other two learning rates, but gradually decreases (following the shape of a cosine in the 0 to $\pi$ rad range) to a value of zero at 50k iterations. 
The results of the different learning rate choices are presented in Table \ref{table:learning_rate}.
From these results, we conclude that different learning-rate decay schemes do not significantly affect the results, and so we simply proceed to evaluate our network using a constant learning-rate for 20k iterations. 

\begin{table}
\caption{Maximum catalogue accuracy attained on the large and bright source components by using different image pre-processing approaches.}             
\label{table:viridis}      
\centering                          
\begin{tabular}{c c c}        
\hline\hline                 
Image preprocessing & Train accuracy & Val. accuracy \\    
\hline                        
Viridis no-scaling & $85.8\%\pm0.7$ & $80.7\%\pm0.2$ \\
Viridis 1-30 sigma & $87.5\%\pm0.2$ & $81.5\%\pm0.2$ \\
Custom three-channel & $89.0\%\pm0.3$ & $84.3\%\pm0.4$ \\
\hline                                   
\end{tabular}
\end{table}

We also tested whether our three-channel pre-processing actually benefits the model.
In one experiment, we simply converted the FITS-cutouts to a three-channel (red-green-blue) PNG image using the univariate `viridis' colormap.
In a second experiment, we scaled the FITS-cutout such that the image only displays values within the 1-30 sigma range, again divided over the three-channel PNGs using the `viridis' colour map.
We compared these results to the custom three-channel image pre-processing that we use throughout this work that does not rely on a colour map.  
As mentioned in Section \ref{sec:prepro}, we filled one channel with the 1-30 sigma values of the cutout, we filled the second channel with a three-sigma filled contour, and we filled the third channel with a five-sigma filled contour. 
Using the adapted Fast R-CNN with the FPN-ResNet-50 backbone with a constant learning rate, we trained for up to 50k iterations and three different random seeds and report the values at the number of iterations that results in the highest catalogue accuracy on the validation set for each experiment.
The results in Table \ref{table:viridis} show that our custom three-channel approach is indeed beneficial.

\subsection{Ablation study}
\label{sec:ablation}
\begin{figure}\begin{center}
\includegraphics[width=0.99\columnwidth]{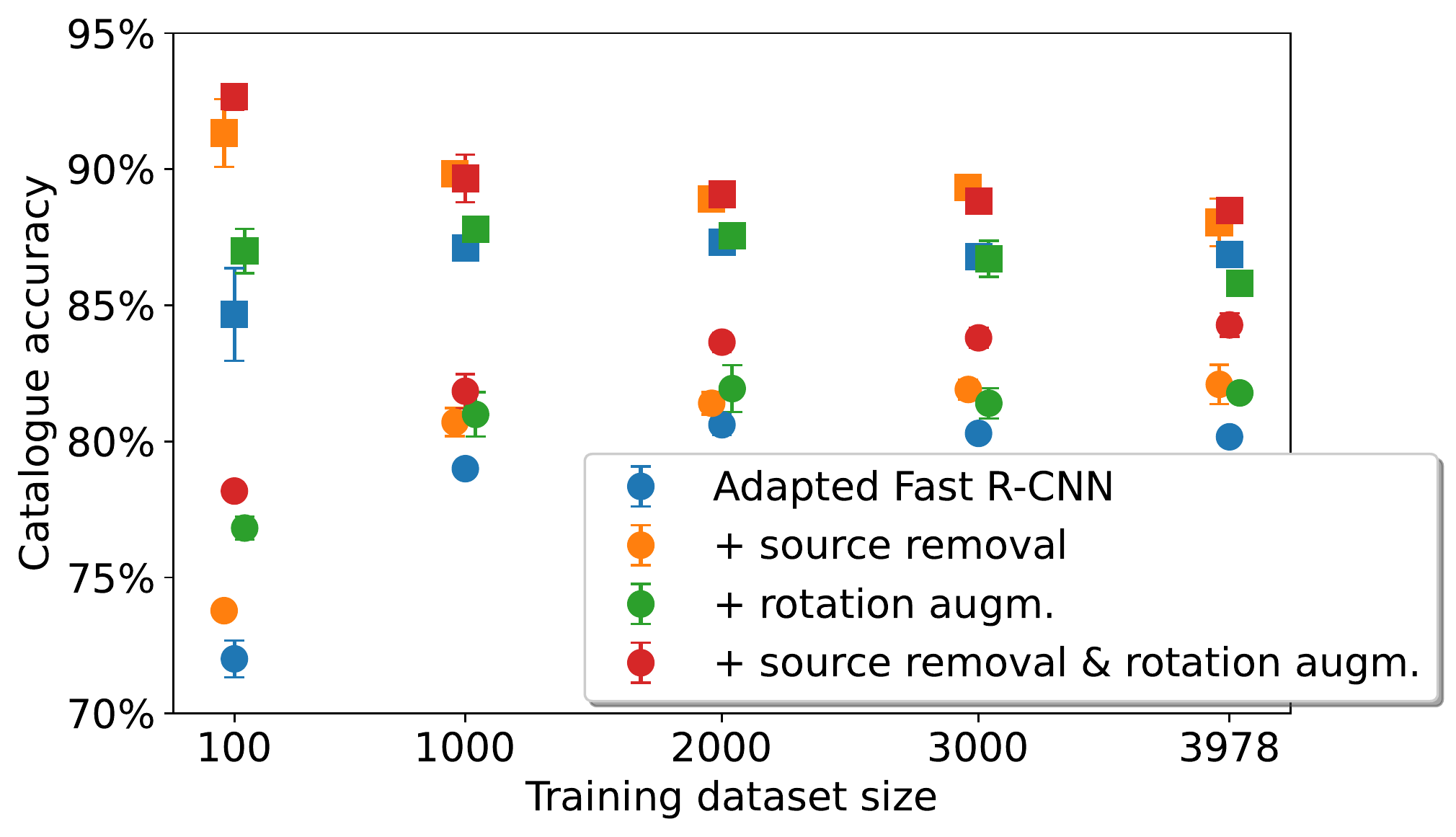}
\caption{Ablation study of source removal and rotation augmentation for the large and bright source component training datasets of increasing size. Round (squared) markers show performance on the full validation (partial training) dataset. Data points and their error bars are the mean and standard deviation of three training runs with different random seeds and otherwise equal setup. The horizontal scatter of the data points around each x-axis tick mark is artificially created to prevent overlap.}
\label{fig:ablation}
\end{center}\end{figure}

Figure \ref{fig:ablation} shows the effect of source removal (Section \ref{sec:remove}) and rotation augmentation (Section \ref{sec:augmentation_method}) on the catalogue accuracy.
Comparing the accuracy with (orange data points) and without (blue data points) source removal, we see that source removal systematically improves our catalogue accuracy, on both our training set and our validation set, for all training dataset sizes. 
Using the full training dataset, source removal increases the accuracy on the validation set 
from $80.2\%\pm0.2$ to $82.1\%\pm0.7$.

Comparing the performance with (green data points) and without (blue data points) data augmentation, we see that the augmentation also systematically improves our catalogue accuracy on the validation set for all training dataset sizes. 
Rotation augmentation does not significantly affect the results on the training set, but it reduces the gap between between the validation and training set.
This indicates that rotation augmentation successfully prevents the network from over-fitting on our training data: the predictions are better generalised (more accurate for images outside of the training dataset). 
Using the full training dataset, rotation augmentation increases the accuracy on the validation set 
from $80.2\%\pm0.2$ to $81.8\%\pm0.1$.

Without the source removal and the data augmentation, the network benefits from a larger training dataset up to about $2,000$ images, whereas with either source removal or data augmentation, it benefits up to at least the size of our full training set (roughly $4,000$ images).
The combined effect of the two improvements is smaller than their summed individual effects, but it is still significant.
Specifically, using the full training dataset, the combined effect of source-removal and data augmentation raises the accuracy on the validation set 
from $80.2\%\pm0.2$ to $84.3\%\pm0.4$.

\subsection{Final results}
\label{sec:final}
\begin{figure*}\begin{center}
\includegraphics[width=0.24\textwidth]{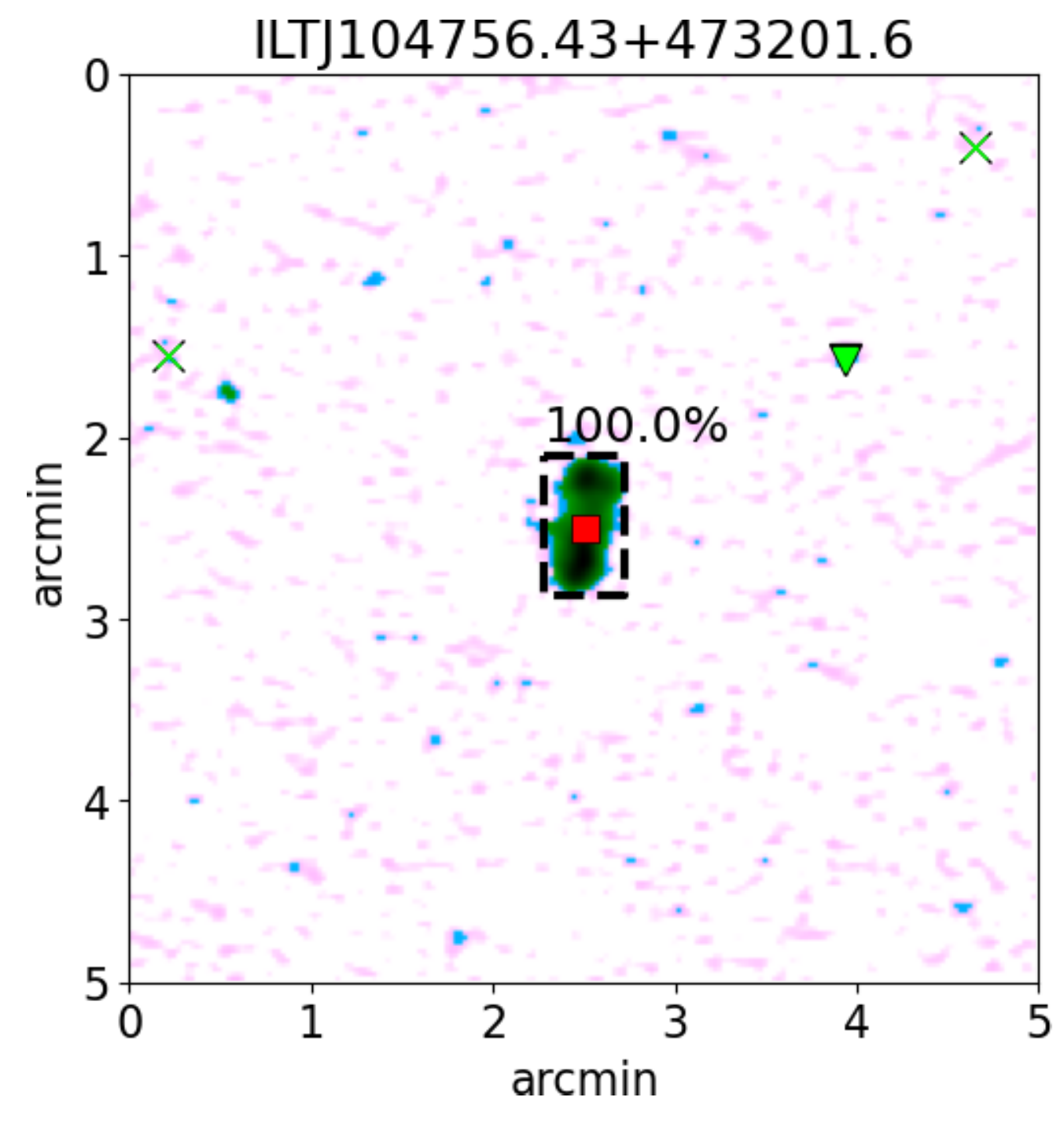}
\includegraphics[width=0.24\textwidth]{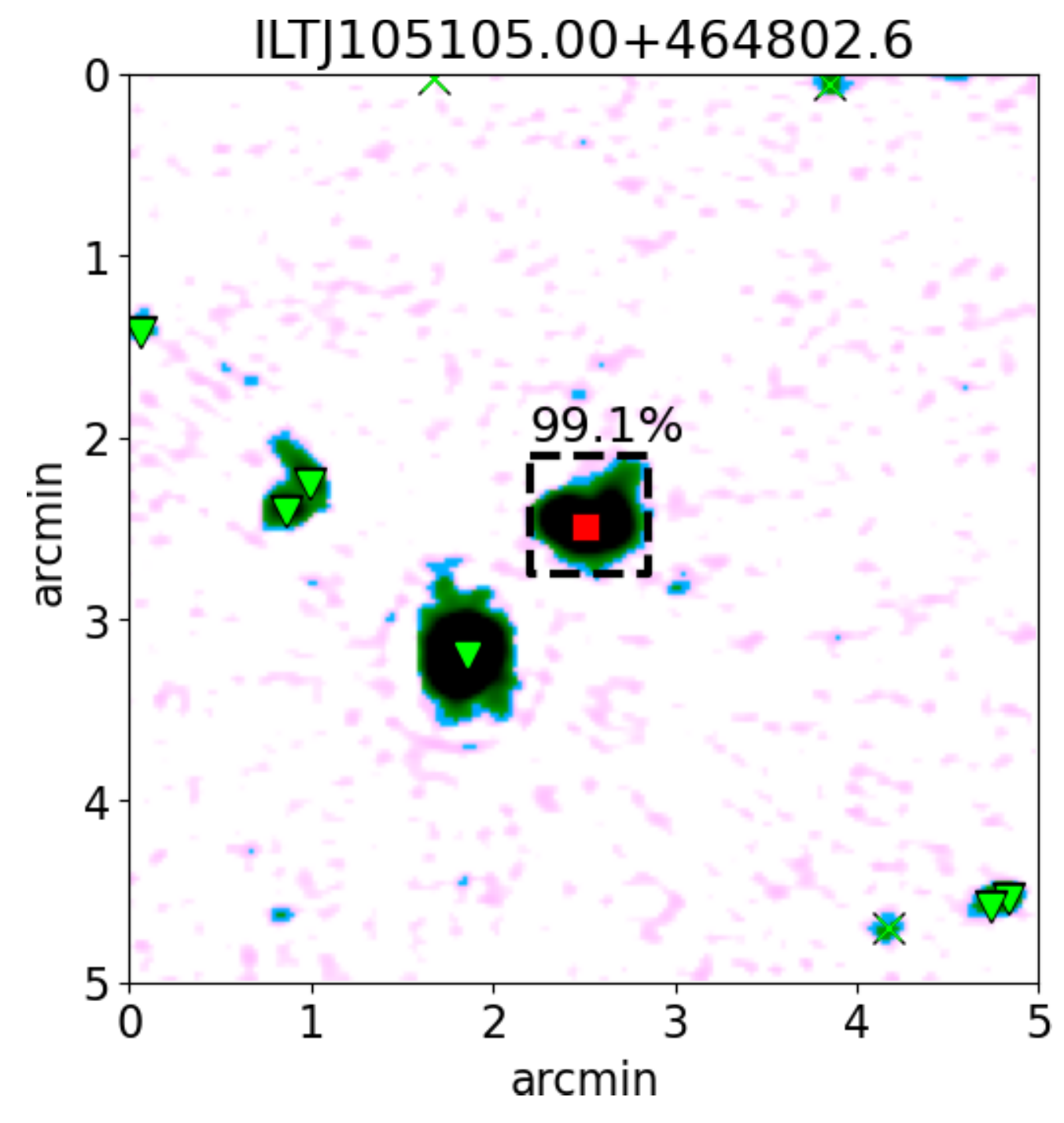}
\includegraphics[width=0.24\textwidth]{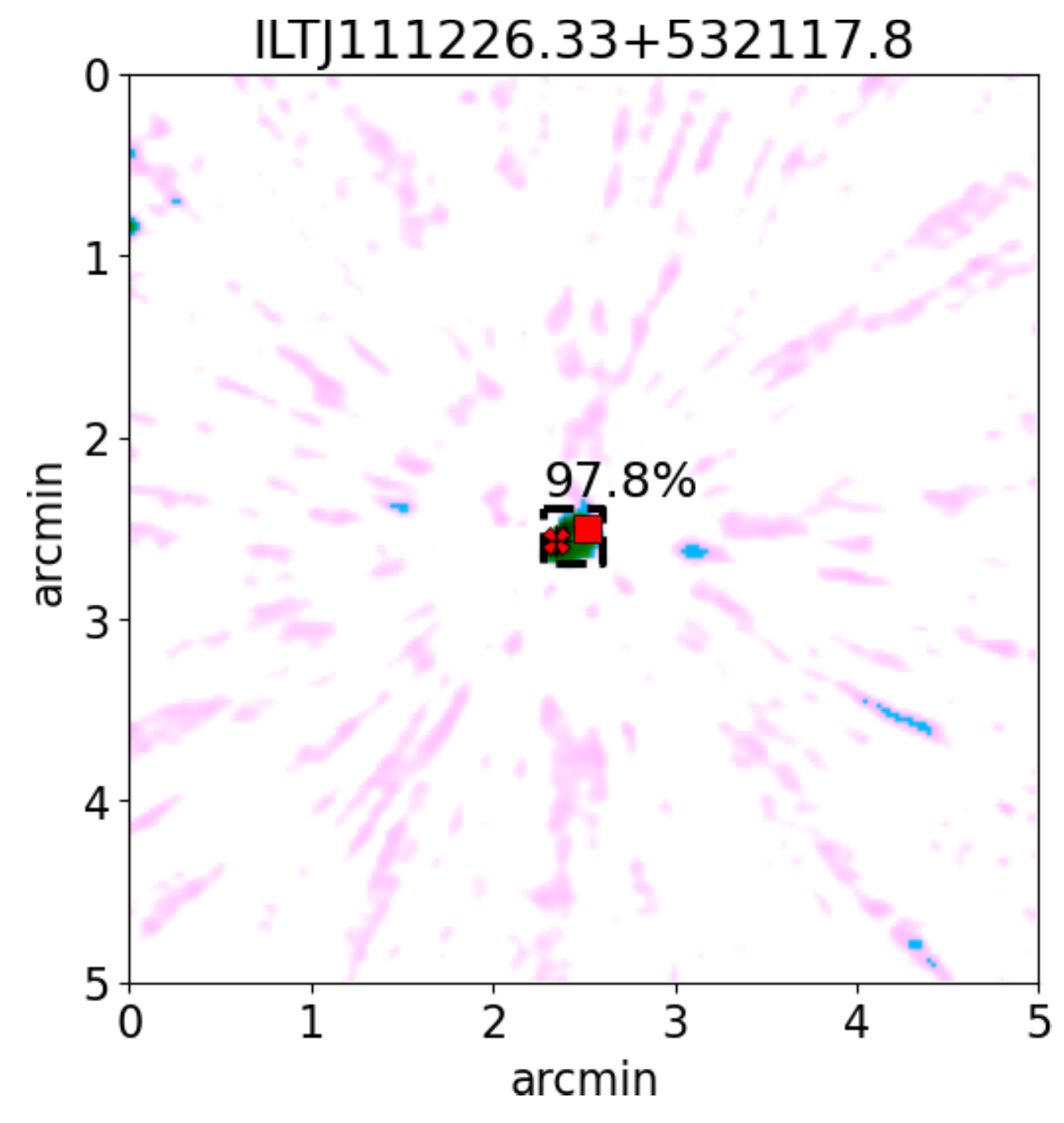}
\includegraphics[width=0.24\textwidth]{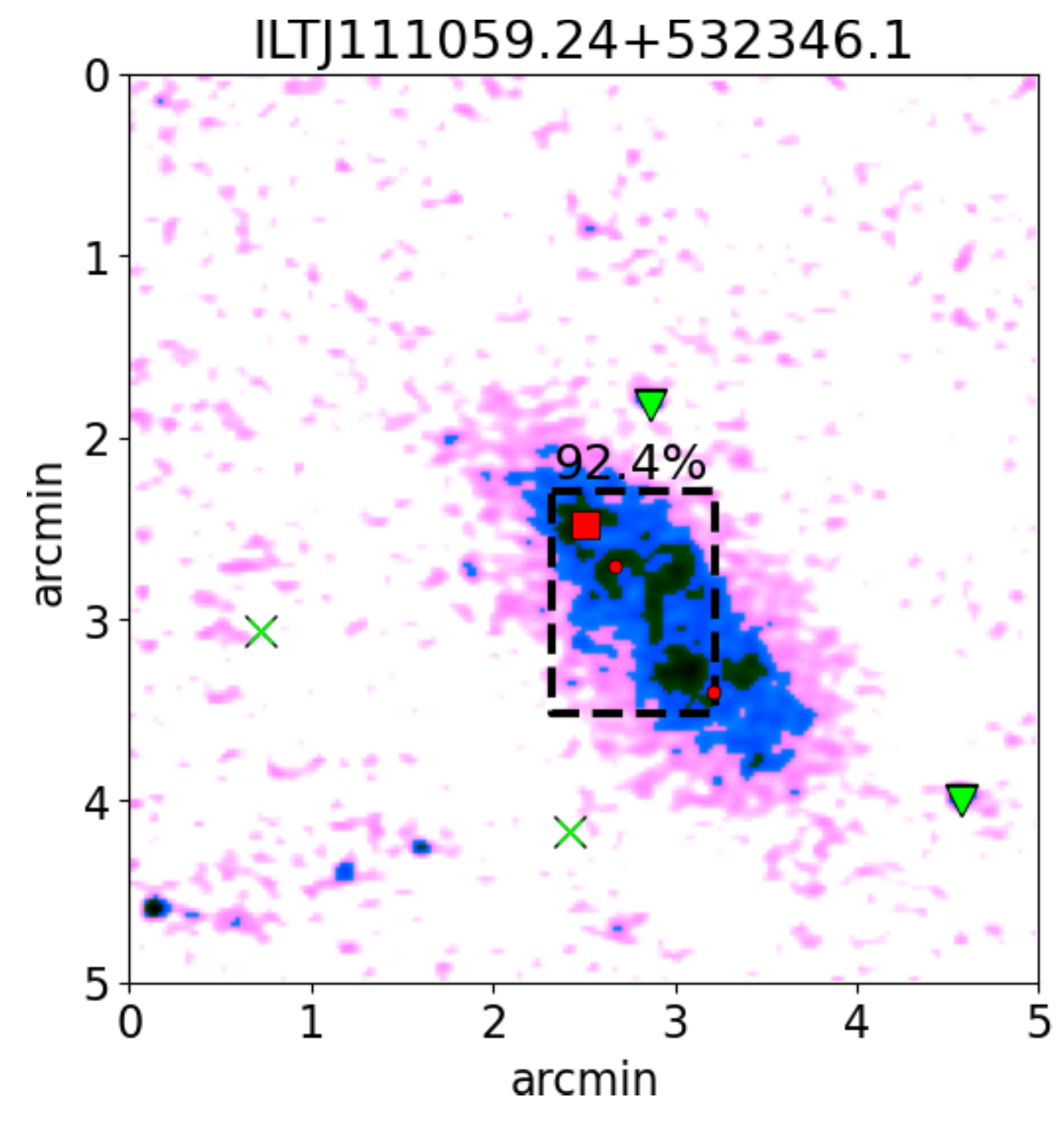}

\includegraphics[width=0.24\textwidth]{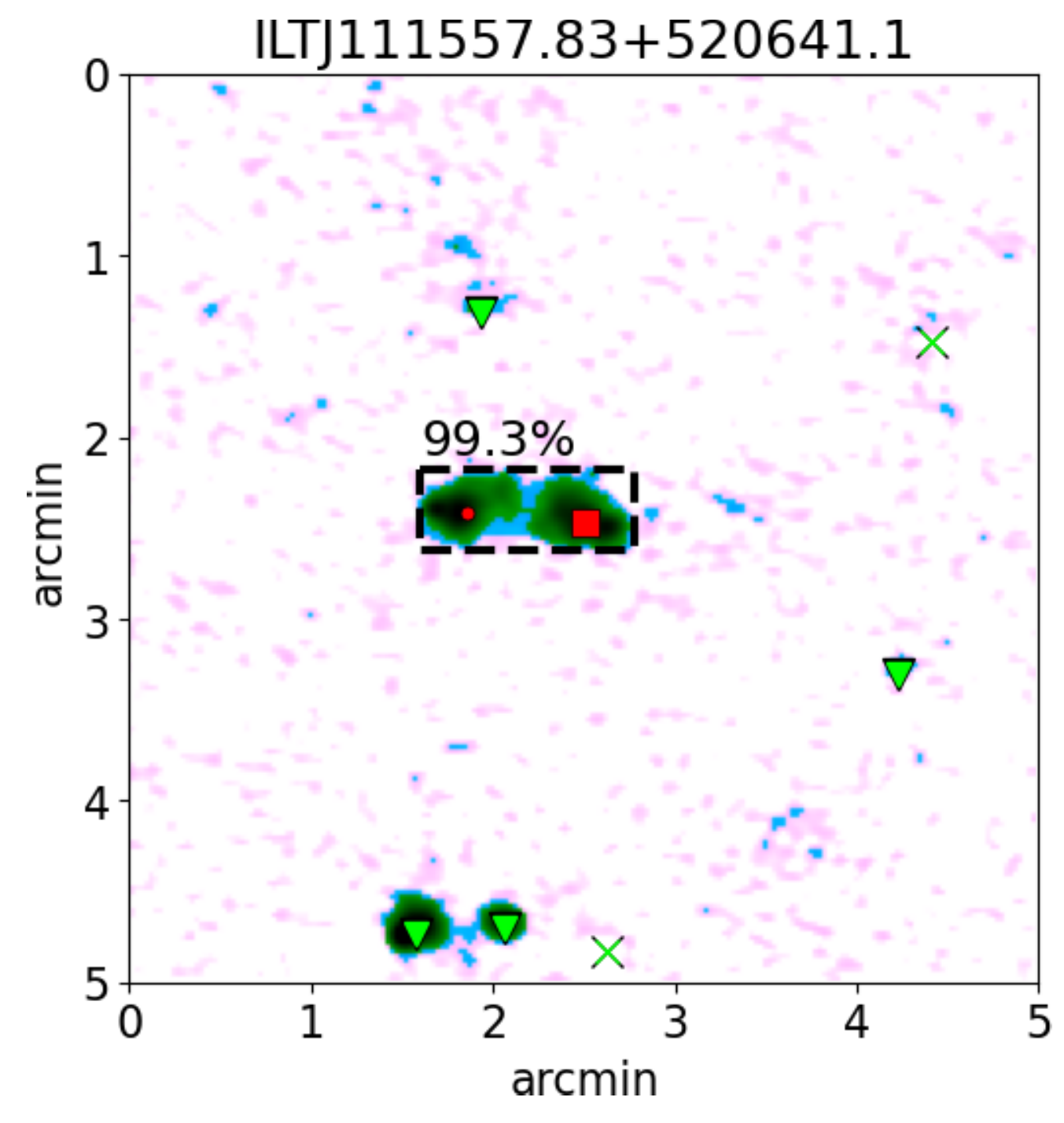}
\includegraphics[width=0.24\textwidth]{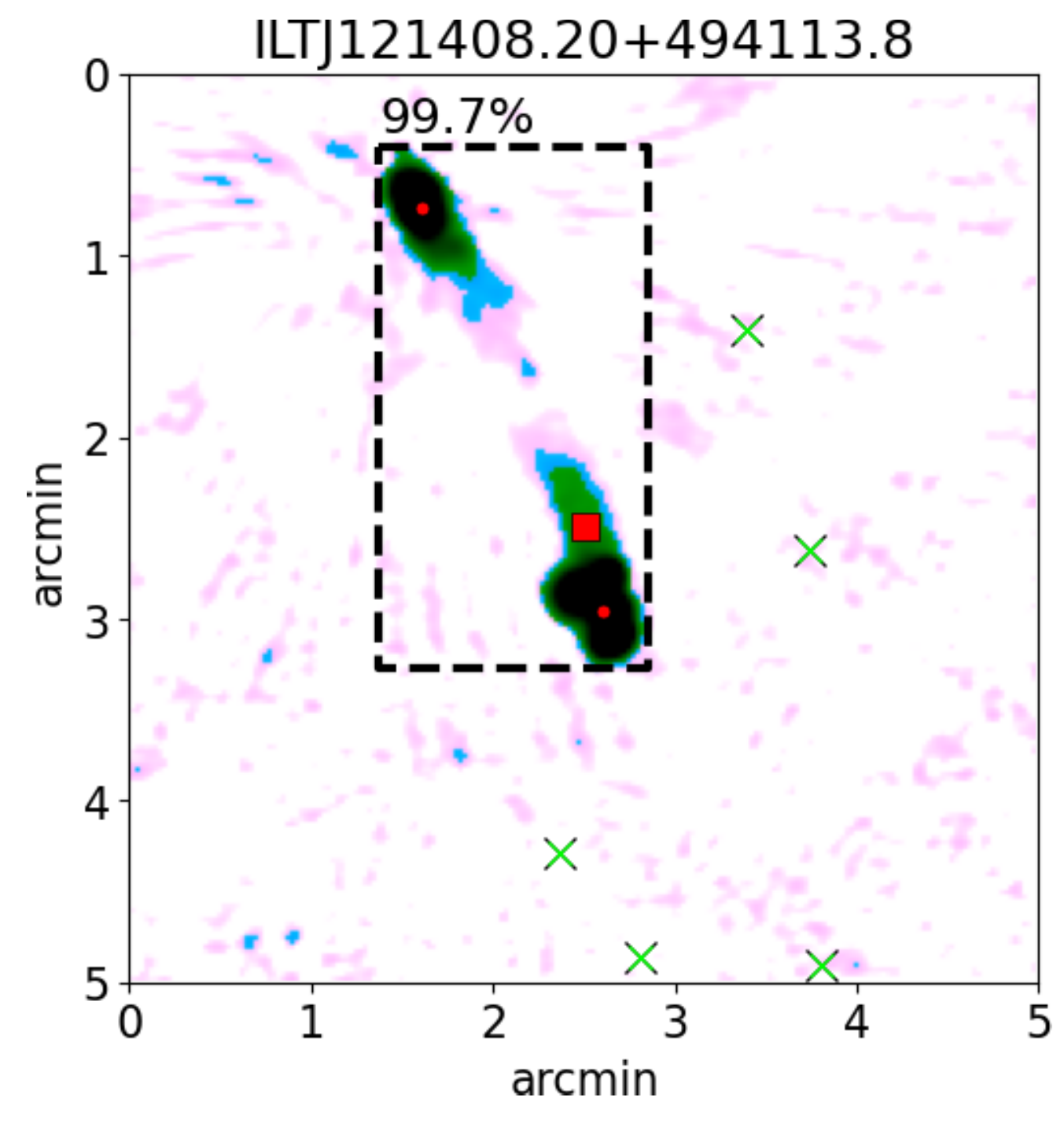}
\includegraphics[width=0.24\textwidth]{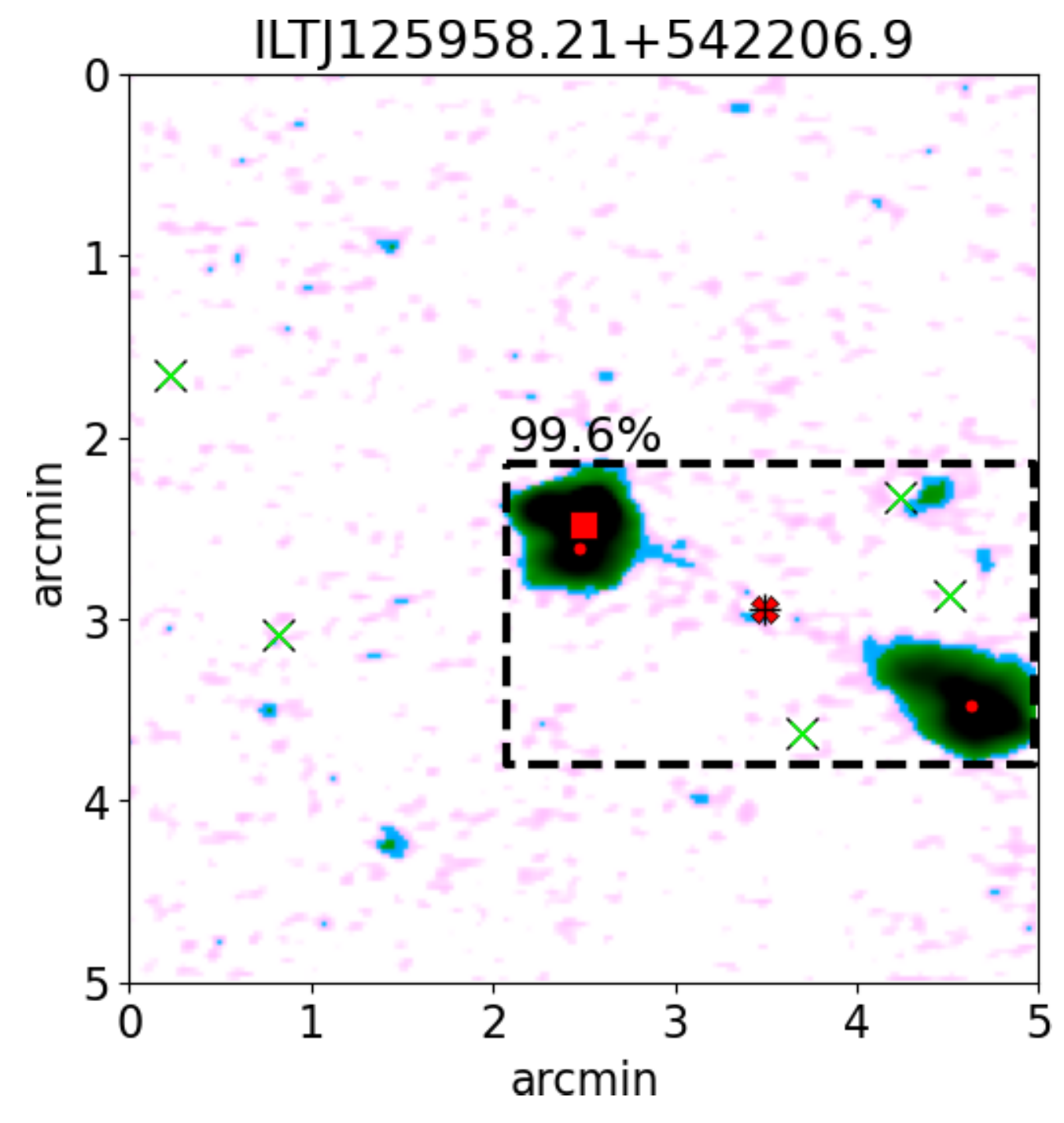}
\includegraphics[width=0.24\textwidth]{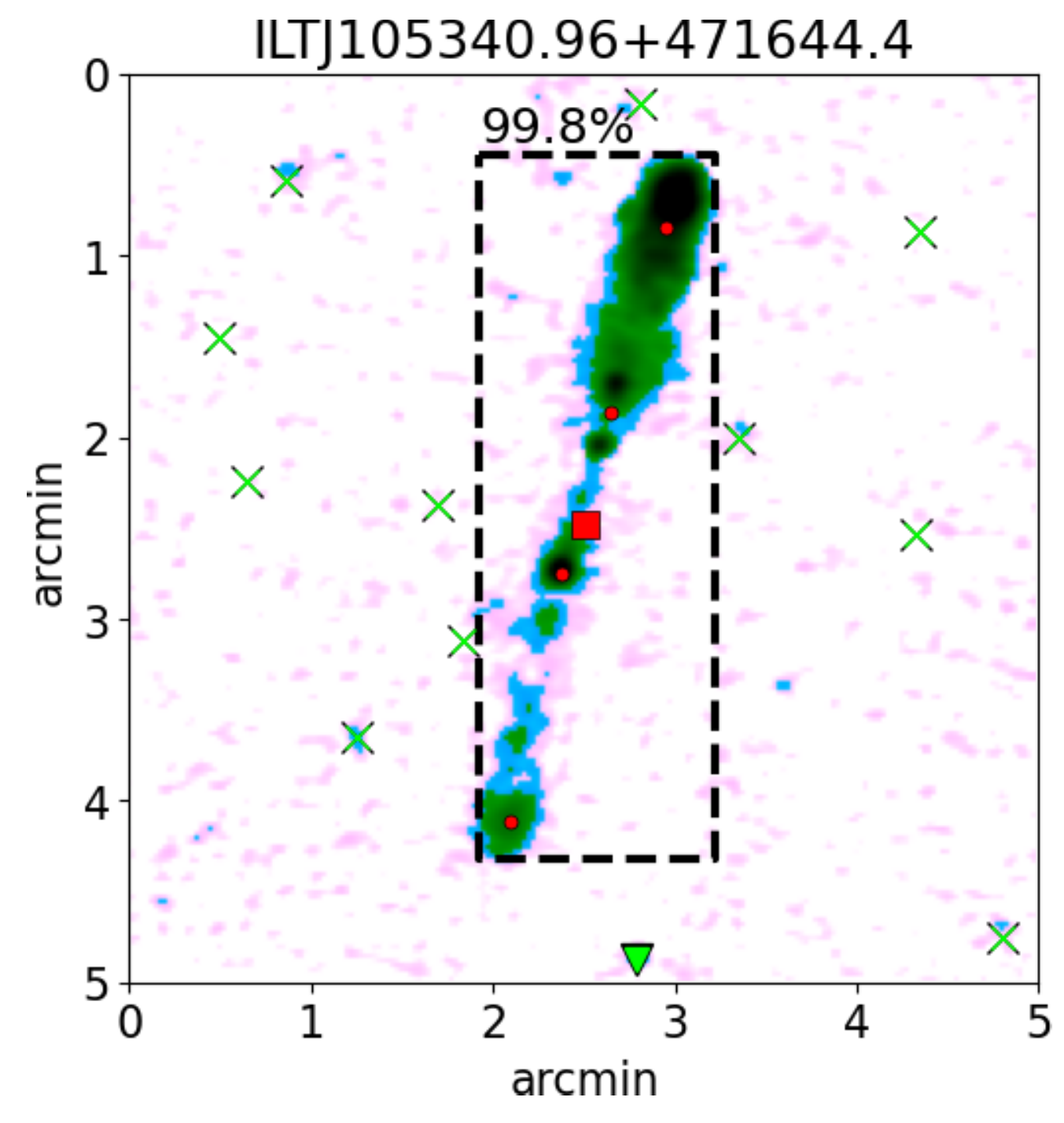}

\includegraphics[width=0.24\textwidth]{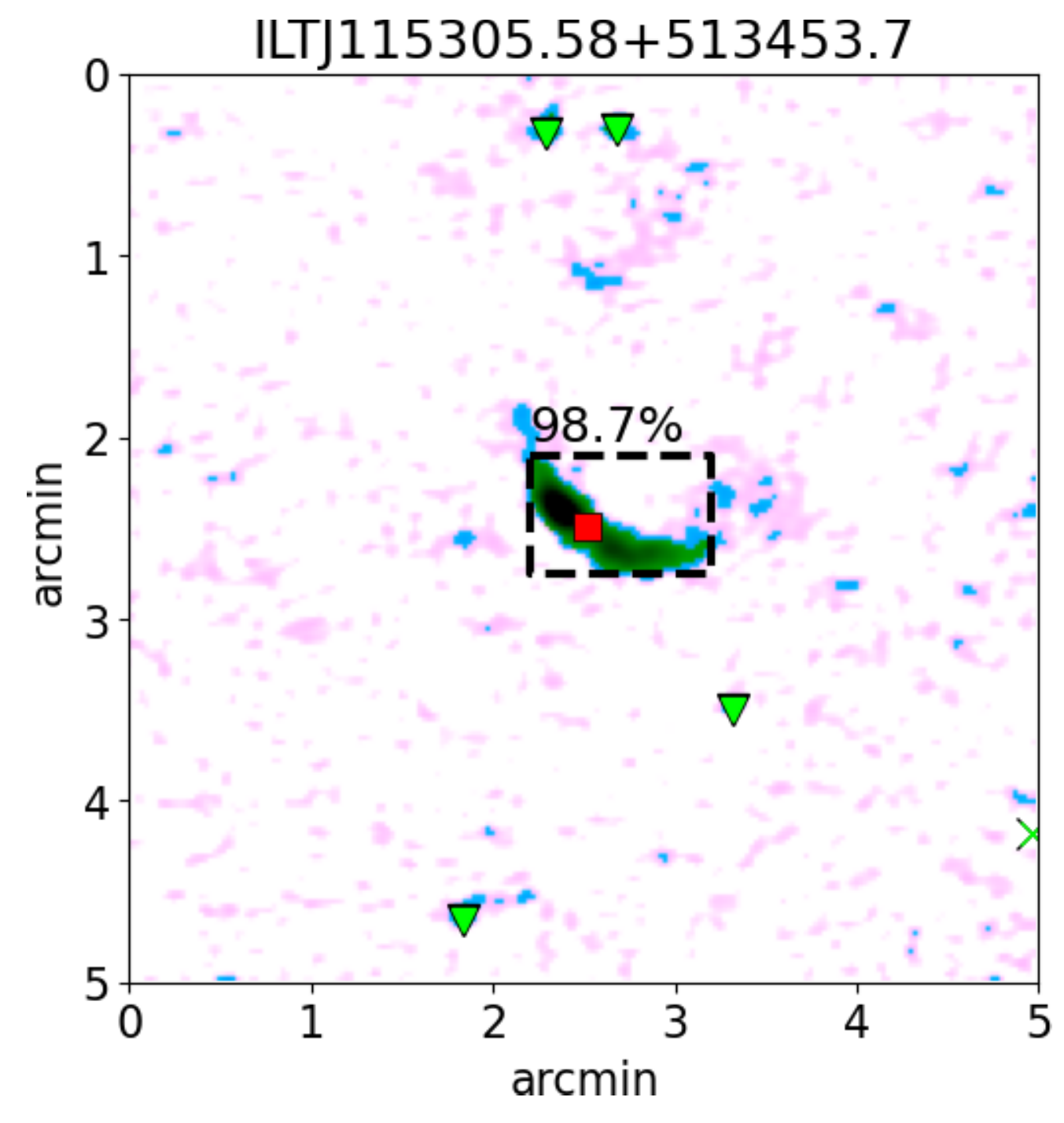}
\includegraphics[width=0.24\textwidth]{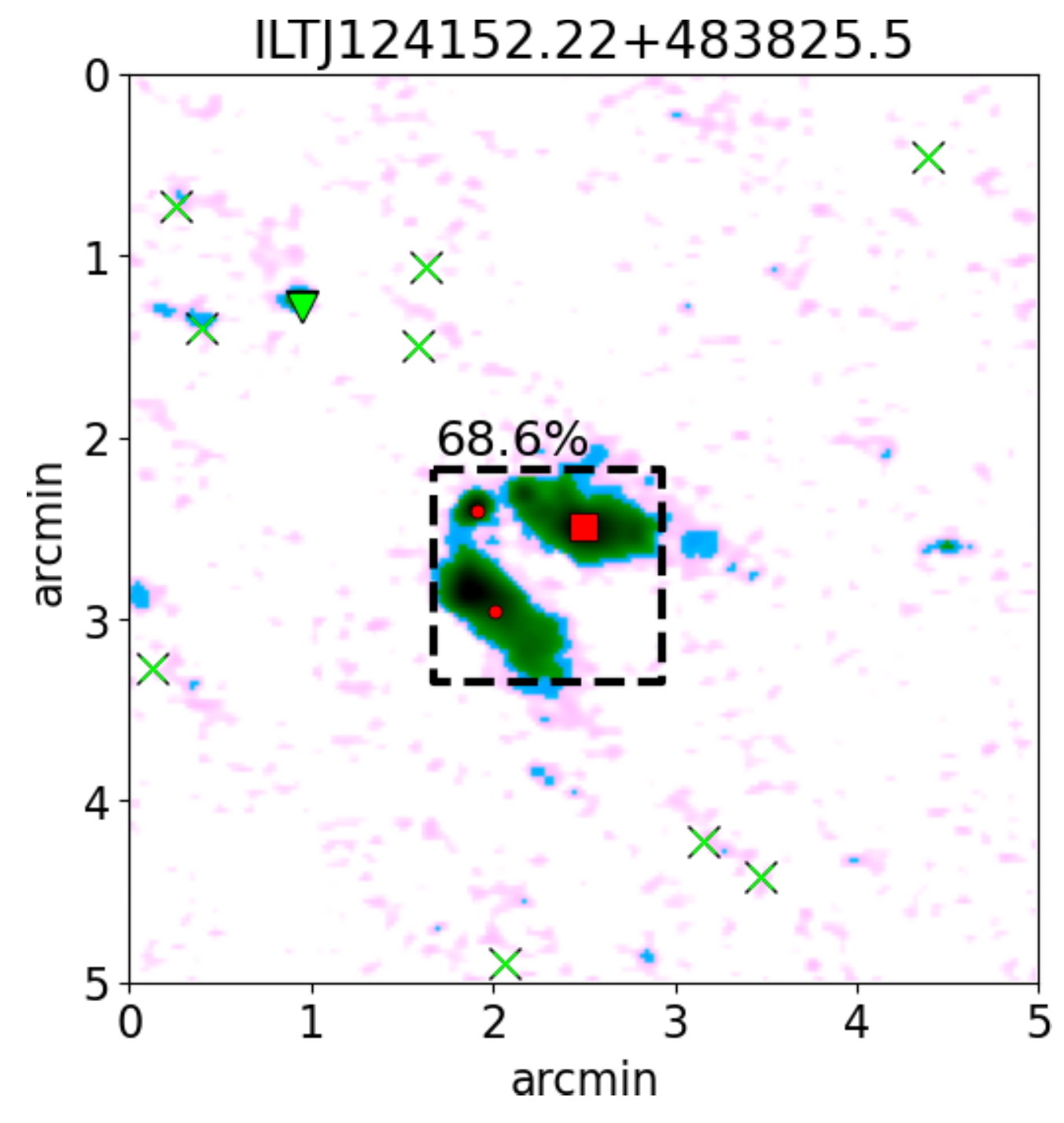}
\includegraphics[width=0.24\textwidth]{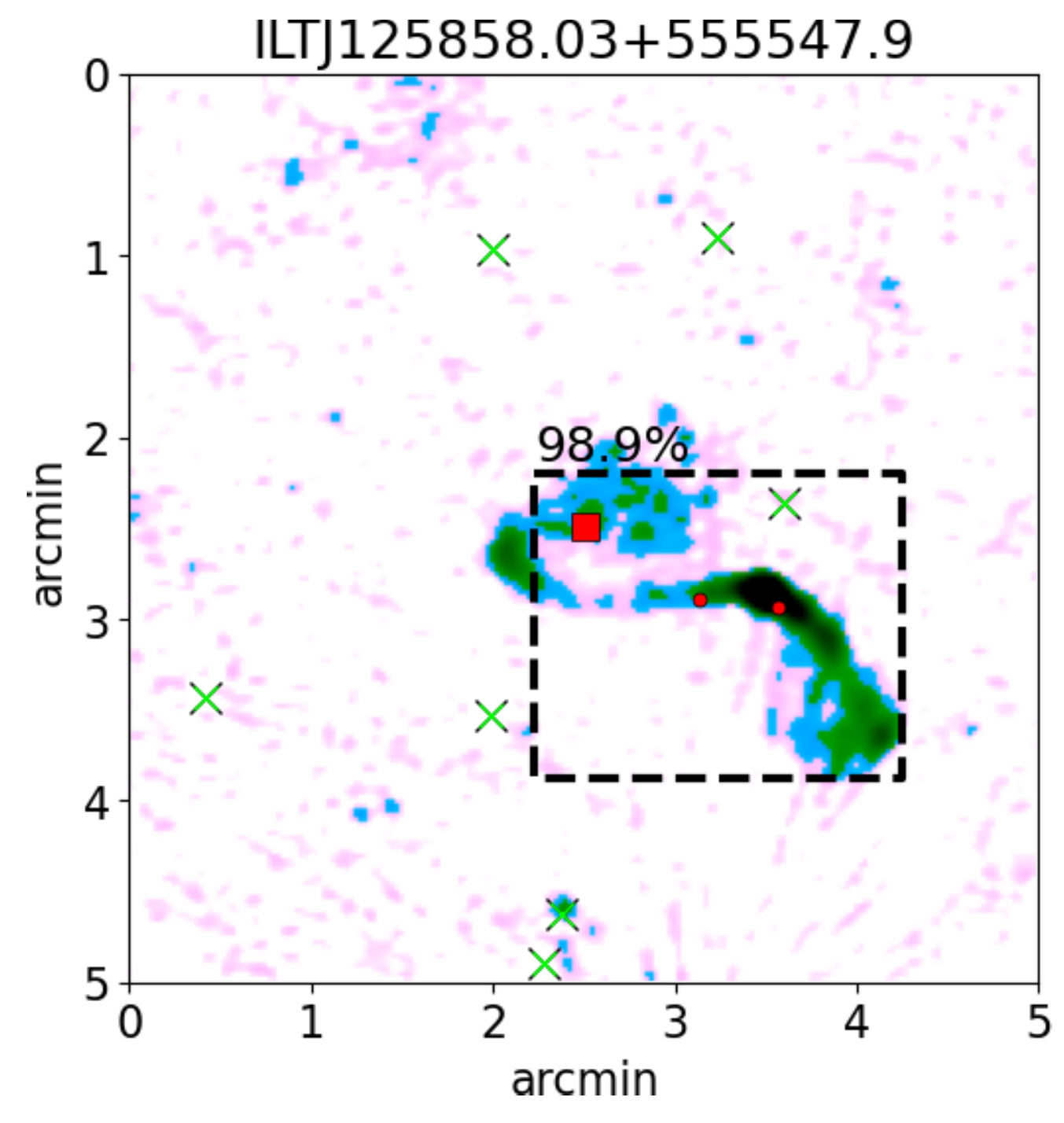}
\includegraphics[width=0.24\textwidth]{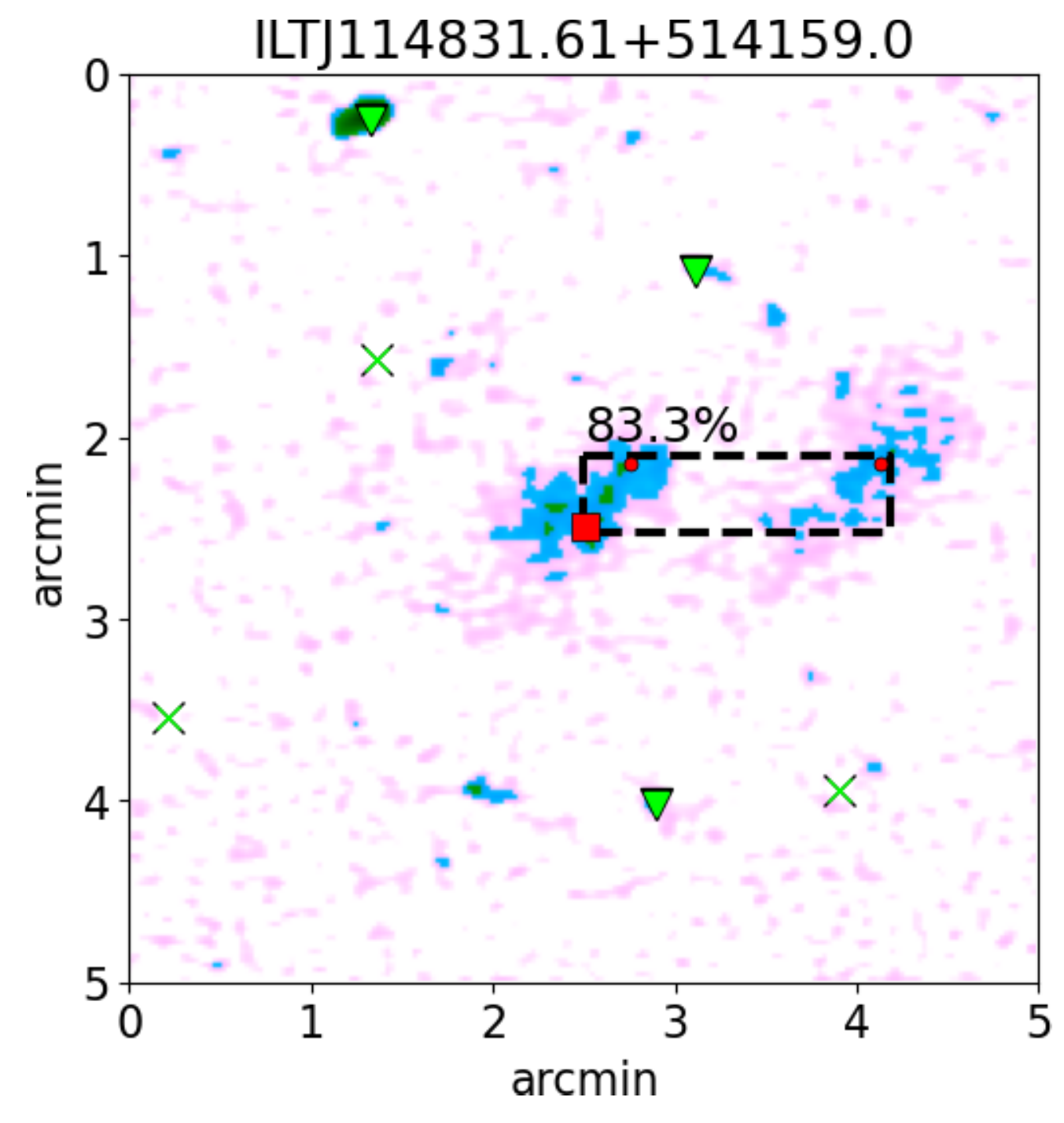}
\caption{Examples of predictions (black dashed rectangles) for images from the validation set that match the manually created catalogue. These examples are curated to show the model predictions for a wide range of source morphologies. Each image is a $300\times300$ arcsec cutout of LoTSS-DR2 Stokes-I, pre-processed as detailed in Section \ref{sec:prepro}. The red square indicates the position of the focussed \textit{PyBDSF} radio component, and the red dots indicate the position of \textit{PyBDSF} radio components that are related to the focussed component according to the corrected LOFAR Galaxy Zoo catalogue. The green triangles indicate the position of components that are unrelated to the focussed component. `x's (thick marker if related, thin if unrelated) indicate components that we removed, and a black cross on top of an `x' means the component was automatically reinserted after the  prediction. In our method, all components that fall inside the predicted black rectangular box (and are not removed) are combined into a single radio source.}
\label{fig:good}
\end{center}\end{figure*}

Up to this point, we present the results on the validation dataset for three models trained with different random seeds.
On the basis of these tests, we opted to use the adapted Fast R-CNN with a FPN-ResNet-50 classification backbone, rotation data augmentation, and unresolved source removal, training with a constant learning rate for 20k iterations.
Using this model and settings, we measured a catalogue accuracy of $85.3\%\pm0.6$ and $88.5\%\pm0.3$ for the large and bright source components on the test and training set, respectively.  
We note that the trained model happens to perform slightly better on the images in the test set than in in the validation set; one may conclude that the performance on the validation and test sets are roughly equal.

The final results we present here and in the discussion Section \ref{sec:discussion} are based on a single model trained with the random seed that resulted in the best performing model on our test set.
In Fig. \ref{fig:good}, we show examples of successful associations.
Going from left to right in the top row of the figure, we see correct predictions that tightly encompass physically related radio emission in a number of different situations, including the following: a single-component radio source while no other emission is nearby; a single-component radio source when another compact source is nearby; a bright unresolved source and the nearby artefact it produces; and all components of a nearby star forming galaxy. 
In the second row, the predictions correctly encompass the following multi-component sources: a typical double-lobed radio source where all related radio emission is clearly connected and above the noise; a double-lobed radio source where the jet-related emission fell below the noise; a radio source with two clear but separated lobes and a radio core that got successfully reinserted; a very elongated radio source with a core and two (fragmented) lobes.
In the third row, we show examples of FRI or edge-darkened radio sources of different sizes, different signal-to-noise ratios, and different bending angles.

\begin{figure*}\begin{center}
\includegraphics[width=0.24\textwidth]{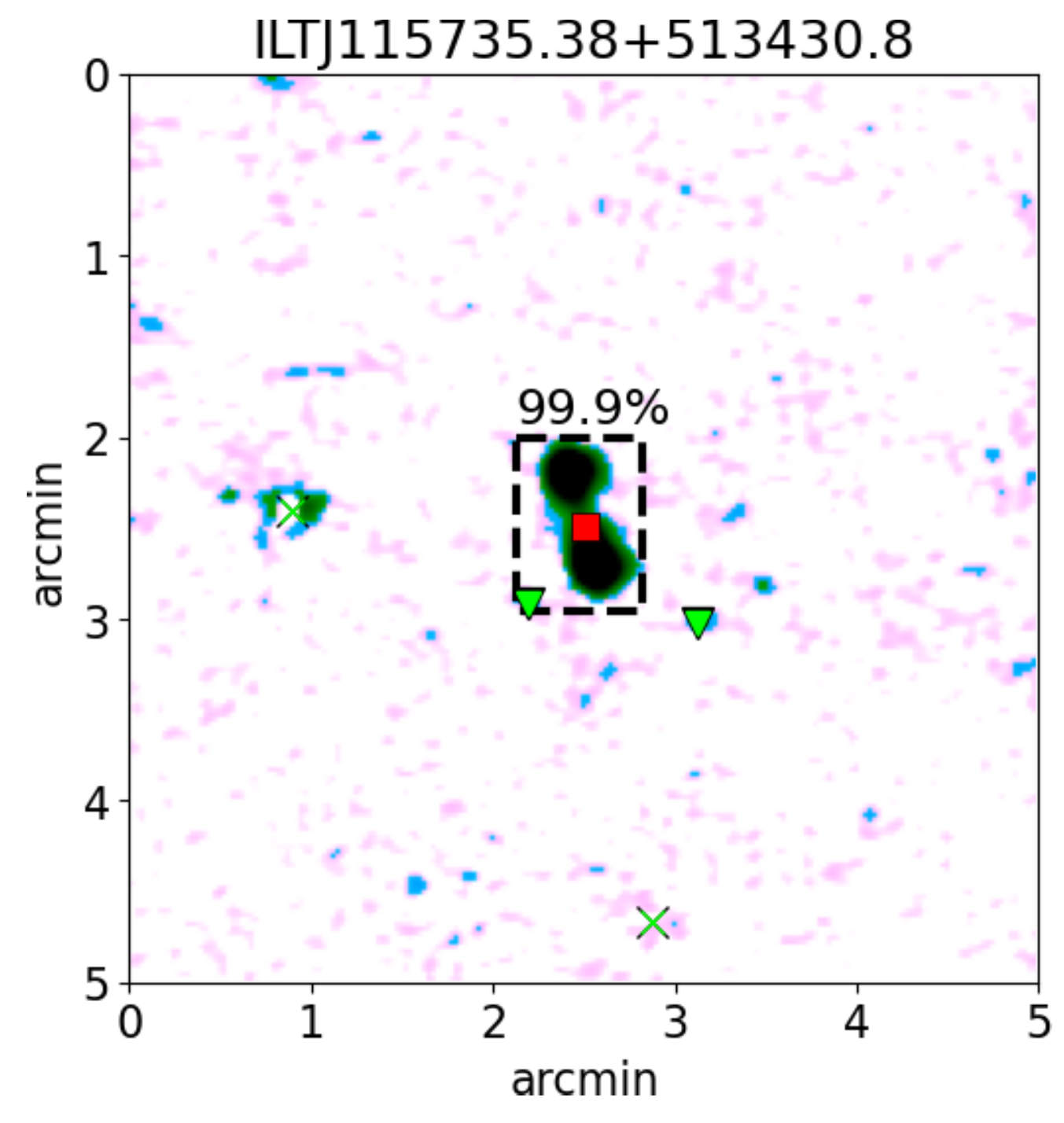}
\includegraphics[width=0.24\textwidth]{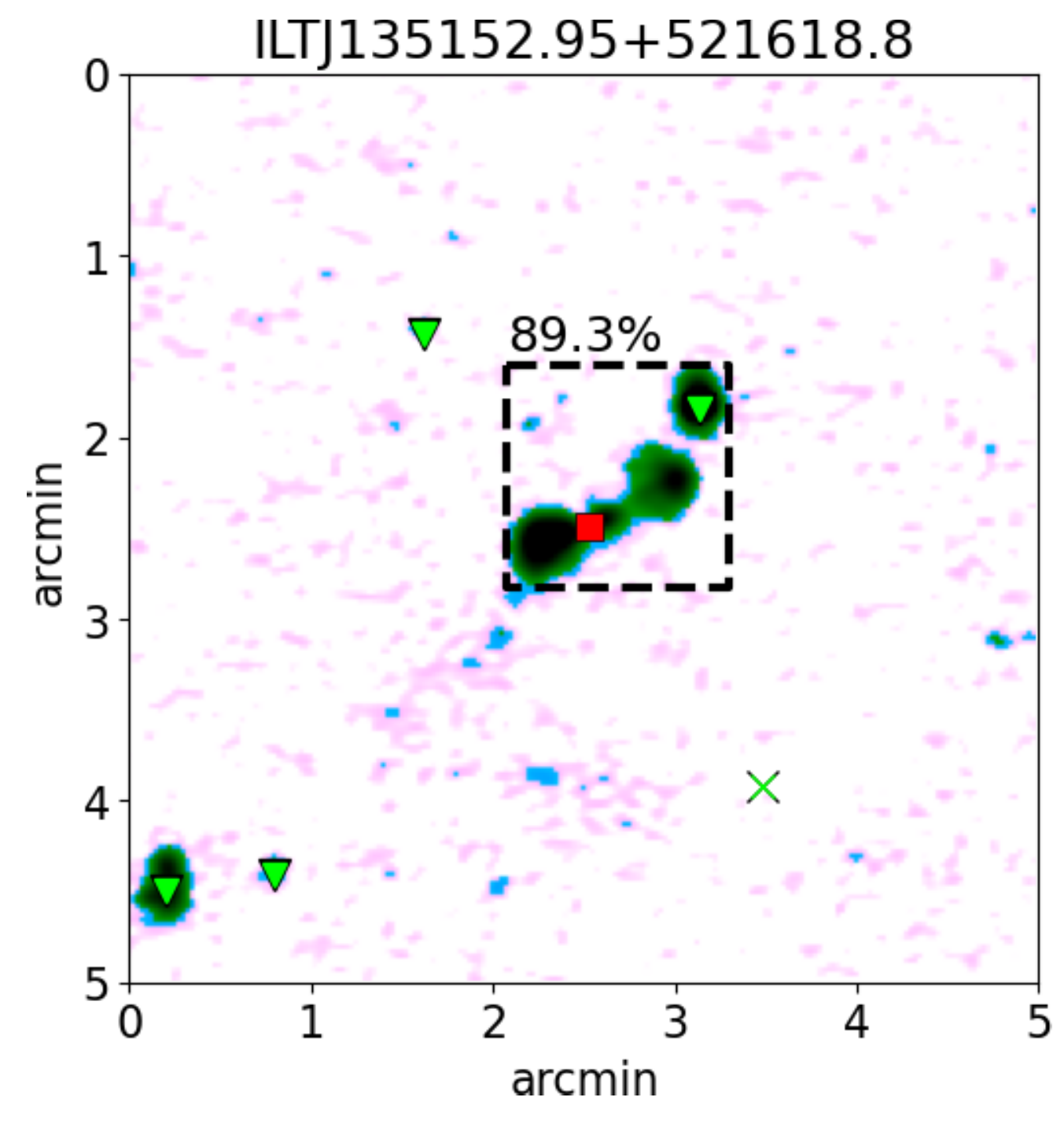}
\includegraphics[width=0.24\textwidth]{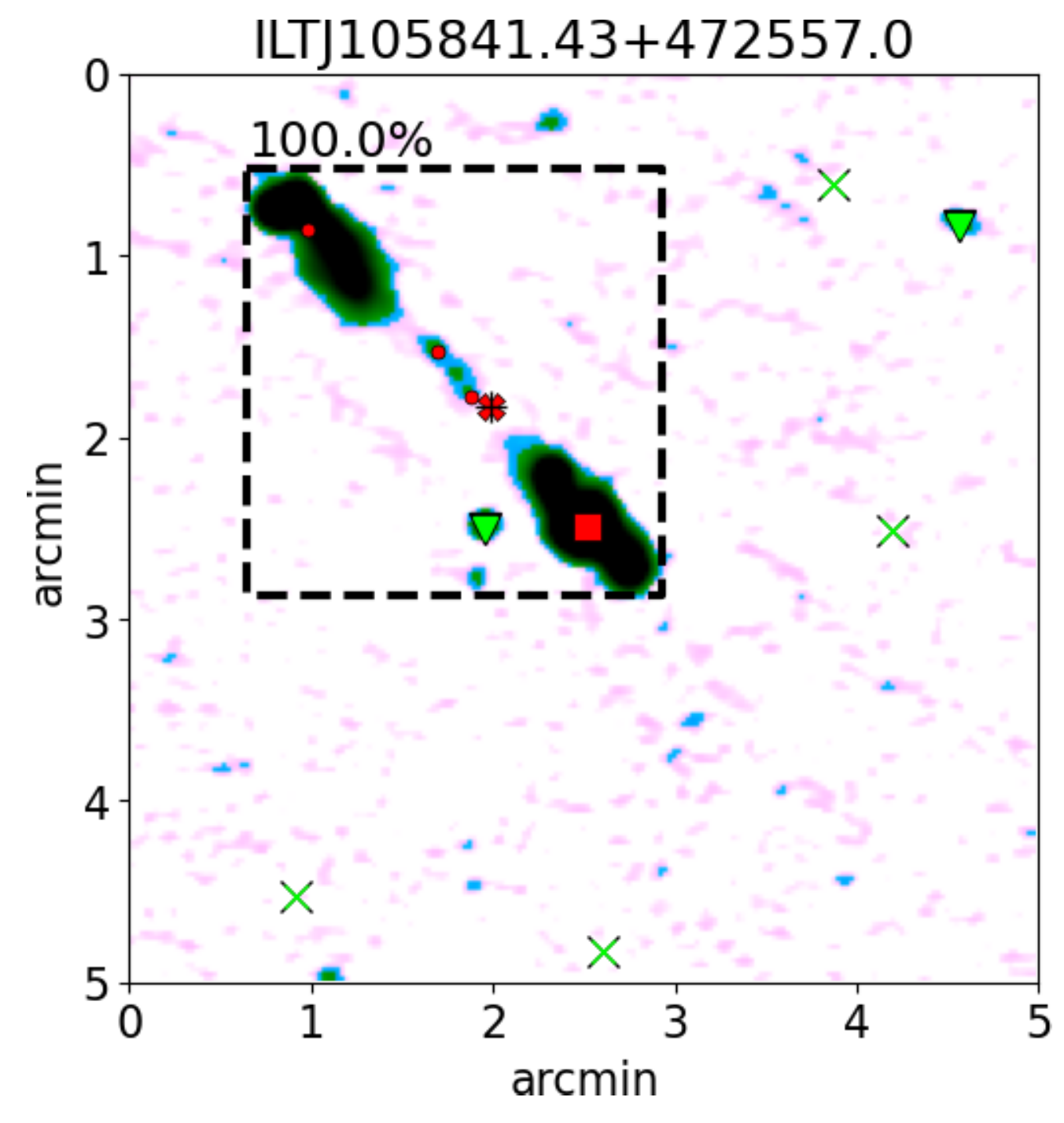}
\includegraphics[width=0.24\textwidth]{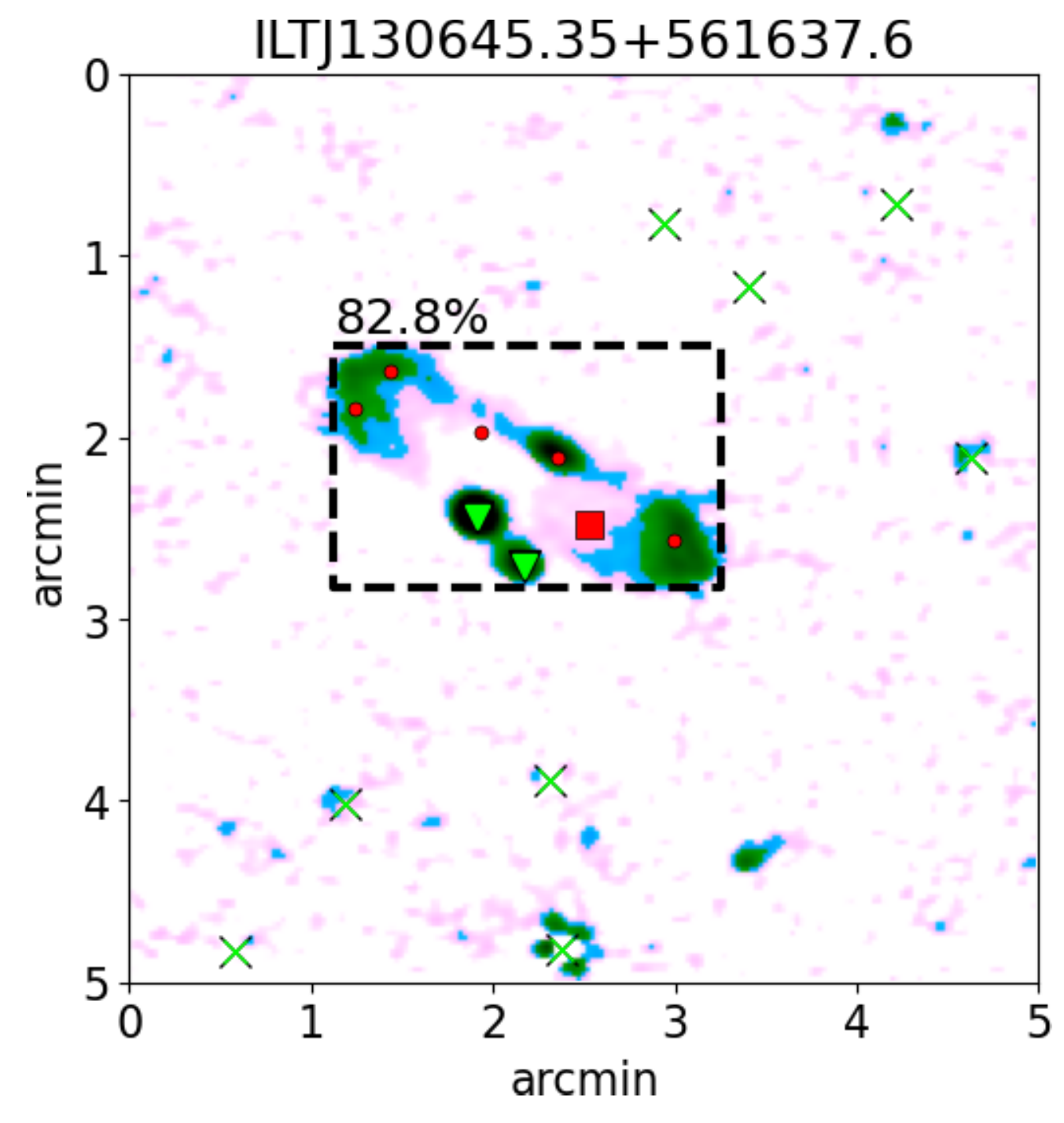}

\includegraphics[width=0.24\textwidth]{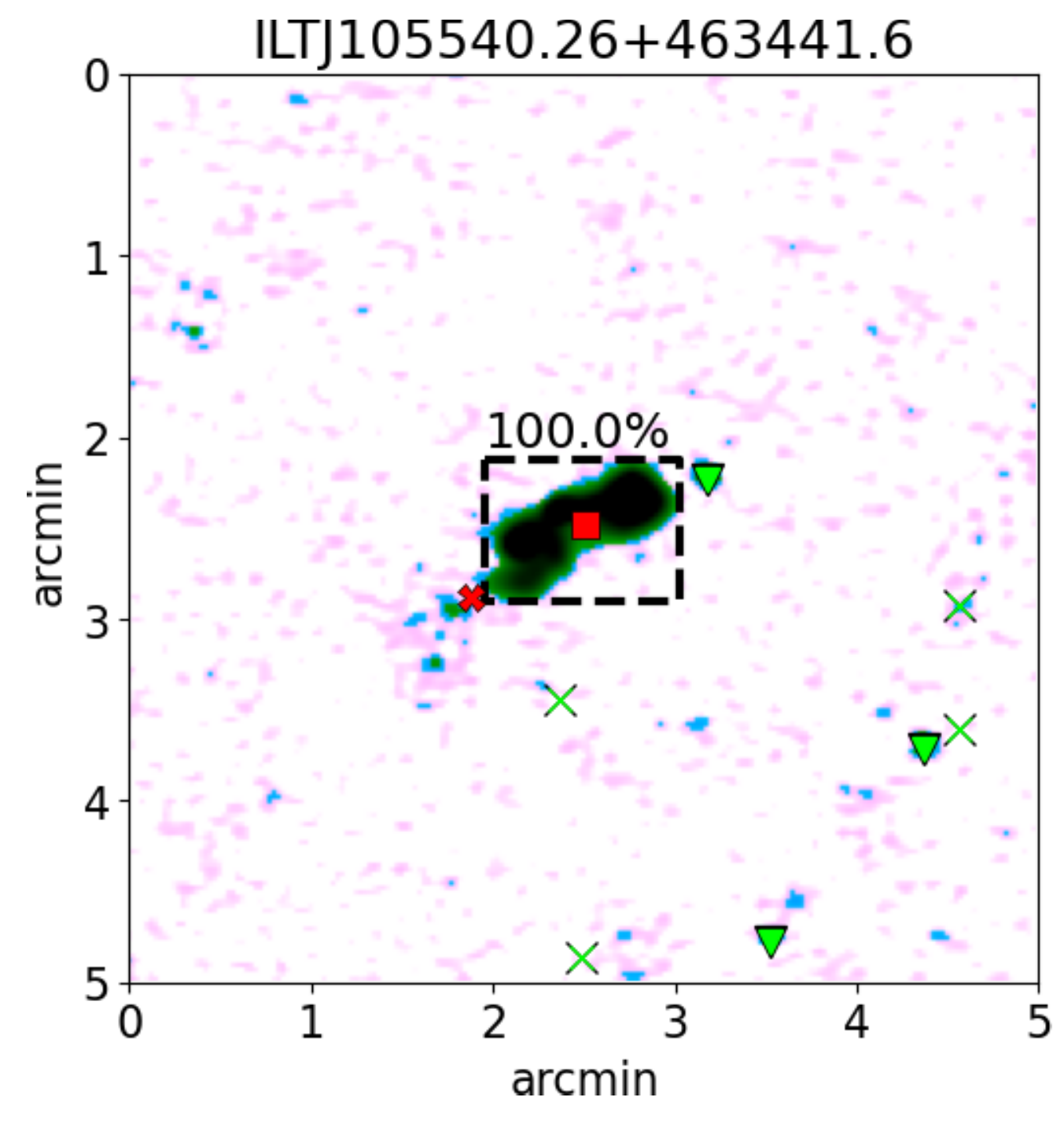}
\includegraphics[width=0.24\textwidth]{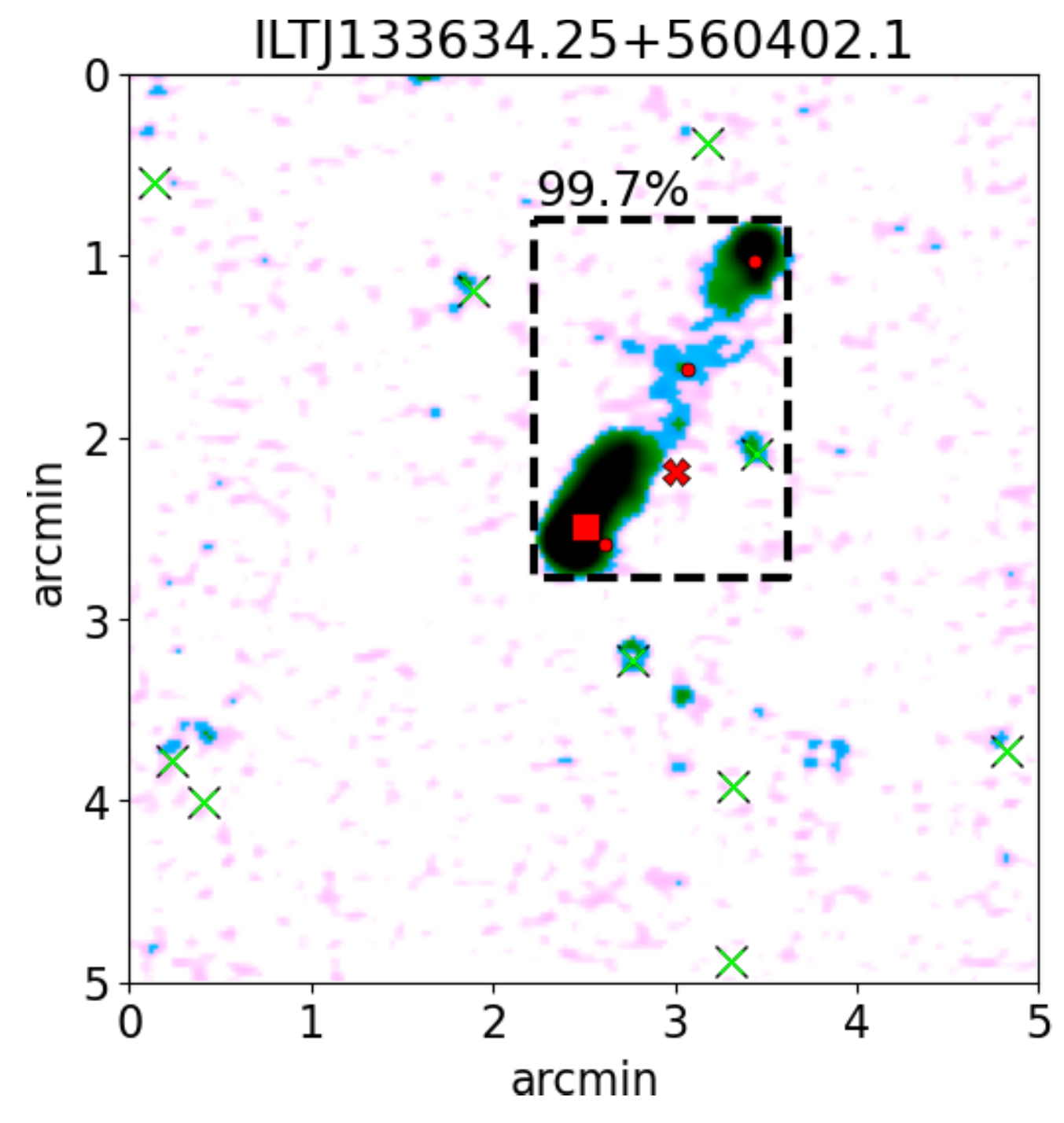}
\includegraphics[width=0.24\textwidth]{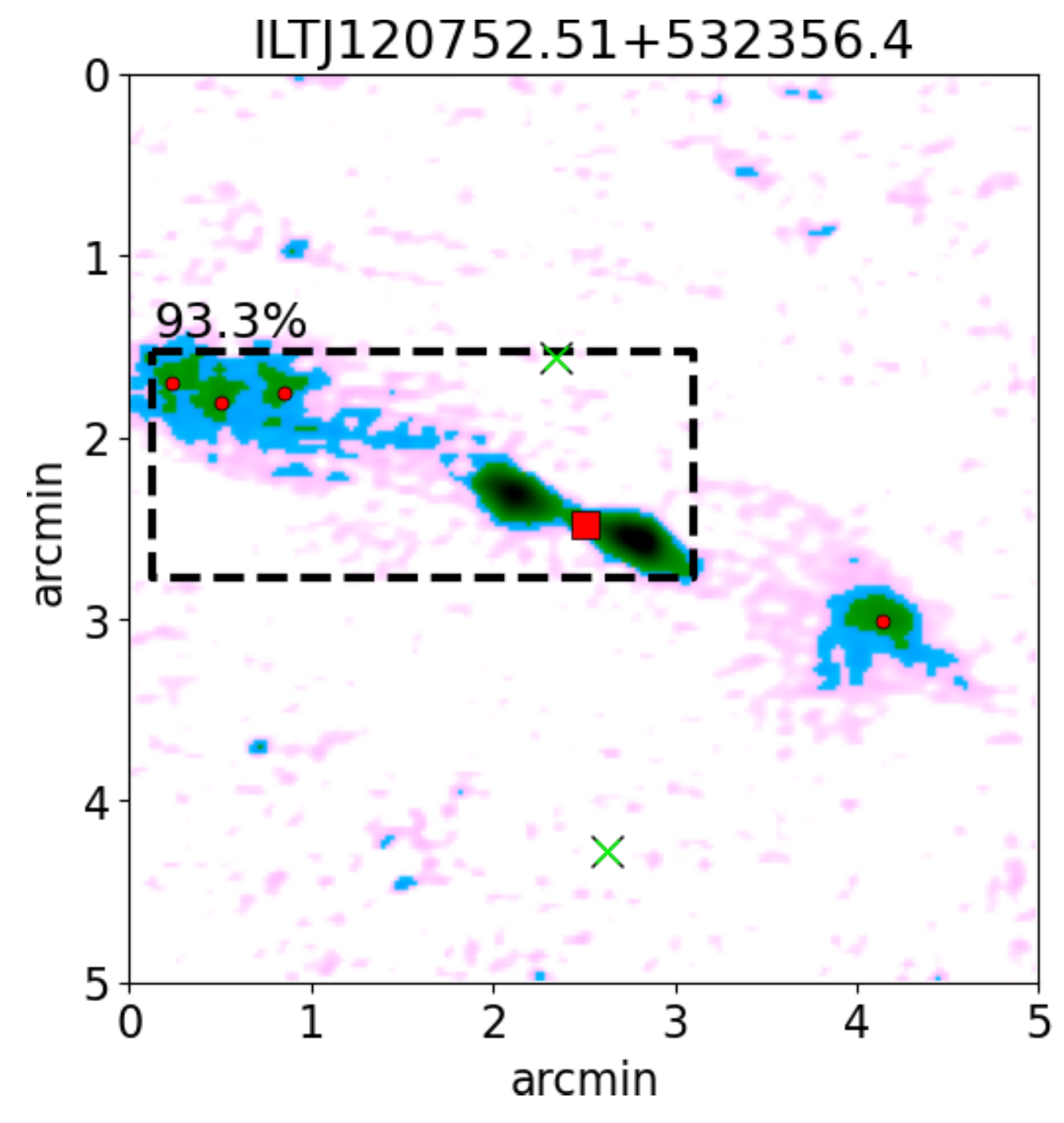}
\includegraphics[width=0.24\textwidth]{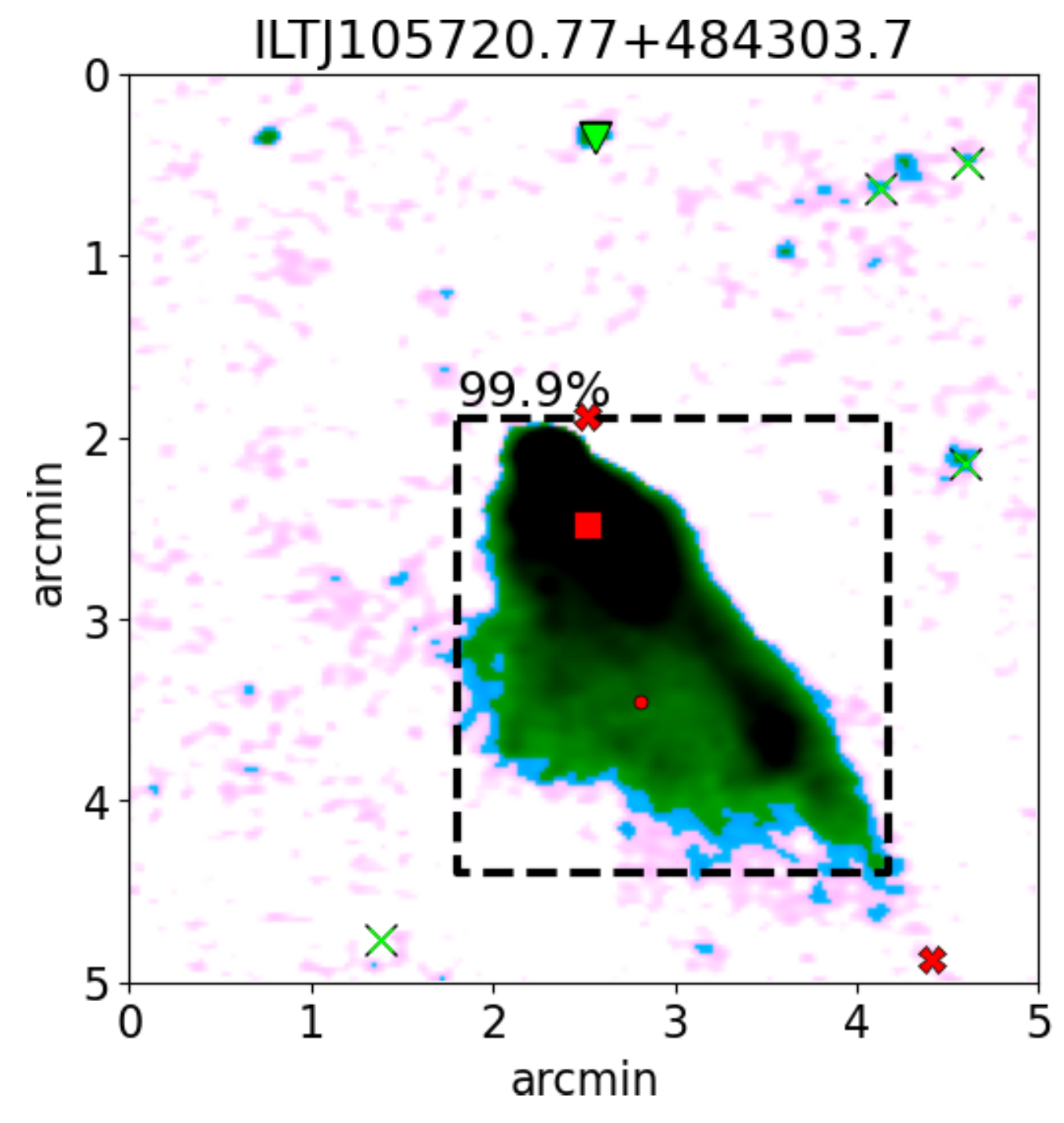}

\includegraphics[width=0.24\textwidth]{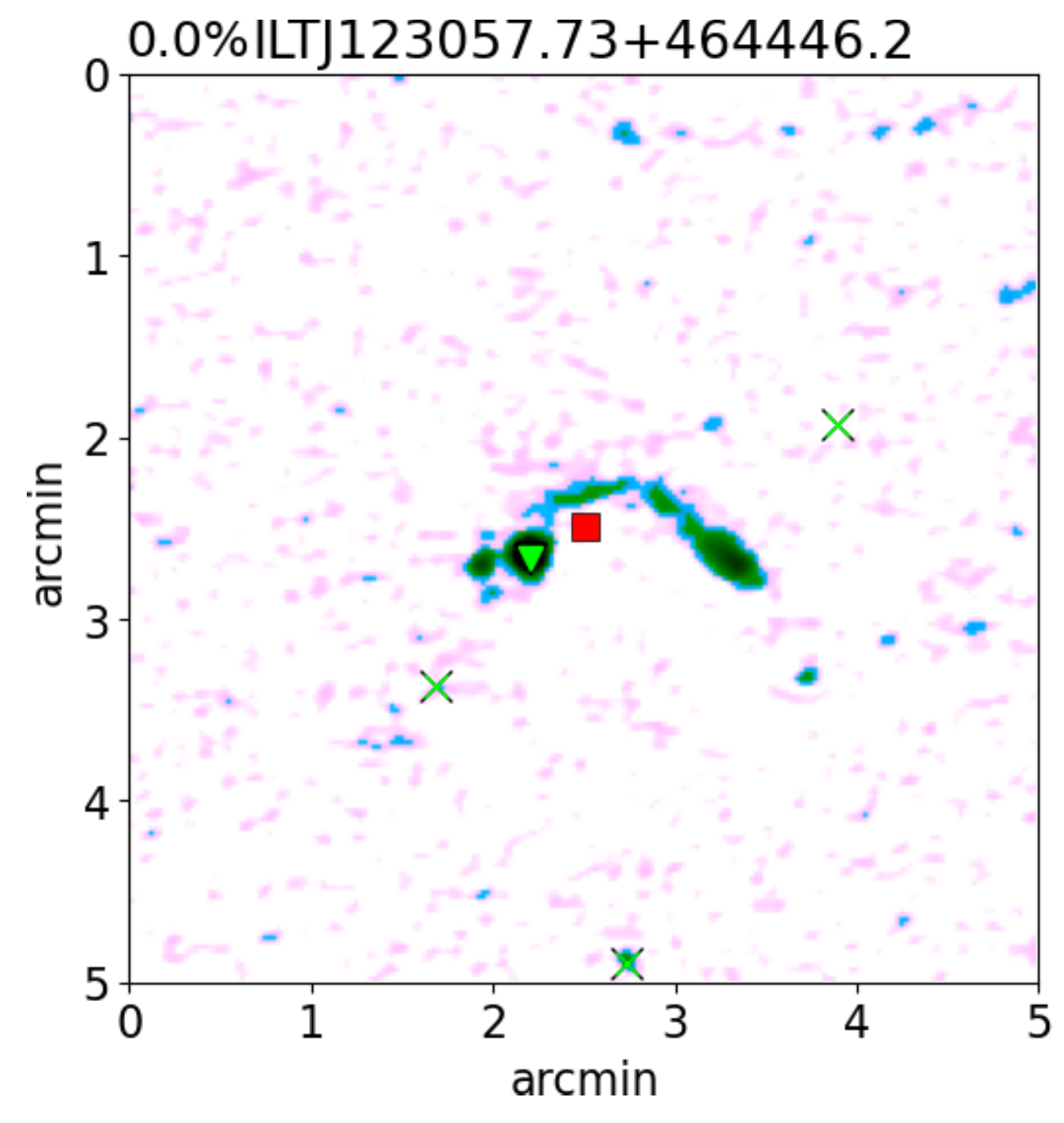}
\includegraphics[width=0.24\textwidth]{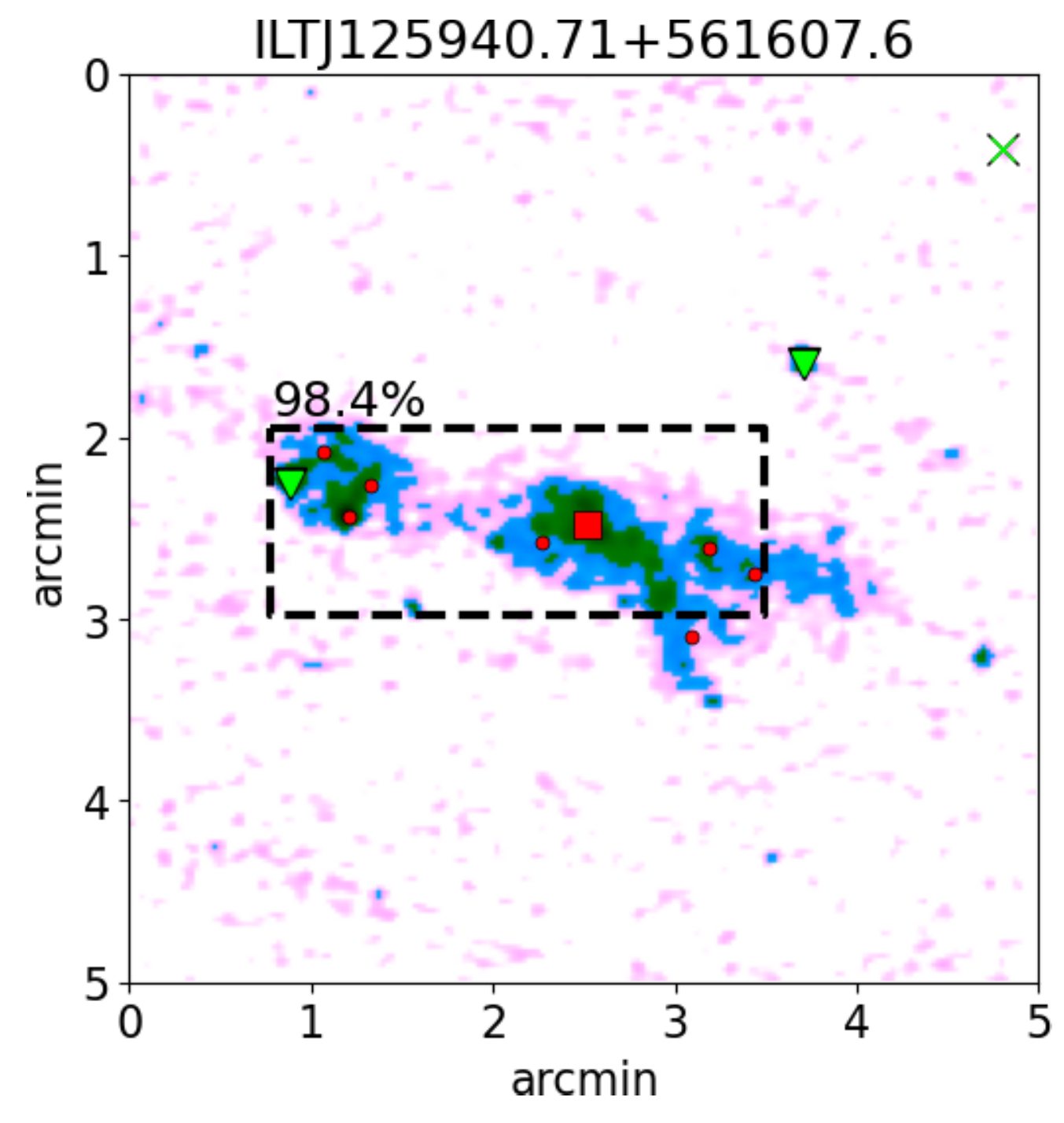}
\caption{Examples of predictions (black dashed rectangles) for images from the validation set that do not match the manually created catalogue. These examples are curated to show the model predictions for different source morphologies and both single- and multi-component sources. Each image is a $300\times300$ arcsec cutout of LoTSS-DR2 Stokes-I, pre-processed as detailed in Section \ref{sec:prepro}. The red square indicates the position of the focussed \textit{PyBDSF} radio component, the red dots indicate the position of \textit{PyBDSF} radio components that are related to the focussed component according to the corrected LOFAR Galaxy Zoo catalogue. The green triangles indicate the position of components that are unrelated to the focussed component. `x's (thick marker if related, thin if unrelated) indicate components that we removed, and a black cross on top of an `x' means the component was automatically reinserted after the  prediction. In our method, all components that fall inside the predicted black rectangular box (and are not removed) are combined into a single radio source.}
\label{fig:bad}
\end{center}\end{figure*}
        
\begin{table*}
\caption{Results on the large and bright source component test dataset using our trained model. The number of components in each category are also given as a percentage of the total number of components in the test dataset.
The false positives (FPs) are broken down further into two categories. If the prediction for a component simultaneously includes both too many and too few components, it is grouped in the `too many' category.}             
\label{tab:results}      
\centering
\begin{tabular}{c c c c}
\hline\hline
                & All components        & Single component source       & Part of a multi-component source\\
Total           & 1121 (100.0\%)                & 701 (62.5\%)          & 420 (37.5\%)\\
\hline
Prediction is correct (TP)      & 962 (85.8\%)          & 650 (58.0\%)          & 312 (27.8\%) \\
Prediction includes too 
few components (FP+FN)  & 63 (5.6\%)            & 1 (0.1\%)             & 62 (5.5\%) \\
Prediction includes too 
many components (FP)    & 96 (8.6\%)            & 54 (4.8\%)            & 42 (3.7\%) \\
\hline
\end{tabular}
\end{table*}

Table \ref{tab:results} summarises the results as a percentage of the total number of components in the test dataset.
In Fig. \ref{fig:bad}, we show examples of associations that do not match our manual annotations. 
Going from left to right in the top row of the figure, the first two examples show cases where we associate more than one component with a single-component radio source. This happens for $7.7\%$ of single-component radio components.
The second two examples show cases where we associate too many components with a multi-component radio source (for example in the case where nearby unrelated unresolved radio sources are not removed or reinserted). This happens for $10.0\%$ of the multi-component radio components.
The second row shows examples of multi-component sources where we fail to include all related radio components. This is the case for $14.8\%$ of the multi-component radio components. The first two examples of the second row show cases where we removed emission that turned out not to be an unrelated background source (the location of this removed component is indicated with a thick red `x'). 
The third example from the second row shows that the predicted region does not encompass both outer lobes of this double-double radio source, illustrating that the model is less likely to be correct for rarer morphologies. 
The fourth example from the second row shows a prediction that fails to include all related radio components because the radio source is too large to fit inside our $300\times300$ arcsec image.
The last row shows an example of a single-component radio source where a predicted region is entirely lacking. 
This rare case of a false negative happens for $0.1\%$ of the single-component radio components; 
our pre-processing code did not create a bounding box for the source if there is no five-sigma emission overlapping the central coordinate of the detected \textit{PyBDSF} component.
The final image shows an example where we have a multi-component radio source for which we simultaneously miss some related components and include some unrelated ones (this is more prone to happening for fainter and bent sources). Cases in this category are included in the $13.1\%$ statistic mentioned above.

\begin{figure}\begin{center}
\includegraphics[width=0.99\columnwidth]{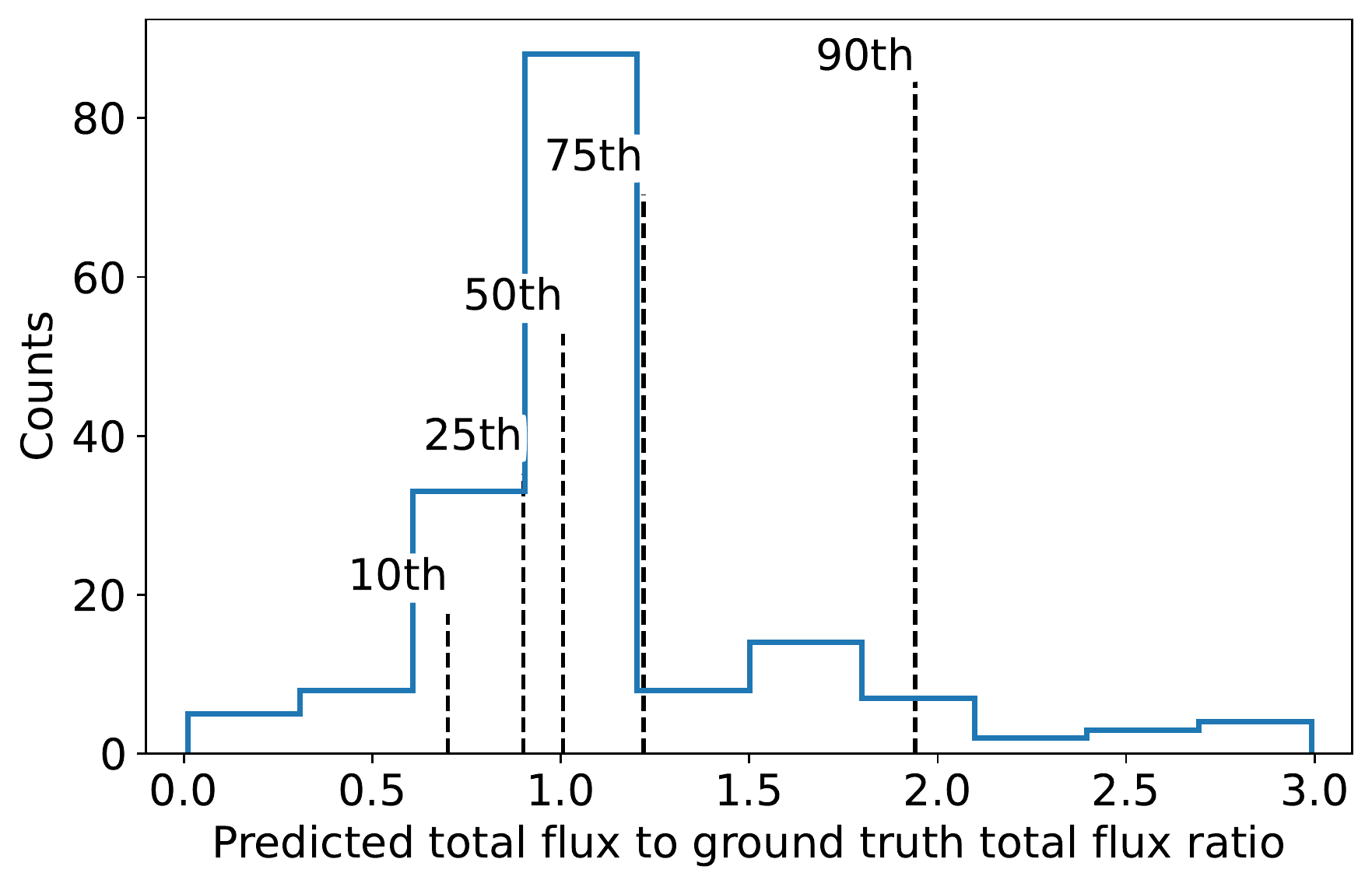}
\caption{Histogram of total flux densities resulting from the incorrect predicted associations versus the total flux densities resulting from the ground truth associations. The black dashed lines indicate the 10th, 25th, 50th, 75th, and 90th percentiles of the flux-density ratios. For legibility, the x-axis shows values up to a ratio of $3$, but $3.9\%$ of the incorrect predicted associations have ratios higher than $3$, with the maximum being $35.9$.}
\label{fig:fluxratios}
\end{center}\end{figure}

Another way to quantitatively inspect our incorrectly predicted associations is to plot the ratios of the total flux densities that result from these incorrect predicted associations versus the ground truth total flux densities.
Figure \ref{fig:fluxratios} shows that the median flux ratio for the incorrect predictions is well centred: close to $1$. The 25th and 75th percentiles with ratios of $0.9$ and $1.2$ show that half the incorrect predictions do still lead to reasonable total flux-density values. 
The ratios at the 10th and 90th percentiles, with values of $0.7$ and $1.9,$ show that the worst incorrect predictions lead to total flux densities that are farther off the mark when they over-predict than when they under-predict. 
Total flux-density over-prediction happens in $52.5\%$ of the incorrect associations and is thus slightly more likely to happen than under-prediction is.
\section{Discussion}
\label{sec:discussion}
We set out to decipher the optimal catalogue accuracy we can achieve irrespective of the technique used to associate the radio components.
Using the Stokes-I images of LoTSS in combination with images from optical and infrared (IR) surveys, experts might agree for up to about $95\%$ ($100\%$ minus the $3.3\%$ hard-to-judge category and minus the $8.68\%\times16\%=1.39\%$ human error, Section \ref{sec:manual_process}) as the rest is difficult or impossible to judge given the information available.
No automated method will be able to surpass this level of accuracy given the same input information.
Given our results, the difference between this expert-attained accuracy and the accuracy on our training set (known as `avoidable bias') is of about six percentage points ($95\% - 88.5\%$). 
The difference between the accuracy on our training set and that on our test set (known as `variance') is of about three percentage points for our final model ($88.5\% - 85.3\%$).
High variance is a sign of over-fitting, while high bias is a sign of under-fitting. 

Given large neural networks such as the ones used in this work and enough regularisation, avoidable bias might be decreased without a strongly increasing variance, by increasing the model size \citep{Krizhevsky2012,resnet,nakkiran2021deep}.\footnote{Assuming we are not in the `double-dip' critical regime where performance first gets worse and then better with increasing model size \citep{Belkin2019,nakkiran2021deep}.}
However, Section \ref{sec:backbone_experiments} demonstrates that both the catalogue accuracy on the validation set and the variance (manifest in the difference between the accuracy on the training and that of the validation data) do not significantly increase or decrease when using the larger CNN backbones.
This might be explained by the observation that deep (many-layered) neural networks, trained with stochastic gradient descent, seem to have critical layers (usually the layers closest to the input) that have a lot of impact on the output and many more layers that do not \citep{zhang2019all}.

Variance can be addressed by increasing the regularisation; we might, for example, add in more data augmentation during training (see Section \ref{sec:future}).
Both avoidable bias and variance can be addressed by increasing the number of images we use for training and testing \citep{sun2017revisiting}.
Generally, the accuracy attained on a test set will not surpass that of the training set.
Given the deep neural networks of the size that we use in this work, plus source removal, we can easily over-fit our training data; thus, increasing the size of the training set will simultaneously increase the attained accuracy on the test set and decrease the attained accuracy on the training set (see Fig. \ref{fig:ablation}).
By extrapolating the final results on our training and test set as a function of training dataset size, using a logarithmic curve fit, we estimate that we would need at least an additional eight thousand training sources (twice the size of our current training set) to potentially realise a two-percent increase in the catalogue accuracy of our test set.

A different way to reduce avoidable bias and variance is by modifying the input features or modifying the model.
A sensible modification of our model input features would be to incorporate optical and or IR data.

\subsection{Prediction score versus catalogue accuracy}
\label{prediction_score}
\begin{figure}\begin{center}
\includegraphics[width=\columnwidth]{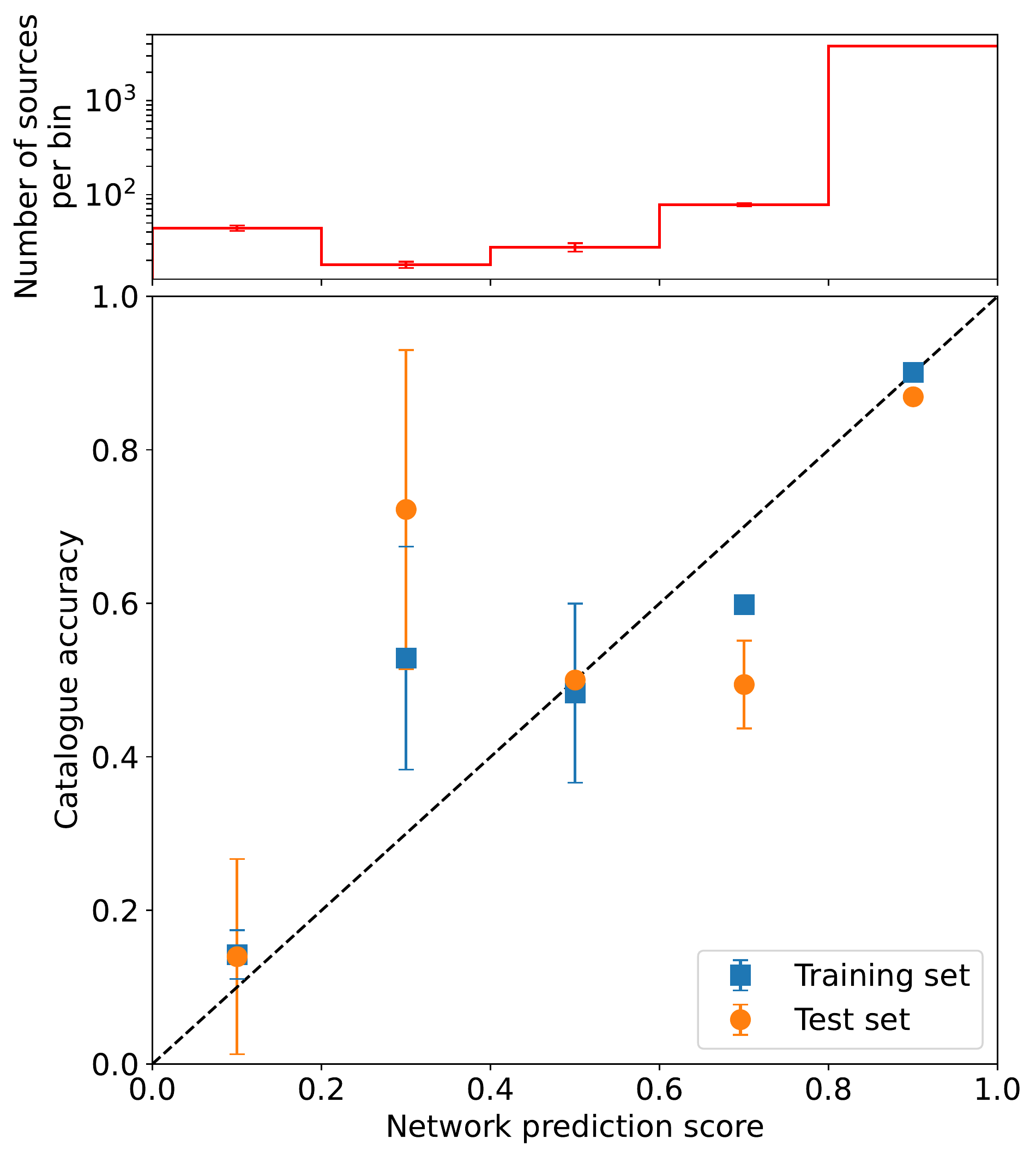}
\caption{Catalogue accuracy of our final model binned by its prediction scores. Error bars indicate the standard deviation between three independent runs initiated with different seeds.}
\label{fig:prediction_scores}
\end{center}\end{figure}
Predictions using R-CNNs do not only provide a predicted region and a class label (the single class `radio object' in our case), but also a prediction score. 
This score is a number between 0 and 1, and it indicates how strongly the input activates the neural network for a certain class. 
These prediction scores can be compared to the actual catalogue accuracy attained. 
In Fig.  \ref{fig:prediction_scores}, we plot the catalogue accuracy for subsets of our data, binned according to their prediction score.
If these data points lie along the diagonal dashed line, our model would be well calibrated.
Figure  \ref{fig:prediction_scores} shows that our model is only well calibrated for components with prediction scores below $0.2$ and above $0.8$.
This means that, for the model we trained, we cannot generally use the prediction score of a single prediction to obtain a good estimate of the probability that this particular prediction is correct.
Nevertheless, we can still predict the catalogue accuracy of our model over an aggregated sample of sources. 

Depending on our science case, we could accept only the associations that surpass a certain prediction score at the cost of leaving more sources to manual association.
For example, the test set indicates that if we only accept associations with a prediction score that surpasses $0.2$ (or $0.8$), we can associate $99.1\%$ (or $96.4\%$) of the large and bright components and achieve a level of accuracy of $86.3\%$ (or $87.4\%$), which is marginally better than the $85.3\%$ accuracy level for all large and bright components in the test set.

Upon inspection, we see that components receive a low prediction score when (i) the source is too large to fit inside our $300$ arcsec cutout; (ii) the source is diffuse and the region, which only encloses the signal exceeding five times the noise level, does not capture the full shape of the source; and (iii) source removal erroneously removes a lobe of a double-lobed source.
We invite the reader to consult Appendix \ref{app:low_score} for examples of sources with a prediction score below $0.5$.

\subsection{Scope of usability and limitations}
\label{sec:scope}

Two notable limitations to our method are the impact of imaging artefacts in the input images, which propagate through to the final source catalogue, as well as the deblending of sources.
As a temporary solution, we trained our network to associate imaging artefacts with the bright radio sources that caused them to appear, such that they would not distort source-density counts too much. 
In future work, a dedicated supervised approach -- such as a decision tree  -- could be prepended to our pipeline to identify and discard imaging artefacts.

Deblending individual radio components is also outside the scope of this paper.
Deblending entails separating a radio component into two or more unique sources.
In LoTSS-DR1, from the $15,806$ radio components that went to LOFAR Galaxy Zoo, $386$ ($2.4\%$) were flagged as `blended'. 
This indicates that for these components, with the given settings, PyBDSF combined radio emission from multiple physically distinct radio sources into a single radio component.
However, for deeper surveys \citep[e.g. the LoTSS Deep Fields;][]{Kondapally2020} or lower resolution surveys \citep[e.g. MIGHTEE;][]{HeywoodMightee}, the percentage of radio components flagged as `blended' is generally higher.
Individual \textit{PyBDSF} radio components are themselves composed of a single or multiple 2D Gaussian.
To adapt our method to deeper or lower resolution surveys, one would first need to set \textit{PyBDSF} parameters such that the threshold for merging multiple 2D Gaussians into a single radio component is higher.
Furthermore, for these types of surveys, adding optical information to our input images might be more crucial to performing accurate source-component associations than in LoTSS.

A third limitation of our method is that source size estimates for some large FRI sources will be underestimated. 
Our method of pre-computed regions that encompass the emission that surpasses the local noise more than five times means that we miss out on parts of large diffuse sources for which individual pixels do not surpass this signal to noise threshold, while the total flux density in this large patch of emission is significant. In practise, this means that for certain large FRI sources, the outer parts of the lobes will not be associated with the rest of the source. 

\subsection{Prediction for large and faint radio components}
\label{sec:discussion_faint}
Radio association of large and faint sources ($>15$ arcsec and $<10$mJy) is possible with the method presented in this paper.
We chose to train our neural network on extended sources with a high signal-to-noise ratio ($>15$ arcsec and $>10$mJy) as we expect these sources to clearly show the characteristic shapes that jetted RLAGNs exhibit.
The larger size and high signal-to-noise ratio makes it easier for humans to do the association, which means that we can have more confidence in the crowd-sourced labels that we have for these sources.
For sources with a lower signal-to-noise ratio, the associations can be determined with less certainty, but this is not well reflected in the crowd-sourced labels; users had no option to indicate when they were uncertain about their component association.
We therefore did not use these labels for training or validation.
Instead, we ran our model, trained on large and bright sources only, directly on the large and faint radio components. 
We observed that the predicted association for radio components for which the predicted region had a prediction score below $0.1$ were mostly incorrect, and we decided that these radio components would not be associated with any other radio component.
We visually inspected the resulting component association prediction of our model for all $553$ large and faint sources in three random pointings of our `testing' dataset (see Section \ref{sec:data}) and tentatively judged $80\%$ to be correct and $11\%$ to be incorrect, and we labelled the remaining $9\%$ as `hard to judge'.
These percentages could vary within a few percentage points as the visual inspection is increasingly subjective for fainter sources.
Depending on subsequent science cases, this performance may or may not be sufficient.
As is, our Fast R-CNN model trained on the large and bright radio components attains a catalogue accuracy on the large and faint radio components that is similar to the catalogue accuracy of a random forest trained specifically to associate the large and faint radio components.
The random forest is at a relative advantage here as our Fast R-CNN is trained on the large and bright components and then applied to the large and faint test set, while the random forest is separately trained on the large and faint components (including a hyper-parameter search of the large and faint component validation set) before running inference on the large and faint test set.
 
Expanding the scope of our model to unresolved and barely resolved radio components ($<15$ arcsec) is of limited additional value. 
As mentioned in the introduction, associating the large and bright lobes will already include $48.7\%$ of the components of $<15$ arcsec that are barely resolved and not correctly associated by \textit{PyBDSF}.
However, this will be different for radio observations with surveys that are less sensitive to diffuse emission, such as FIRST, where FRIIs with two lobes often appear as two unresolved point-like blobs with no emission in between.

\subsection{Comparison to the public LOFAR Galaxy Zoo}
\label{sec:public_zoo}

\begin{figure}\begin{center}
\includegraphics[width=\columnwidth]{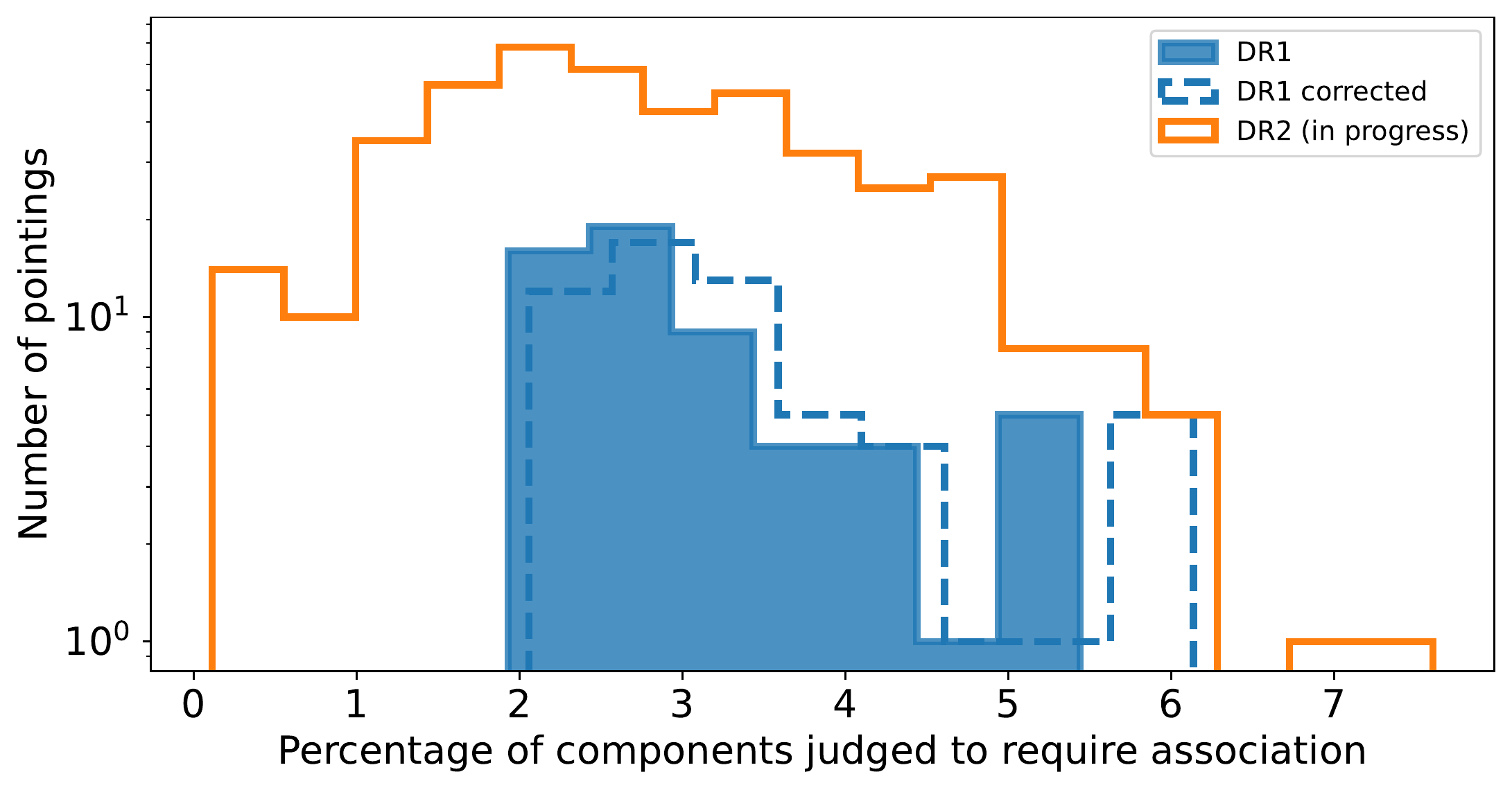} 
\includegraphics[width=\columnwidth]{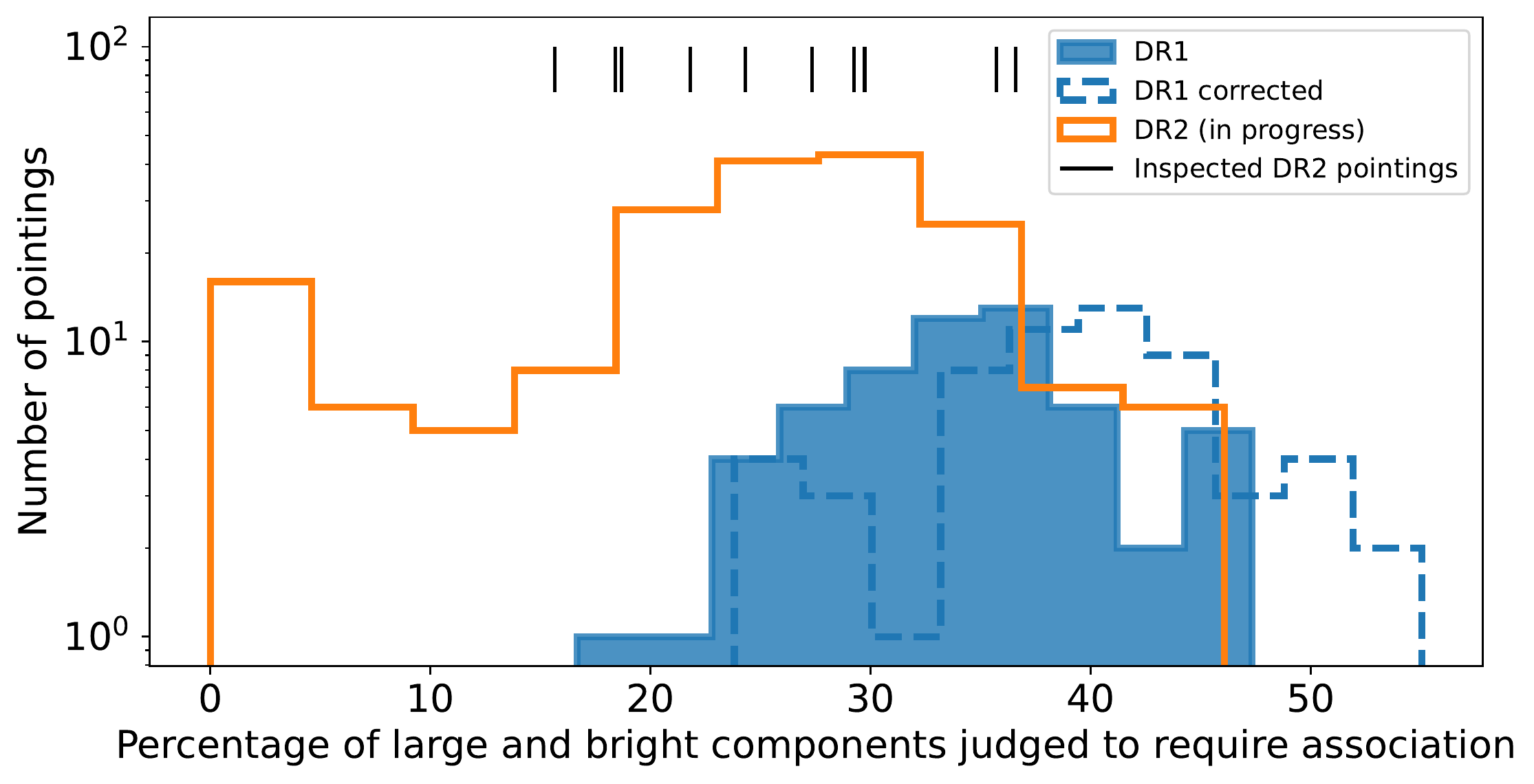} 
\caption{Percentage of radio components that are manually associated with other radio components. The top panel shows the percentage for all radio components per pointing, while the bottom panel shows the percentage of just the large ($>15$ arcsec) and bright ($>10$mJy) components. The vertical black lines in the bottom panel show the association percentage for the ten DR2 pointings that we selected for further inspection.}
\label{fig:internal_vs_public}
\end{center}\end{figure}

In the results section of this paper, we compare the performance of our method against (LoTSS-DR1) manual annotations done by astronomers in the LOFAR collaboration. 
In this sub-section, we assess how well the general public is able to perform manual annotations (for LoTSS-DR2).
We show that the accuracy of our automated associations is on par with that of the public.
We also show that there is a larger variance in the quantity of associations per pointing that the public detects compared to the astronomers.
These two observations indicate that expanding our training data by including the associations done by the public  will not necessarily improve our model.

The first data release (annotated by astronomers) contains $323,343$ radio components, of which $3,674$ ($1.14\%$) were combined into multi-component sources. 
In our manual correction of the DR1 annotations (see Section \ref{sec:manual_process}), we combined $3,982$ components ($1.23\%$) into multi-component sources. 
For DR2 (annotated by the public), $1,838,763$ components have been processed up to this point, and $17,993$ ($0.98\%$) of those were labelled as multi-component sources. 
The top panel of Fig. \ref{fig:internal_vs_public} shows that our manual correction of DR1 increased the percentage of multi-component sources as we added missing associations. The top panel also demonstrates that DR2 annotations show a large variation between the percentages of multi-component sources in different pointings compared to the DR1 associations. 
We excluded pointings that contain fewer than $1000$ components, so a small number of sources per pointing cannot explain this effect. 

This effect is more clear if we focus on just the large ($>15$ arcsec) and bright ($>10$mJy) components.
The number of components associated into multi-component sources is $34.36\%$, $39.95\%,$ and $24.70\%$ for DR1, DR1-corrected, and DR2, respectively.
The lower number of associations in DR2 compared to DR1 is a sign that astronomers are able to recognise associations that some lay volunteers miss.
The bottom panel of Fig. \ref{fig:internal_vs_public} shows that the variation in the number of associations across different pointings is also larger in DR2 than in DR1.

\citet{northcutt2021pervasive} show that it is crucial to verify public annotations of a dataset before assessing the accuracy and design aspects of a model on the training, validation, and test data.
Therefore, we inspected the large ($>15$ arcsec) and bright ($>10$mJy) components of ten DR2 pointings in more detail -- their percentages of multi-component sources per pointing are visible in the bottom panel of Fig. \ref{fig:internal_vs_public}.
Sorting the $1,097$ components from these pointings into the categories `association seems correct', `association is hard to judge', and `association seems incorrect' (as we did for DR1 in Section \ref{sec:manual_process}), we ended up with $82.8\%$, $3.5\%,$ and $13.8\%$ in each category, respectively. 
That means we observe more components in the public `seemingly incorrect' category compared to the $8.68\%$ in the `seemingly incorrect' category of the expert-annotated DR1. Especially as $57\%$ of this $8.68\%$ could be attributed to not having access to better-calibrated images.
There are multiple reasons why volunteers could miss an association, but one reason we spot by our visual inspection of this data is that volunteers did not flag a number of large radio components that fell out of the variable-sized cutouts that were presented in the LOFAR Galaxy Zoo project as `too zoomed-in'. 
Both in the internal and the public LoTSS Galaxy Zoo project, if more than two out of five of the users clicked the `too zoomed-in button' in the annotation process, the source would be associated by a single expert using an interface that allowed panning and zooming. 

We manually corrected the annotations in the `seemingly incorrect' category for the ten DR2 pointings and compared this corrected catalogue to the predictions from our Fast R-CNN with rotation augmentation and unresolved sources removed. 
The baseline assumption in which we do not combine any radio component would lead to a catalogue that is correct for $67.4\%$ of the components. 
Predicted associations of our trained model give a catalogue accuracy of $84.0\%$. 
This number falls within the range of seemingly correct public associations ($82.8\%$) and `seemingly correct' plus `hard to judge' public associations ($82.8\%+3.5\%=86.3\%$). 
We can therefore claim that, given the project design and aggregation choices made in the public LoTSS Zooniverse project, our automated method yields associations that are similar in quality to the associations obtained through this public crowd-sourcing effort.

\subsection{Future work}
\label{sec:future}
As our network accuracy is already comparable to that attained by public crowd-sourcing, increasing the quality of the annotated data is more prescient even if that comes at the cost of the quantity of the annotated sources. 
The quality of the manual annotation can be improved by increasing the number of viewers per source and increasing the weight of the annotations performed by vetted persons (be they experienced volunteers or astronomers).
As discussed in Section \ref{sec:discussion}, even additional high-quality (expert-annotated) data on the full large and bright component population would not drastically improve our network accuracy.
However, the increased number of annotations per source would allow us to predict a calibrated posterior on possible source-component associations on a source-by-source basis.
\citet{Walmsley2020} demonstrated this possibility for galaxy morphology classifications using a Bayesian CNN.
In our current work, we were able to judge the accuracy of our predicted associations for the full sample, but as Fig. \ref{fig:prediction_scores} implies, not on a source-by-source basis.

A complementary approach would be to focus the manual association effort on the radio components that are poorly classified by our neural network and thus efficiently target the deficiencies of the network.
Ideally, a future pipeline suggests radio sources for manual inspection -- for example by suggesting more sources that receive a low prediction score from our R-CNN or sources that are automatically flagged as exhibiting unusual morphology \citep{Mostert2020} -- and uses the obtained annotations to retrain its association network and suggest further sources for manual annotations. This practice, known as `active learning', has proven itself across many research fields where manual annotation is costly (see \citet{settles2009} for a review). 
Most notably, \citet{Walmsley2020} applied an active learning technique described by \citet{Houlsby2011} to reduce the number of annotations required by the Galaxy Zoo 2 project \citep{Willett2013}.

On a technical level, given more training labels, the flexible Detectron2 framework \citep[Yuxin][]{wu2019detectron2} allows us to replace the feature-extraction backbone with a future state-of-the-art backbone. 
Complementarily to this, \citet{cs231n_ensemble} writes that training an ensemble of models and averaging their predictions at test time will generally improve performance of classification networks by a few percent.\footnote{This technique is employed by all winners of the ImageNet Object Detection competition since 2014 (\url{https://image-net.org/challenges/LSVRC}).}
As there are not many different automated techniques for radio component association as of yet, we believe exploring new techniques and heuristics will be more fruitful than making significant effort to combine the existing techniques into an ensemble of models.
A sensible modification of the techniques used in this paper would be to move from the rectangular bounding boxes to arbitrarily shaped segmentations \citep[e.g.][]{he2017mask}.

Our model's performance on large and faint sources is likely to improve by training the network with (correlated) noise augmentation. 
One could train the network on copies of the images with the large and bright radio components, for which we artificially raised the noise level. 
The benefit of noise augmentation is that it requires no additional manual labelling or label-checking as it relies on the labels of large and bright components that we already have: labels of which we can be more confident as they are based on high signal-to-noise ratio emission.

On the input side, we could augment our data with optical or IR information.
The most straightforward way to do so would be to swap one of the three radio channels in our current images for (normalised) image cutouts of a single band extracted from an optical or IR survey; for example, $z$-band from the DESI Legacy Imaging Surveys \citep{Dey+19} or unWISE band $1$ \citep{Meisner+18}.
A more involved approach would be to swap one of our current channels for an image that represents some of the information present in the optical or IR source catalogue within the cutout. We could plot the optical or IR sources as 2D Gaussians. The location and variance of these Gaussians could reflect the optical source's coordinates and apparent size, and the amplitude of the 2D Gaussians could reflect a property of the optical source. The benefit of this last approach would be the absence of image artefacts including diffraction spikes, trails, and ghosts; the ability for us to filter out unrelated objects such as stars; and the flexibility to let the amplitude of the 2D Gaussian reflect either a source's apparent magnitudes, its colour, or any other sensible feature available in the optical or IR catalogue.
The source density in optical surveys is generally much higher than in radio surveys, which is a potential source of confusion for both inexperienced volunteers and automated methods. Chen \citet{Wu2019} showed that masking the optical information where there is no significant radio emission works best to mitigate this problem.

The work in this paper paves the way for three avenues of further application. The first is that of morphological classification or the clustering of jetted RLAGNs. Given that the application of our method results in the mostly correct association of well-resolved radio components, subsequent classification into FRI or FRII objects becomes a lot easier -- both for simple, feature-based classifications and for supervised deep neural networks. Correct associations also simplify the process of automatically determining radio-lobe bending angles, allowing large samples of (non)bent objects to be studied, for example, in relation to their (cluster) environment.
The second avenue opened up by the results in this work leads us towards finding the galaxies (in the optical or infrared regime) that host the RLAGNs from which the radio emission originates. Except for cluster or RLAGN-remnant-related emission, the origin of most extragalactic radio emission can be found by tracing two related radio lobes to the point where the two radio jets (are projected to) meet \citep{Barkus2022}.
Incorporating morphological outlier detection and host-galaxy identification in a future active learning pipeline would be an efficient replacement of manual pre-filtering of sources for crowd-sourced annotations.
The third avenue opened up by the combination of the previous two avenues it that of assessing the completeness and reliability of different types of extended radio objects in LoTSS and other large-scale sky surveys.
To assess the completeness of a survey for a particular type of radio object, one has to reinsert and detect fainter copies of the corresponding object in the survey images up to and including the parameter space where the reinserted noisier objects are either not detected, misclassified, and/or their components are not correctly associated.

\section{Conclusions}
In this work we adapted a Fast region-based convolutional neural network to perform radio-component association using Stokes-I radio images as the sole input. 
We constructed training labels from manually performed expert radio component associations.
We tested different backbone architectures (Section \ref{sec:architectures}), implemented rotation data augmentation in the pre-processing stage (Section \ref{sec:architectures}), leveraged the radio components from regular radio-source-detection software to create pre-computed regions (Section \ref{sec:pre_computed_roi}), and simplified the association task for large and bright sources by using a gradient boosting classifier trained by \citep{Alegre2022} to remove unresolved and barely resolved sources that are likely unrelated (Section \ref{sec:remove}). 

We conclude that for large ($>15$ arcsec) and bright ($>10$ mJy) radio components, our automated method -- an adapted Fast R-CNN with rotation augmentation and unresolved sources removed, trained on expert annotations from the LOFAR Two-metre Sky Survey (LoTSS) first data release -- performs component associations on the LoTSS second data release with a level of accuracy ($84.0\%$) that is comparable to that attained by public crowd-sourcing efforts ($82.8\sim 86.3\%$). 

We show that with a deep neural network, a similar performance can be achieved using a variety of different settings. We show that stepwise and cosine learning rate decay schemes have similar performances. Also, for our training dataset, deeper networks -- convolutional neural networks with more layers -- do not result in significant performance gains. 

We implemented two features that do improve the performance of our model.
Firstly, overfitting on our training data is successfully decreased by adding rotation augmentation. 
Secondly, we can reliably increase the component association performance on large and bright radio components by removing unresolved and barely resolved radio components in the pre-processing stage. 

As is, our network can be used to replace the crowd-sourced manual radio-component association for large and bright radio components. It can also serve as a basis for either automated radio-morphology classification or automated optical-host identification.

\begin{acknowledgements}
We thank the referee for his constructive and insightful comments.
This research has made use of the python Astropy package \citep{astropy}.
LOFAR is the Low Frequency Array designed and constructed by ASTRON. It has observing, data processing, and data storage facilities in several countries, which are owned by various parties (each with their own funding sources), and which are collectively operated by the ILT foundation under a joint scientific policy. The ILT resources have benefited from the following recent major funding sources: CNRS-INSU, Observatoire de Paris and Université d'Orléans, France; BMBF, MIWF-NRW, MPG, Germany; Science Foundation Ireland (SFI), Department of Business, Enterprise and Innovation (DBEI), Ireland; NWO, The Netherlands; The Science and Technology Facilities Council, UK; Ministry of Science and Higher Education, Poland; The Istituto Nazionale di Astrofisica (INAF), Italy.
This research made use of the Dutch national e-infrastructure with support of the SURF Cooperative (e-infra 180169) and the LOFAR e-infra group. The Jülich LOFAR Long Term Archive and the German LOFAR network are both coordinated and operated by the Jülich Supercomputing Centre (JSC), and computing resources on the supercomputer JUWELS at JSC were provided by the Gauss Centre for Supercomputing e.V. (grant CHTB00) through the John von Neumann Institute for Computing (NIC).
This research made use of the University of Hertfordshire high-performance computing facility and the LOFAR-UK computing facility located at the University of Hertfordshire and supported by STFC [ST/P000096/1], and of the Italian LOFAR IT computing infrastructure supported and operated by INAF, and by the Physics Department of Turin university (under an agreement with Consorzio Interuniversitario per la Fisica Spaziale) at the C3S Supercomputing Centre, Italy.
KJD acknowledges funding from the European Union’s Horizon 2020 research and innovation programme under the Marie Sk\l{}odowska-Curie grant agreement No. 892117 (HIZRAD). LA is grateful for support from UK STFC via CDT studentship grant ST/P006809/1. WLW  acknowledges support from the CAS-NWO programme for radio astronomy with project number 629.001.024, which is financed by the Netherlands Organisation for Scientific Research (NWO).  MJH acknowledges support from the UK STFC [ST/V000624/1]. For the purpose of open access, the author has applied a Creative Commons Attribution (CC BY) licence to any Author Accepted Manuscript version arising from this submission.
\end{acknowledgements}

\bibliographystyle{aa} 
\bibliography{association_bib} 
%
\begin{appendix}
\section{LoTSS-DR2 Zooniverse project interface}
\label{app:lgz2}
\begin{figure}[h]
\begin{center}
\includegraphics[width=0.325\columnwidth]{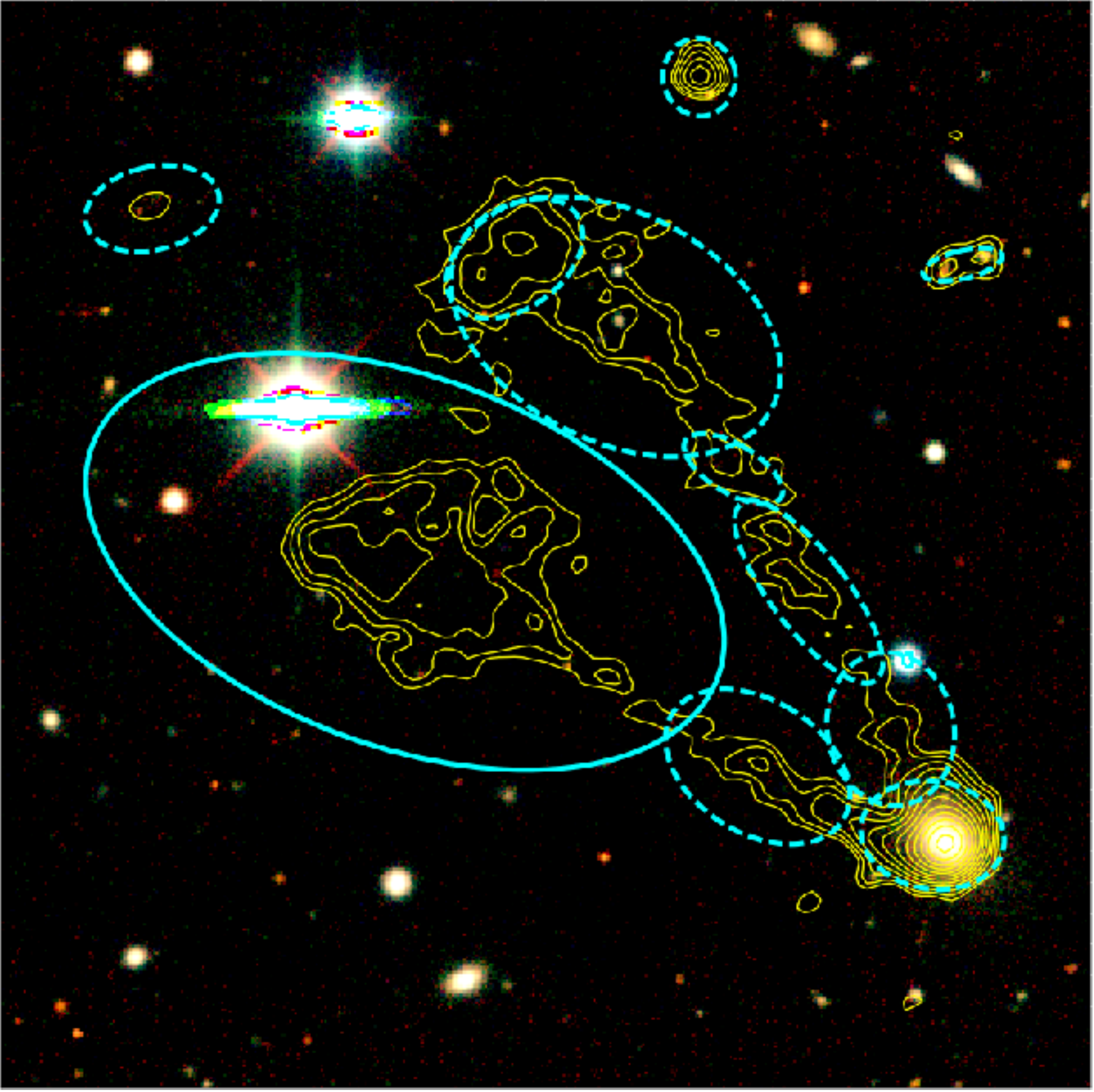}
\includegraphics[width=0.325\columnwidth]{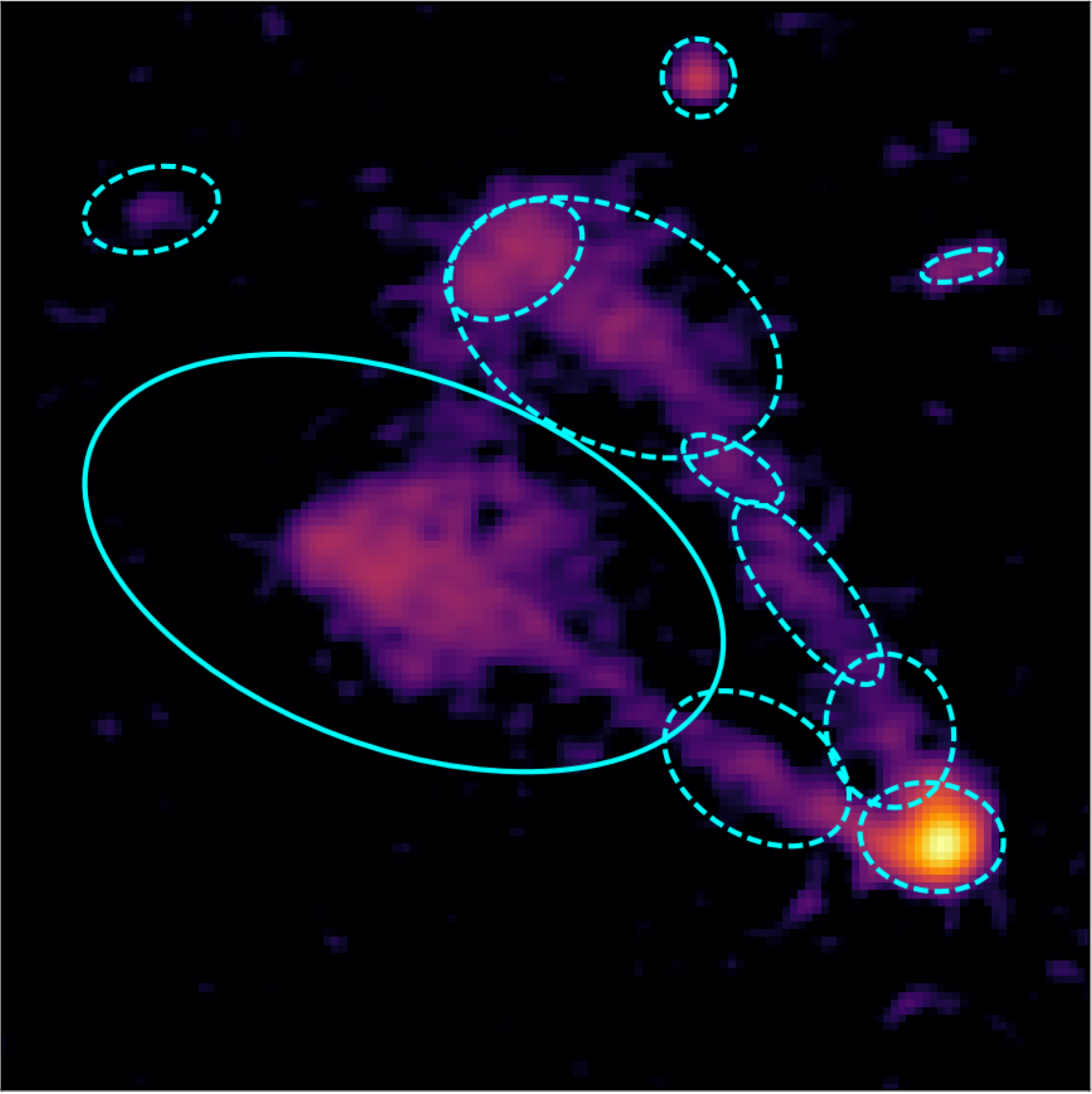}
\includegraphics[width=0.325\columnwidth]{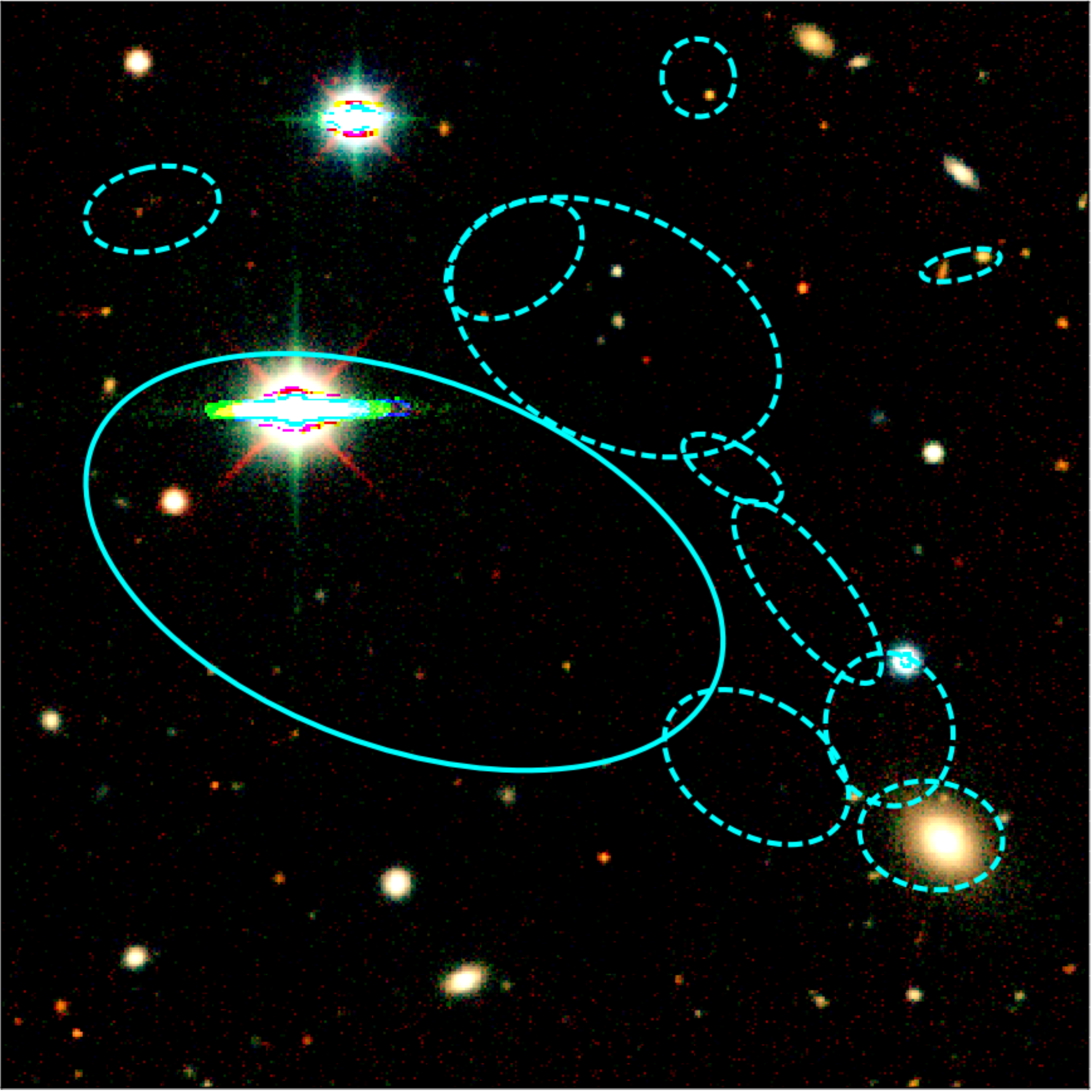}
\caption{Three figures or panels available to the user in the manual association process of LoTSS-DR2 in the public Zooniverse project. The first and third figures show the LEGACY (optical) R-band intensity image and the second figure shows LoTSS-DR2 stokes I intensity. LoTSS-DR2 radio stokes I (yellow) contours are overlaid on the first figure. 
The light blue ellipses show the FWHM of the \textit{PyBDSF}-fitted Gaussians to each LoTSS-DR2 radio component. The light blue ellipse with solid line indicates which component the user should focus on. 
The users are asked to click on the centre of each dashed ellipse that they believe should be associated with the emission behind the full ellipse. The figures are taken from the ongoing public LOFAR Galaxy Zoo project at \mbox{\textsc{lofargalaxyzoo.nl}}.}
\label{fig:lgz_dr2}
\end{center}\end{figure}

For the association of radio components and the cross-identification with host-galaxies in LoTSS-DR2, a public radio galaxy project was set up by the LOFAR collaboration.  
Figure \ref{fig:lgz_dr2} shows three panels of the interface as shown to the user.
The project is live and ongoing.\footnote{\url{https://lofargalaxyzoo.nl}}
A future publication will describe the results of the completed project in more detail.

\section{Manually corrected associations}
\label{app:correct}
\begin{figure}[h]
\begin{center}
\includegraphics[width=0.99\columnwidth]{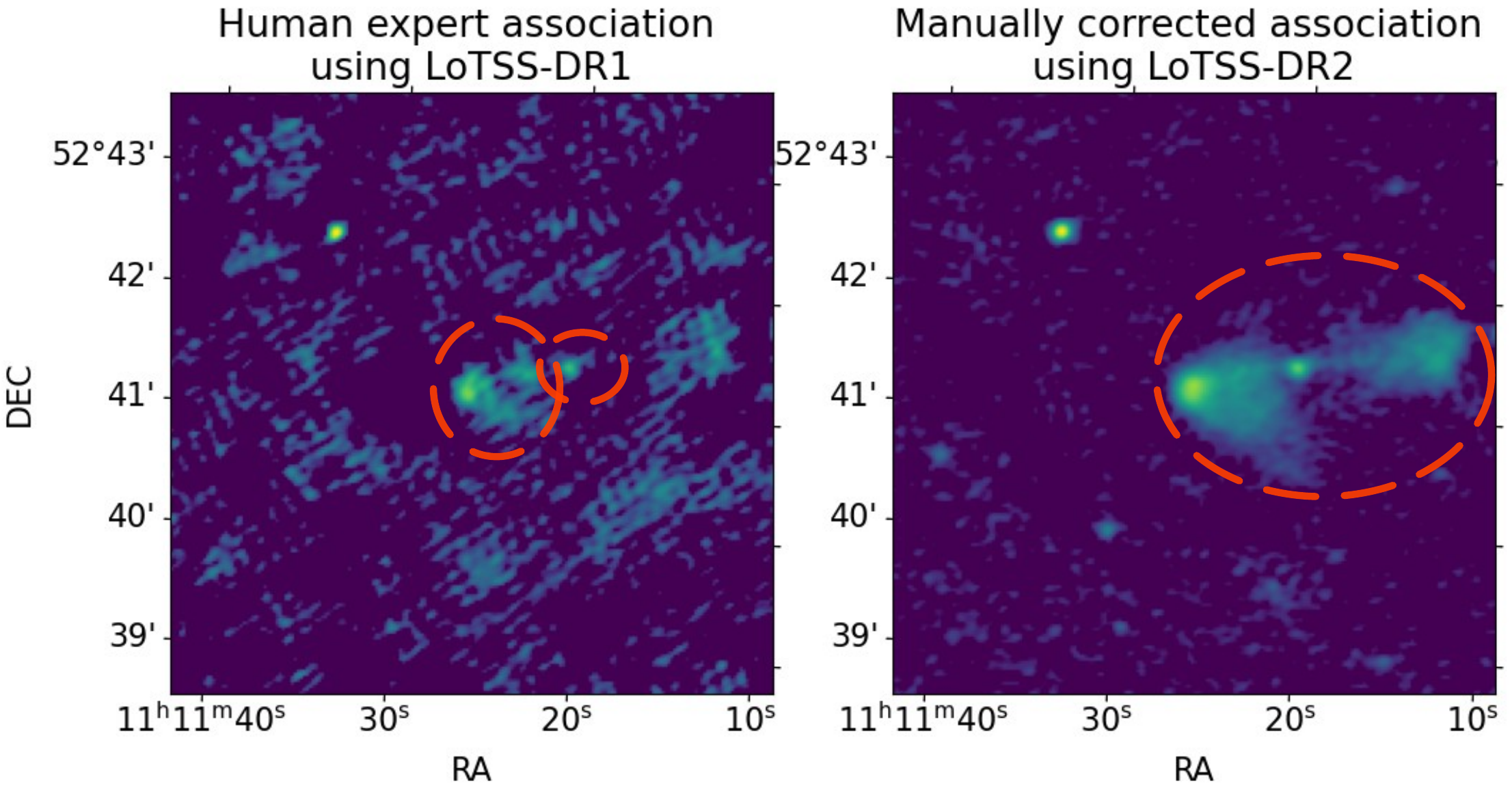}
\includegraphics[width=0.99\columnwidth]{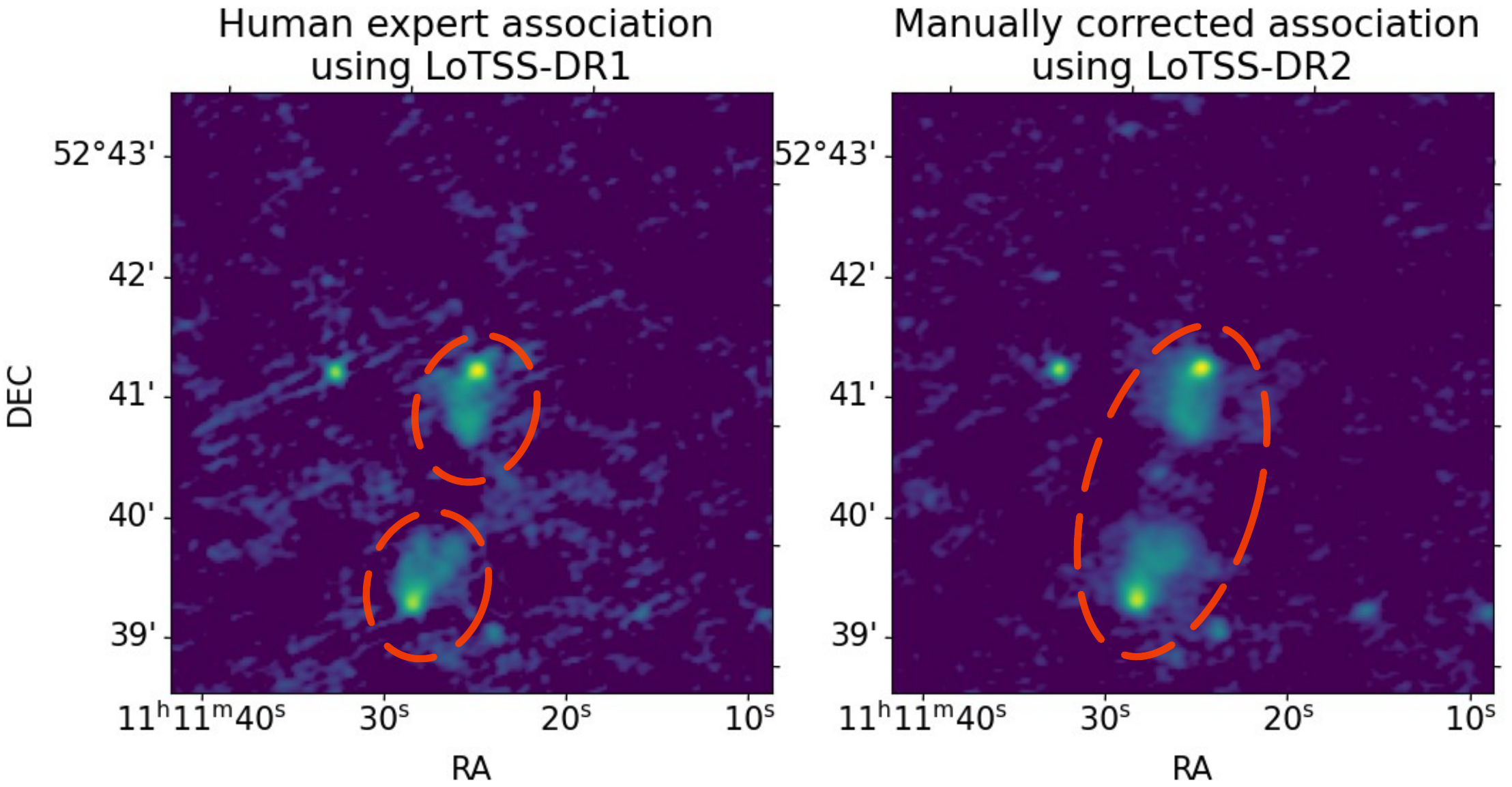}
\includegraphics[width=0.99\columnwidth]{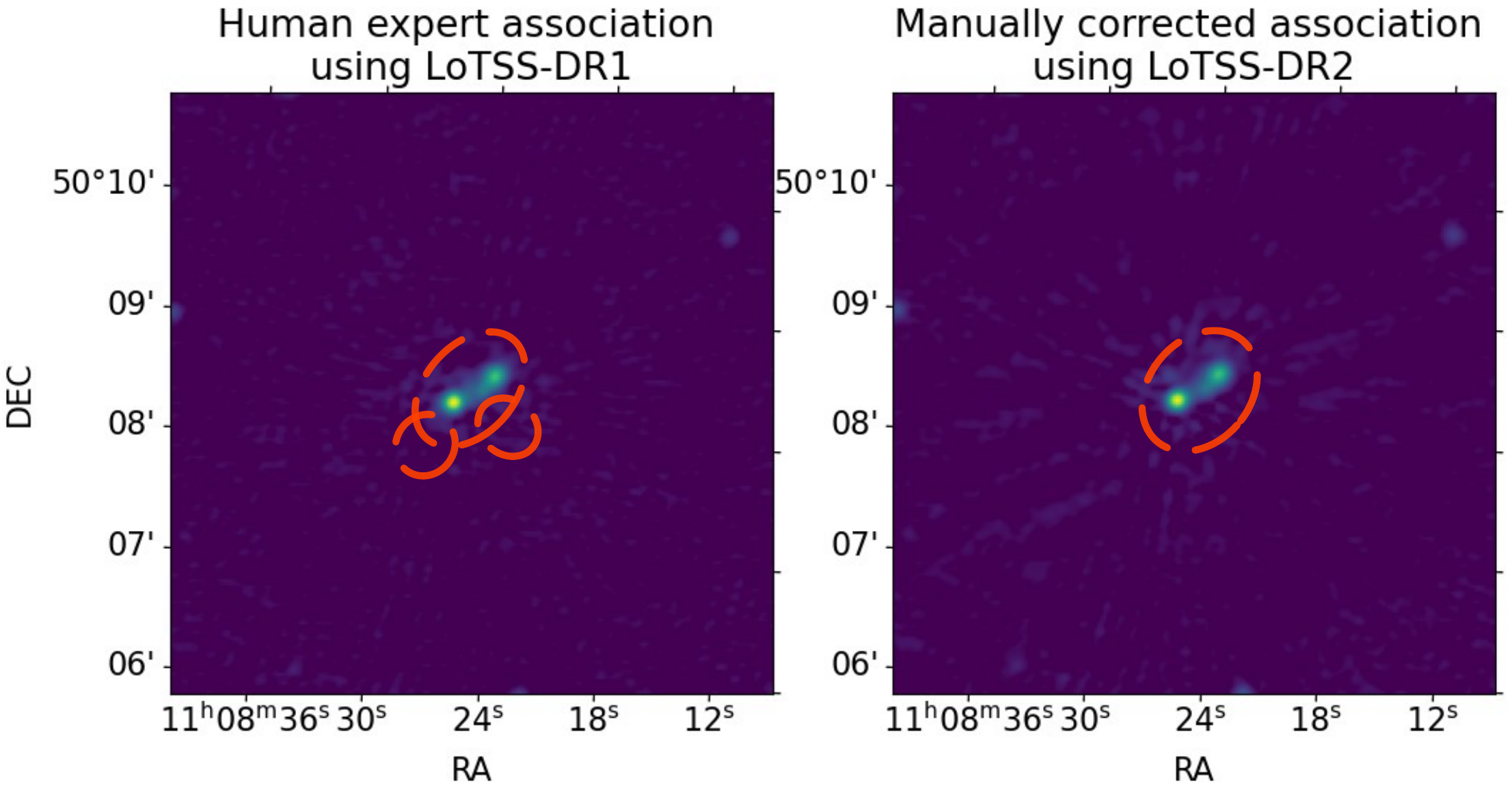}
\caption{Three examples of manually corrected associations before training. Each dashed red circle represents a separate catalogue entry. The left panel shows LoTSS-DR1 Stokes-I image and the initial human expert association, the right panel shows the LoTSS-DR2 Stokes-I image and the manually corrected association. Each row shows a different type of corrected association: the first stems from the improved quality of the LoTSS-DR2 images, the second shows a human error, the third stems from our decision to group image artefacts with the source they originate from.}
\label{fig:correct}
\end{center}\end{figure}

We manually corrected the human expert associations from LoTSS-DR1 that seem incorrect (see Section \ref{sec:manual_process}).
The main reason for the initial incorrect association stems from the fact that the LoTSS-DR2 images with higher dynamic range \citep[LoTSS-DR2;][]{Shimwell2022} were not available to the experts at the time of annotation (this seems to be the case for $58\%$ of the manually corrected associations), or human error ($16\%$ of the manually corrected associations).
We also chose to associate image artefacts that were not flagged and removed in LoTSS-DR1 with the source that created them.
These sources account for the final $26\%$ of the manually corrected components.
As explained in Section \ref{sec:manual_process}, we do so to prevent artificially inflated source counts and to prevent futile attempts at optical host identification for these artefacts by users of our radio source catalogues (see Fig. \ref{fig:correct} for examples of sources in each error sub-category).

\FloatBarrier

\section{Regions with low prediction scores}
\label{app:low_score}
\begin{figure*}\begin{center}
\includegraphics[width=0.32\textwidth]{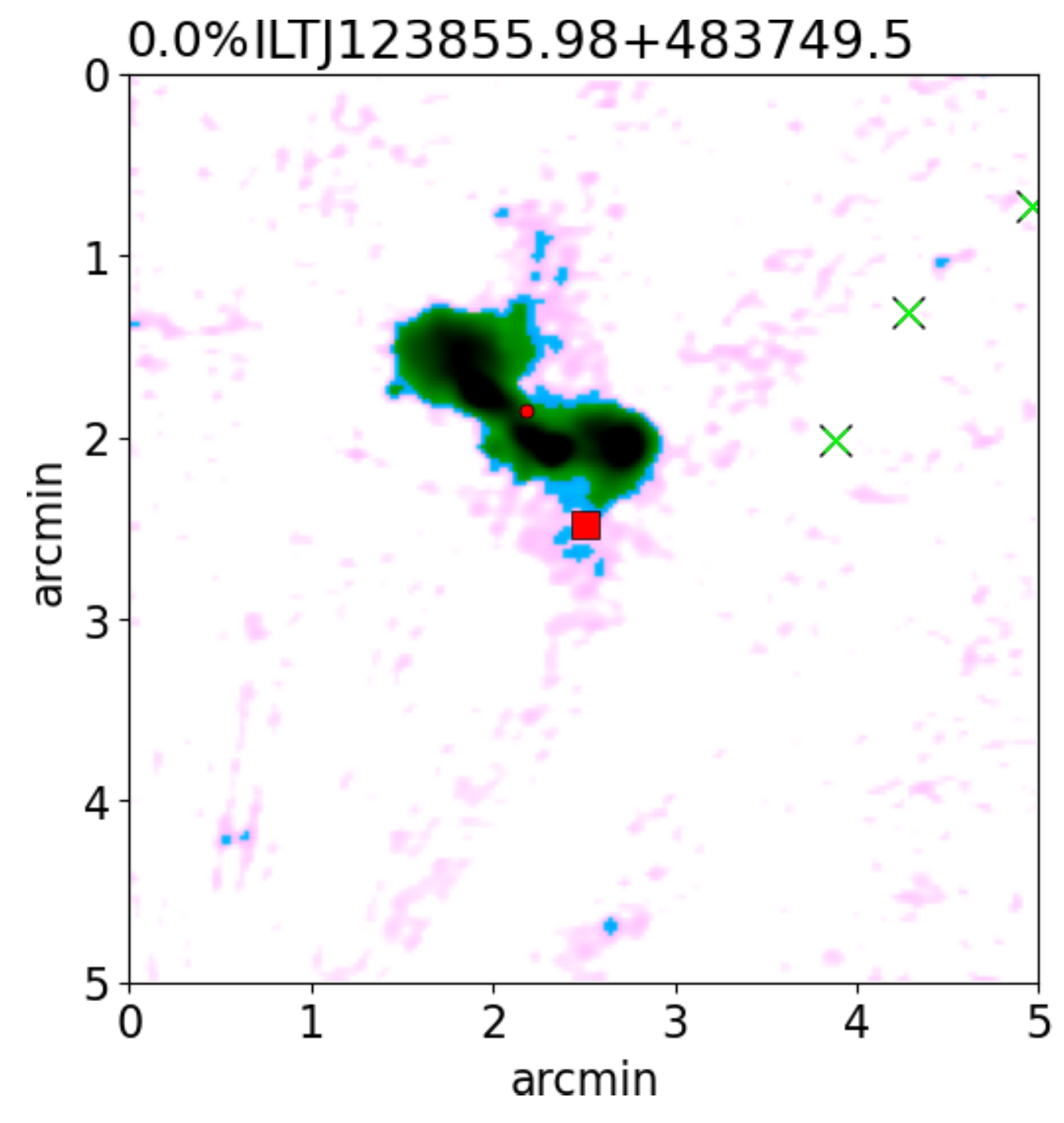}
\includegraphics[width=0.32\textwidth]{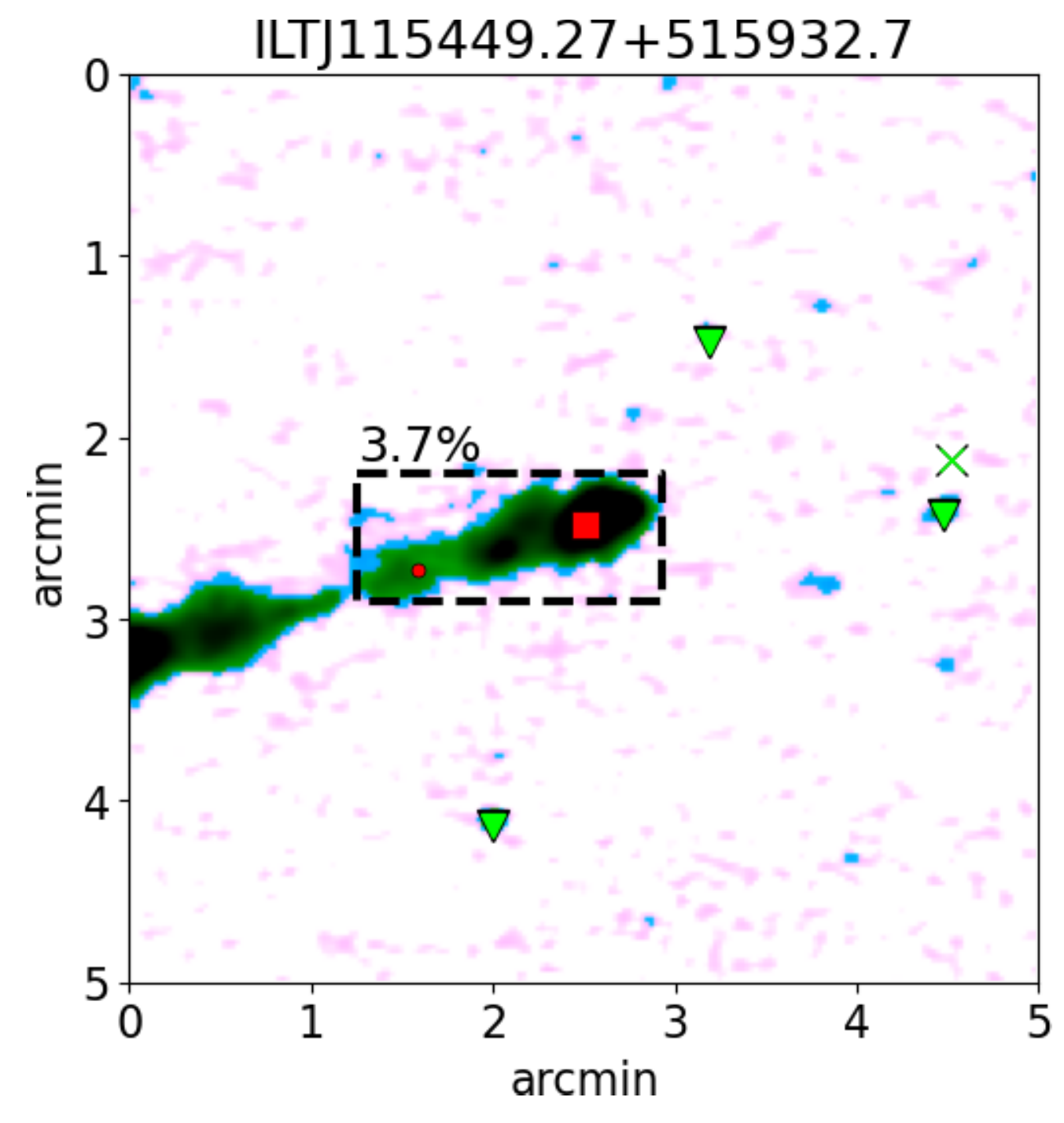}
\includegraphics[width=0.32\textwidth]{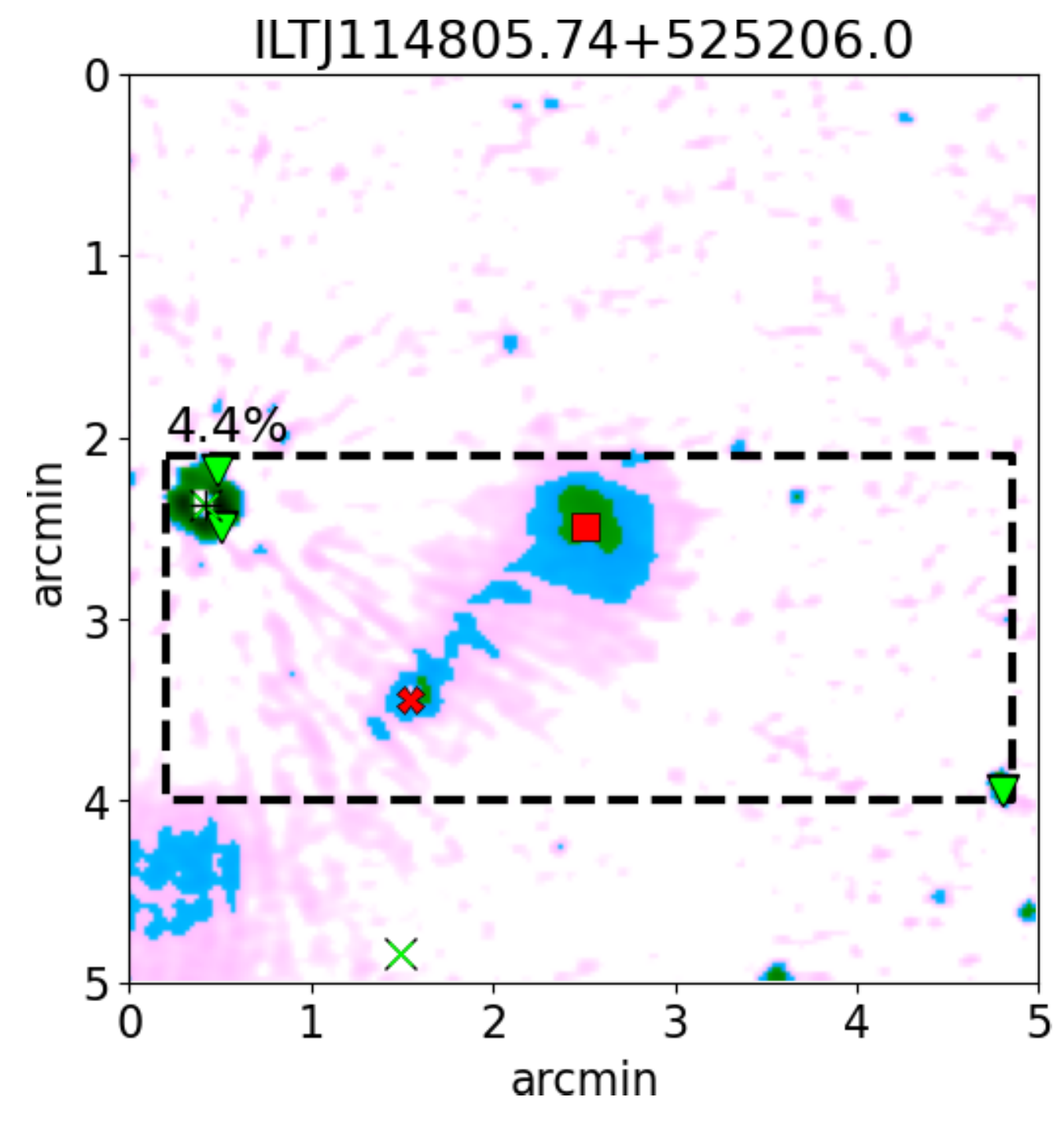}

\includegraphics[width=0.32\textwidth]{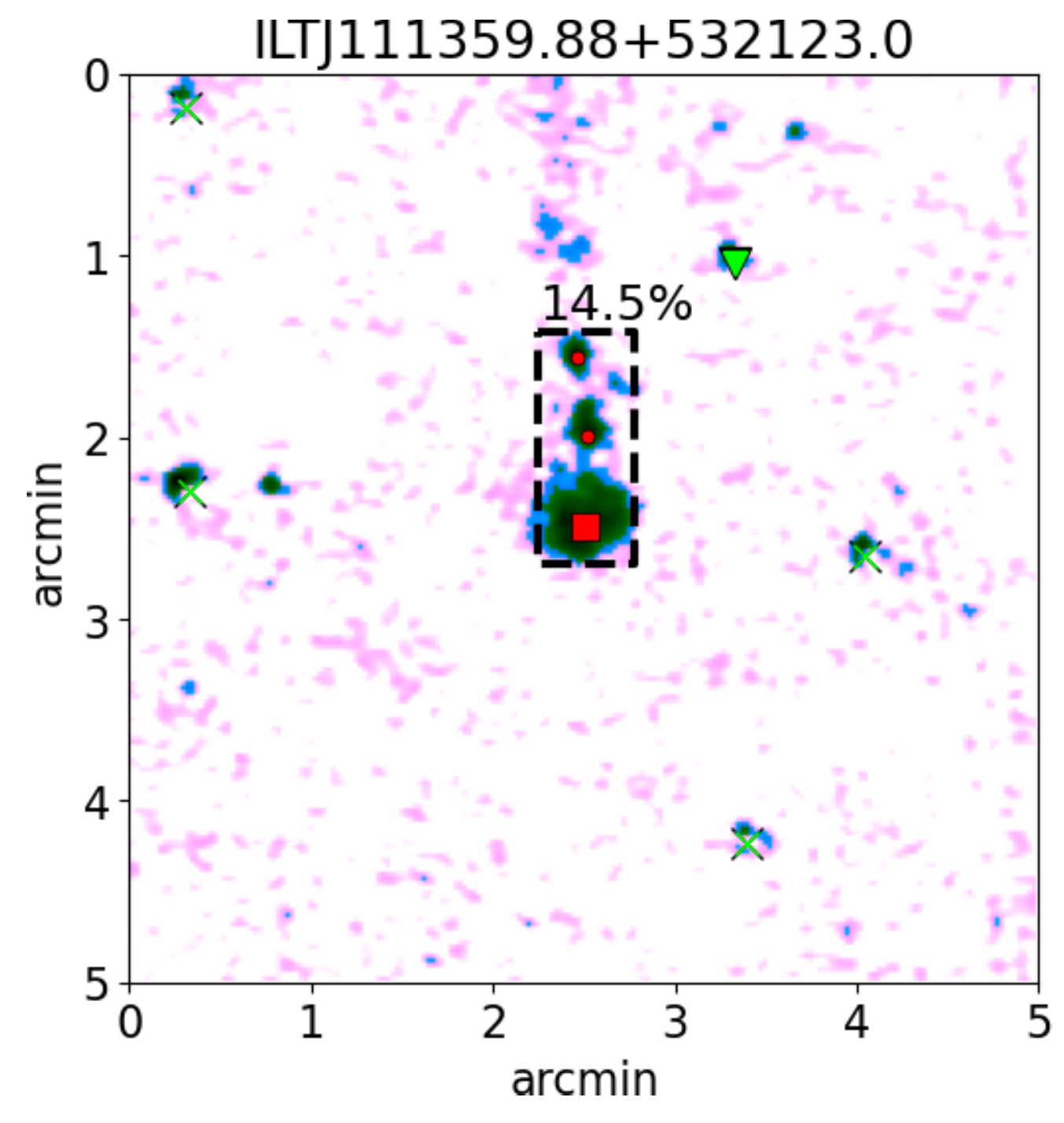}
\includegraphics[width=0.32\textwidth]{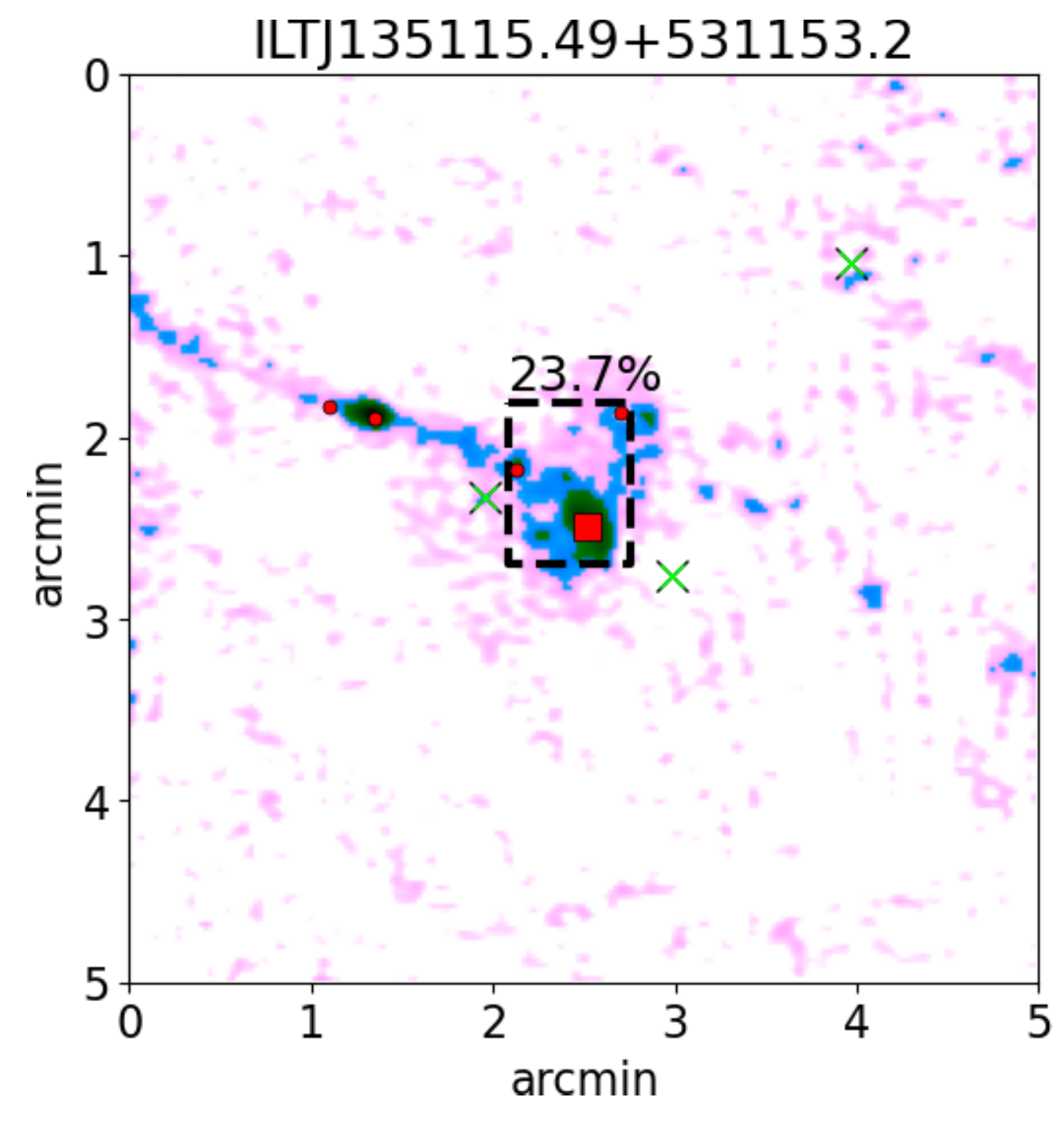}
\includegraphics[width=0.32\textwidth]{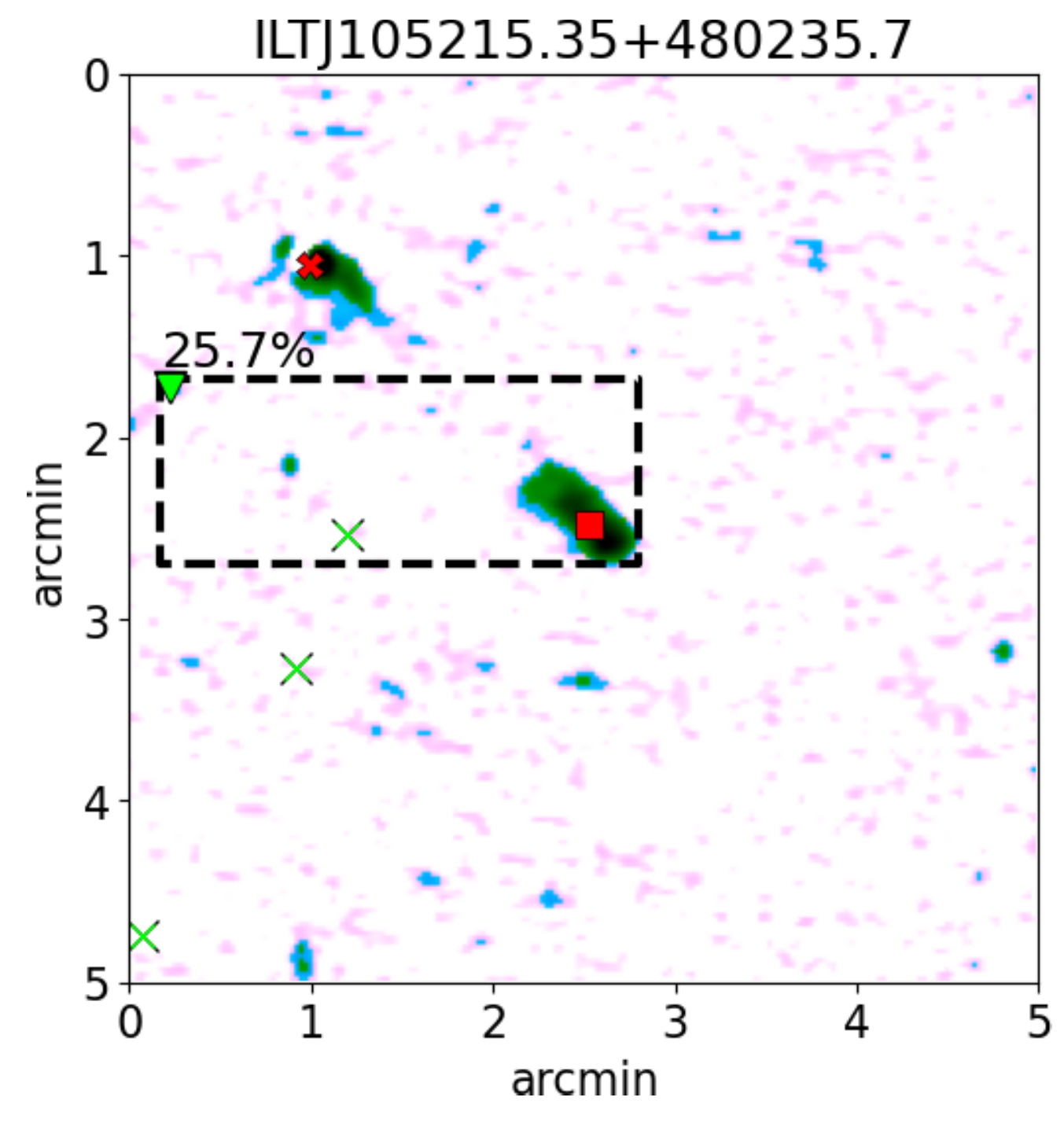}

\includegraphics[width=0.32\textwidth]{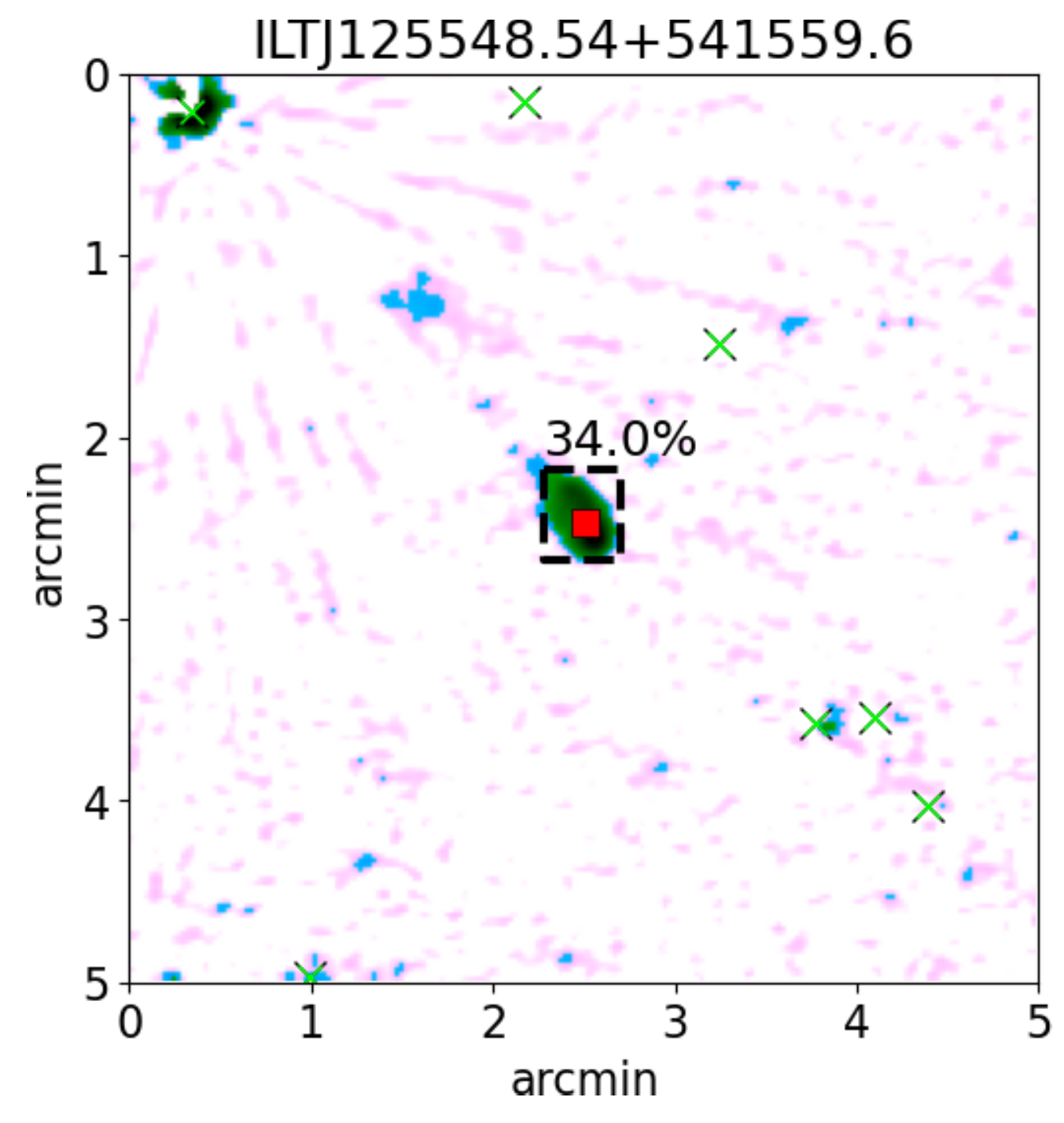}
\includegraphics[width=0.32\textwidth]{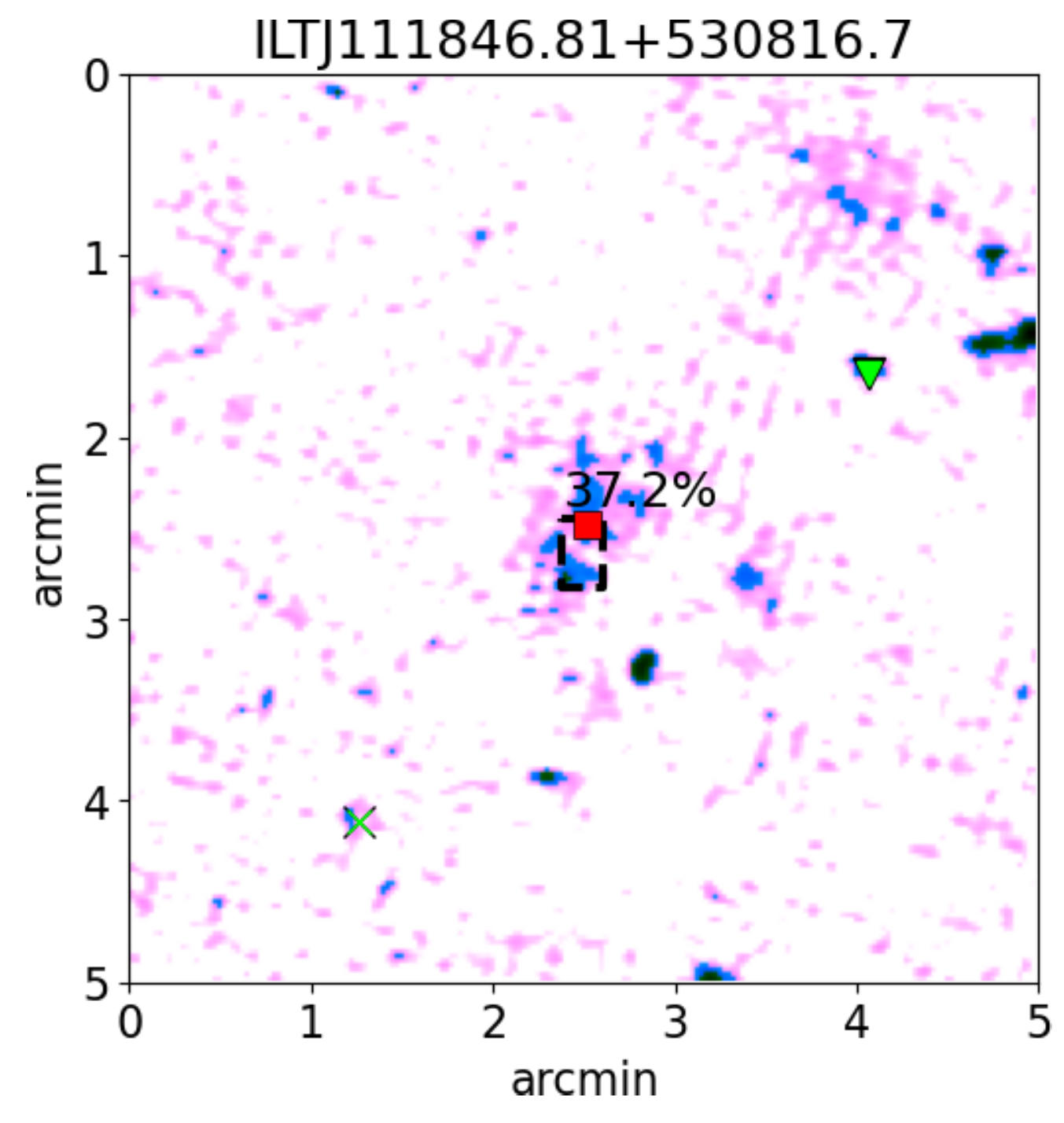}
\includegraphics[width=0.32\textwidth]{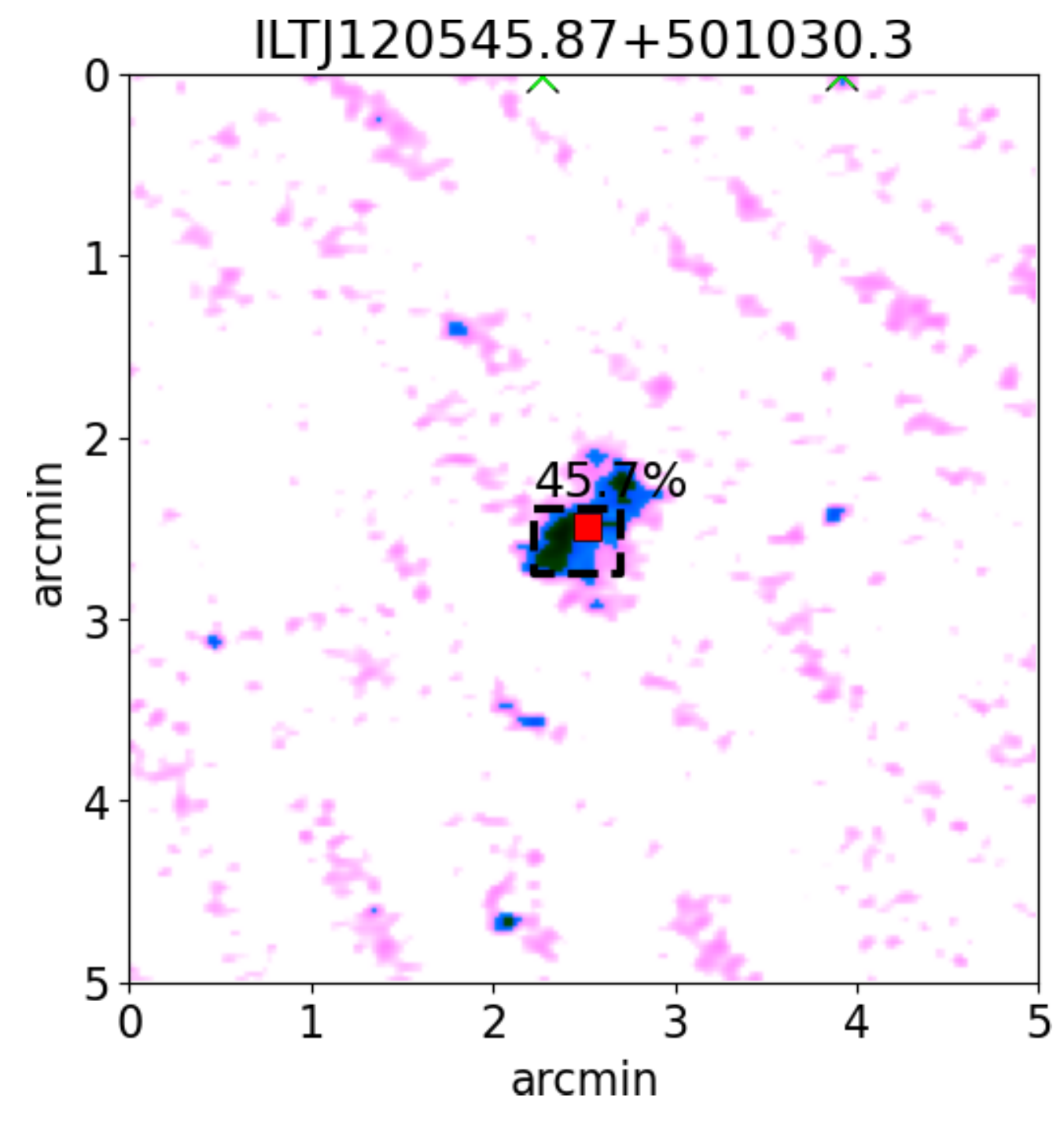}
\caption{Characteristic examples of regions (dashed black rectangles) with a predicted score below $50\%$, sorted by ascending prediction score (see Fig. \ref{fig:good} for an explanation of the markers used).}
\label{fig:low_score}
\end{center}\end{figure*}

Figure \ref{fig:low_score} shows nine sources with regions that have a predicted score below $50\%$. 
These images highlight low scores caused by our method (fixed cutout size, which can be smaller than the entire source; incorrect source removal) and low scores due to faint radio components (for which some components are not detected by \textit{PyBDSF}).
For $1.4\%$ of the radio components in our test set, a score below $50\%$ was predicted.

\end{appendix}

\end{document}